\documentclass[aps,10pt,pre,nofootinbib,citeautoscript,longbibliography,superscriptaddress,twocolumn]{revtex4-2}
\usepackage[T1]{fontenc}
\usepackage[utf8]{inputenc}
\usepackage{graphicx}
\usepackage{amsmath,amssymb,amsfonts,bm,dsfont,units}
\usepackage{hyperref}
\hypersetup{colorlinks=true,citecolor=blue,linkcolor=magenta,urlcolor=blue,pdfstartview=FitH}
\usepackage{braket}
\usepackage[normalem]{ulem} 
\usepackage{units}
\usepackage{soul}
\usepackage{tgtermes}

\usepackage{colortbl}
\usepackage{color}
\usepackage[dvipsnames,table]{xcolor}
\usepackage{bbding}
\usepackage{rotating}
\usepackage{multirow}
\usepackage{tabularx}
\usepackage{enumitem}

\setlist{nosep}
\usepackage{dsfont} 
\usepackage{pifont}
\newcommand{\xmark}{\ding{55}}%

\DeclareMathAlphabet{\mathcald}{U}{dutchcal}{m}{n}
\SetMathAlphabet{\mathcald}{bold}{U}{dutchcal}{b}{n}
\DeclareMathAlphabet{\mathalt}{U}{dutchcal}{b}{n}

\DeclareFontFamily{U}{BOONDOX-calo}{\skewchar\font=45 }
\DeclareFontShape{U}{BOONDOX-calo}{m}{n}{
  <-> s*[1.05] BOONDOX-r-calo}{}
\DeclareFontShape{U}{BOONDOX-calo}{b}{n}{
  <-> s*[1.05] BOONDOX-b-calo}{}
\DeclareMathAlphabet{\mathcalb}{U}{BOONDOX-calo}{m}{n}
\SetMathAlphabet{\mathcalb}{bold}{U}{BOONDOX-calo}{b}{n}
\DeclareMathAlphabet{\mathbcalbx}{U}{BOONDOX-calo}{b}{n}

\usepackage{titlesec}
\titleformat{\title}
	{\bfseries\Huge}
	{}
	{}
	{\MakeUppercase}

\titleformat{\section}[hang]
     {\bfseries}
	 {\thesection.	}
     {1em}
     {\MakeUppercase}
\titlespacing{\section}{0pt}{10pt}{8pt}[0pt]

\titleformat*{\subsection}{\bfseries}
\titlespacing{\subsection}{0pt}{6pt}{6pt}[0pt]

\titleformat{\subsubsection}[hang]
     {\itshape}
     {\thesubsubsection.	}
     {1em}
     {}
\titlespacing{\subsubsection}{0pt}{6pt}{6pt}[0pt]

    {\clearpage\pagebreak[4]\global\pdfpageattr\expandafter{\the\pdfpageattr/Rotate 90}}%
    {\clearpage\pagebreak[4]\global\pdfpageattr\expandafter{\the\pdfpageattr/Rotate 0}}%

\definecolor{orange}{rgb}{1,0.5,0}

\definecolor{dkgreen}{rgb}{0,0.5,0}
\definecolor{midnightblue}{rgb}{0.39,0.58,0.93}

\definecolor{kspink}{RGB}{200,0,200}

\newcommand{\comment}[1]{}{}

\renewcommand{\v}{\boldsymbol}
\newcommand{\id}{\mathds{1}}

\allowdisplaybreaks

\newcommand{\nt}{\notag\\}
\newcommand{\abs}[1]{\left| #1 \right|}

\newcommand{\cCom}{\mathbin{\raisebox{0.5ex}{,}}}
\newcommand{\vk}{{\boldsymbol{k}}}
\newcommand{\vK}{{\boldsymbol{K}}}

\newcommand{\vq}{{\boldsymbol{q}}}
\renewcommand{\vr}{{\boldsymbol{r}}}
\renewcommand{\v}{\boldsymbol}

\newcommand{\veo}{{\boldsymbol{a}_1}}
\newcommand{\vet}{{\boldsymbol{a}_2}}
\newcommand{\C}{{\mathcald{C}}}
\newcommand{\M}{{\mathcald{M}}}
\newcommand{\K}{{\mathcal{K}}}

\newcommand{\fl}{\mathit{fl}}
\newcommand{\mBZ}{{\mathcald{m}\mathrm{BZ}}}
\newcommand{\Tns}{{\mathcald{T}}}
\newcommand{\vG}{{\boldsymbol{G}}}

\newcommand{\h}{{\mathcalb{h}}}

\newcommand{\colA}{\cellcolor{blue!40}}
\newcommand{\colB}{\cellcolor{WildStrawberry!50}}
\newcommand{\colC}{\cellcolor{Green!50}}
\newcommand{\colD}{\cellcolor{orange!50}}
\newcommand{\colE}{\cellcolor{cyan!50}}
\newcolumntype{C}{r<{\kern\tabcolsep}@{}}
\newcolumntype{B}{>{\centering\arraybackslash}X}
\newcommand{\AppCaptions}{\fontfamily{lmr}\linespread{1.}\selectfont}

\makeatletter
\def\maketitle{
\@author@finish
\title@column\titleblock@produce
\suppressfloats[t]}
\makeatother

\makeatletter
\newcommand{\settitle}{\@maketitle}
\makeatother

\makeatletter 
    
\renewcommand\onecolumngrid{
\do@columngrid{one}{\@ne}%
\def\set@footnotewidth{\onecolumngrid}
\def\footnoterule{\kern-6pt\hrule width 1.5in\kern6pt}%
}
\makeatother    

\begin{document}

\title{Gate-defined wires in twisted bilayer graphene: from electrical detection of inter-valley coherence to internally engineered Majorana modes}

\author{Alex Thomson}
\thanks{These authors contributed equally to this work.}
\affiliation{Department of Physics, California Institute of Technology, Pasadena, CA 91125, USA}
\affiliation{Institute for Quantum Information and Matter, California Institute of Technology, Pasadena, CA 91125, USA}
\affiliation{Walter Burke Institute for Theoretical Physics, California Institute of Technology, Pasadena, CA 91125, USA}
\author{Ina M. Sorensen}
\thanks{These authors contributed equally to this work.}
\affiliation{Department of Physics, California Institute of Technology, Pasadena, CA 91125, USA}
\affiliation{Institute for Quantum Information and Matter, California Institute of Technology, Pasadena, CA 91125, USA}
\author{Stevan Nadj-Perge}
\affiliation{Institute for Quantum Information and Matter, California Institute of Technology, Pasadena, CA 91125, USA}
\affiliation{T. J. Watson Laboratory of Applied Physics, California Institute of Technology, Pasadena, CA 91125, USA}
\author{Jason Alicea}
\affiliation{Department of Physics, California Institute of Technology, Pasadena, CA 91125, USA}
\affiliation{Institute for Quantum Information and Matter, California Institute of Technology, Pasadena, CA 91125, USA}
\affiliation{Walter Burke Institute for Theoretical Physics, California Institute of Technology, Pasadena, CA 91125, USA}

\date{\today}

\begin{abstract}
Twisted bilayer graphene (TBG) realizes a highly tunable, strongly interacting system featuring superconductivity and various correlated insulating states.  
We establish gate-defined wires in TBG 
with proximity-induced spin-orbit coupling as $(i)$ a tool for revealing the nature of correlated insulators and $(ii)$ a platform for Majorana-based topological qubits.  
In particular, we show that the band structure of a gate-defined wire immersed in an `inter-valley coherent' correlated insulator inherits electrically detectable fingerprints of symmetry breaking native to the latter.
Surrounding the wire by a superconducting TBG region on one side and an inter-valley coherent correlated insulator on the other further enables the formation of Majorana zero modes---possibly even at zero magnetic field depending on the precise symmetry-breaking order present.
Our proposal not only introduces a highly gate-tunable topological qubit medium relying on internally generated proximity effects, but can also shed light on the Cooper-pairing mechanism in TBG.

\end{abstract}

\maketitle

{\bf \emph{Introduction.}}~Twisted bilayer graphene (TBG) has emerged as a strikingly versatile platform for correlated phenomena \cite{ caoCorrelatedInsulatorBehaviour2018,caoUnconventionalSuperconductivityMagicangle2018,BalentsReview,AndreiReview}.  Near the `magic' twist angle of $\sim  1^\circ$, moiré periodicity and interlayer tunneling conspire to generate energetically isolated flat bands that, when partially filled, allow interactions to dominate \cite{bistritzerMoireBandsTwisted2011}.  
To date experiments have resolved correlation-driven insulators at flat-band fillings of $\nu = 0, \pm 1, \pm 2, \pm 3$ electrons per moiré unit cell ($\nu = \pm 4$ represents full filling/depletion) \cite{caoCorrelatedInsulatorBehaviour2018,yankowitzTuningSuperconductivityTwisted2019,sharpeEmergentFerromagnetismThreequarters2019, luSuperconductorsOrbitalMagnets2019, serlinIntrinsicQuantizedAnomalousHallEffectInAMoireHeterostructure, StepanovUntyingInsulatingSuperconductingOrders2020, SaitoIndependentSuperconductorsCorrelatedInsulators2020, aroraSuperconductivityMetallicTwisted2020,pierce2021unconventional,lyuStrangeMetalBehaviourHallAngleTBG,cao2020nematicity,Liu1261, stepanov2020competing}, reflecting symmetry-breaking electronic instabilities \cite{zondinerCascadePhaseTransitions2020,wongCascadeTransitionsCorrelated2019} whose precise nature remains a largely open question.  Superconductivity is additionally often observed adjacent to $\nu = \pm 2$ \cite{caoUnconventionalSuperconductivityMagicangle2018, yankowitzTuningSuperconductivityTwisted2019, luSuperconductorsOrbitalMagnets2019, StepanovUntyingInsulatingSuperconductingOrders2020, SaitoIndependentSuperconductorsCorrelatedInsulators2020,aroraSuperconductivityMetallicTwisted2020,cao2020nematicity,Liu1261,pierce2021unconventional} and in some samples extends to broader fillings \cite{luSuperconductorsOrbitalMagnets2019}.  Crucially, the phase diagram is not only rich but also exquisitely tunable: due to the giant moiré lattice constant $a_M \sim \unit[10]{\text{nm}}$, altering the electron density by a modest value of $\sim\unit[10^{12}]{\text{cm}^{-2}}$  suffices to sweep the system across metallic, band-insulating,  superconducting, and correlated-insulator phases.

Here we address two ostensibly very different key questions for the field: How can one experimentally reveal the symmetry-breaking order underlying the observed correlated insulators?  And can one exploit the richness and tunability of the TBG phase diagram to construct novel quantum devices for technological applications?  To this end we theoretically explore gate-defined wires in TBG supported by a transition metal dichalcogenide (TMD), e.g., WSe$_2$.  Figure \ref{fig_architecture}(a) sketches the architecture, which features a global back gate and a pair of top gates that enable independent tuning of the density in the central `wire' region and the flanking areas. Recent experiments studied related structures in the context of gate-defined TBG Josephson junctions \cite{devries2020gatedefined,rodanlegrain2020highly}.   In our case, the TMD substrate serves to impart appreciable spin-orbit coupling (SOC)---which plays a pivotal role throughout this paper---to the graphene sheets, as seen in many experiments \cite{Avsar,Wang2015,Yang_2016,PhysRevX.6.041020,PhysRevB.96.041409,Ghiasi,PhysRevB.96.125405,PhysRevB.97.075434,Benitez,PhysRevLett.120.106802,islandSpinOrbitdrivenBand2019,wangQuantumHallEffect2019,wakamuraSpinorbitInteractionInduced2019,aroraSuperconductivityMetallicTwisted2020,PhysRevLett.126.096801}.  Notably, Ref.~\onlinecite{aroraSuperconductivityMetallicTwisted2020} established that TBG on WSe$_2$  continues to display correlated insulators and superconductivity (the latter over a very broad twist-angle window).  Our essential idea is that the gate-defined wire's electronic properties depend sensitively on the TBG phases realized in its vicinity via `internal' proximity effects, and can thus be tailored by electrostatically controlling the flat-band filling on either side.

\begin{figure}
    \centering
    \includegraphics[width=0.99\columnwidth]{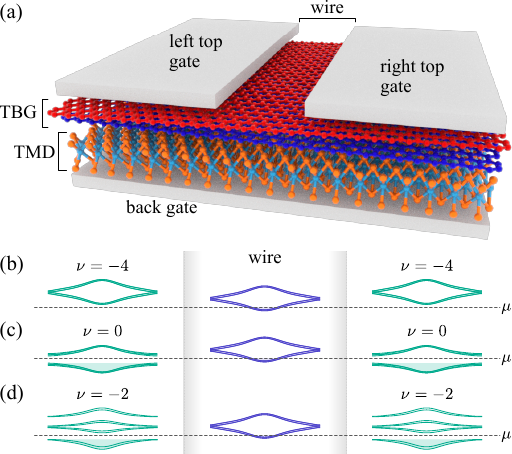}
    \caption{(a) Gate-defined TBG wire architecture, not to scale. (b-d) Schematic flat-band occupations when the wire borders (b) trivial insulators at $\nu = -4$ and correlated IVC insulators at (c) $\nu = 0$ and (d) $\nu = -2$.}
    \label{fig_architecture}
\end{figure}

When immersed within a given correlated insulator, the wire's band structure inherits perturbations that reflect the adjacent symmetry-breaking order. We pay special attention to `inter-valley coherent' (IVC) correlated insulators that are leading candidates for the observed insulating phases at $\nu = 0, \pm 2$ \cite{bultinckGroundStateHidden2020,PhysRevB.102.035136,lian2020tbg,kwan_kekule_2021} and have been proposed as parent states of skyrmion-mediated superconductivity \cite{khalaf2020charged,chatterjee2020skyrmion}; their experimental identification is thus particularly important and promises to illuminate the pairing mechanism in TBG.  In an IVC state, electrons spontaneously develop coherent inter-valley tunneling, thereby breaking translation symmetry on the microscopic graphene (as opposed to moiré) lattice scale \cite{bultinckGroundStateHidden2020}. We show that such ultra-short-scale modulations facilitate generation of band gaps for the wire that would otherwise be forbidden---in turn enabling detection of IVC order via large-scale conductance measurements.  

The presence of IVC order, if indeed confirmed experimentally, further facilitates engineering Majorana zero modes that are widely coveted for fault-tolerant quantum computing \cite{KITAEV20032,RevModPhys.80.1083}.  Majorana zero modes arise at the endpoints of an odd-channel wire gapped via Cooper pairing \cite{Kitaev}.  The well-studied proximitized-nanowire recipe realizes the requisite odd-channel regime through an interplay between Zeeman splitting and SOC that allows gap formation via the proximity effect with a conventional superconductor \cite{PhysRevLett.105.077001,PhysRevLett.105.177002}; in gate-defined TBG wires, valley degeneracies must be removed as well \emph{in a manner conducive to pairing}, posing a nontrivial challenge. The wire band gaps facilitated by proximate IVC order provide precisely the degeneracy lifting needed to open such an odd-channel regime.  Gating one side of the wire into a superconducting phase can then stabilize Majorana zero modes, eschewing the need for `external' superconducting proximity effects almost universally employed in engineered Majorana platforms.  Remarkably, gate-defined TBG wires can potentially harbor Majorana modes even at zero magnetic field depending on details of the IVC order parameter.

\begin{figure*}[t]
\centering
\includegraphics[width=\textwidth]{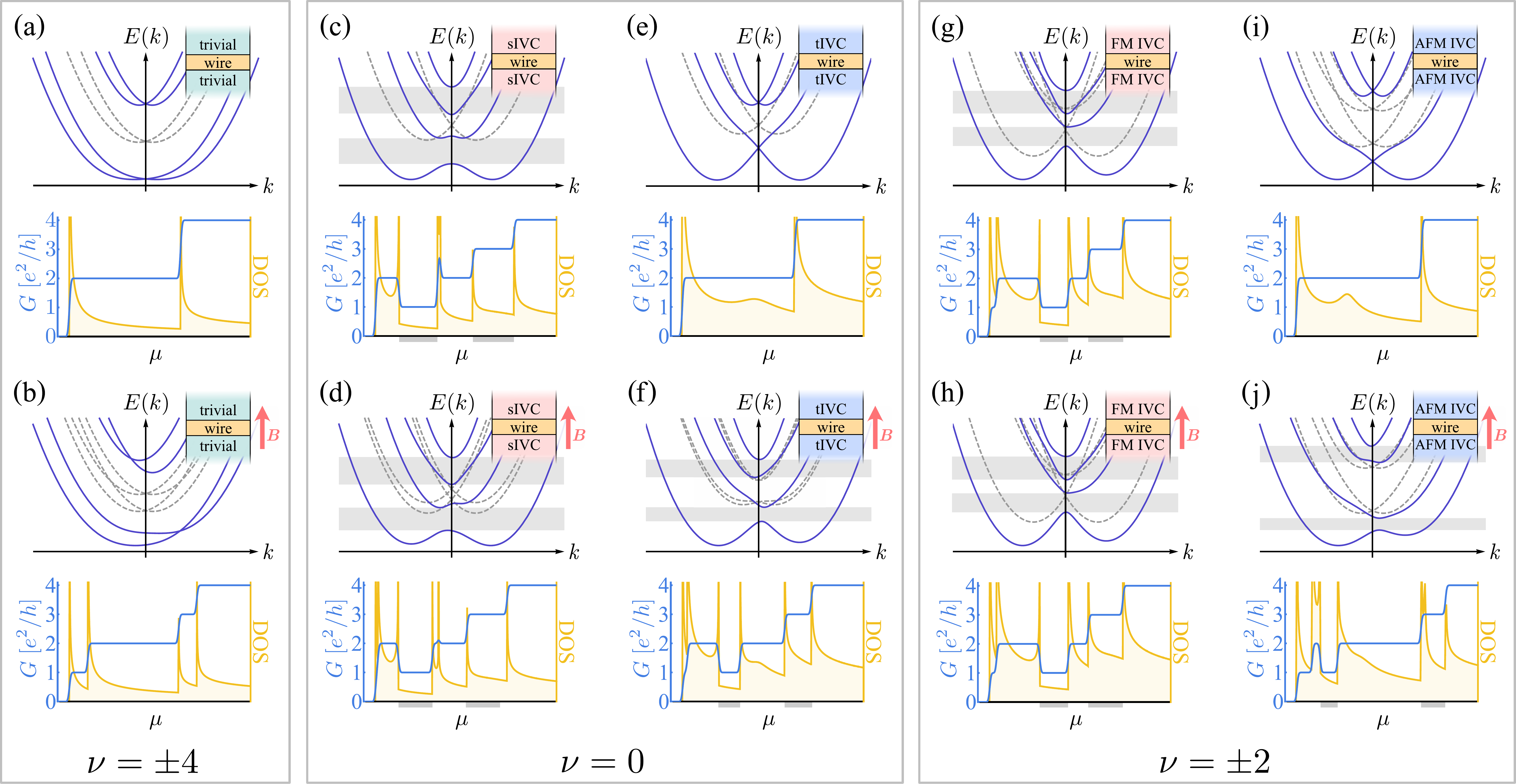}
\caption{Band structure, conductance $G$, and density of states (DOS) for a gate-defined TBG wire immersed in (a,b) trivial band insulators and (c-j) IVC orders.
Insets: top view of wire and proximate phases.  All plots include SOC except dashed-line band structures.  The upper and lower halves respectively correspond to zero and non-zero in-plane magnetic fields. Proximate IVC order facilitates band gaps (shaded rectangles) that manifest as conductance dips 
and associated DOS features within chemical potential windows indicated by grey bars on the $\mu$ axes.
The energy window (vertical axis) shown in each top panel  is equal to the chemical potential interval (horizontal axis) plotted in the corresponding bottom panel.   
}
\label{fig:ExampleBands}
\end{figure*}

{\bf \emph{Trivial wire.}}~We first examine a gate-defined wire surrounded on both sides by trivial $\nu = -4$ band insulators that do not spontaneously break any symmetries (similar results of course hold for $\nu = +4$).  In the wire region the chemical potential resides near the flat-band bottom centered around the $\gamma$ point of the moiré Brillouin zone; see Fig.~\ref{fig_architecture}(b).  Guided by symmetry, we derive a minimal model for the lowest wire subband.  
The TMD substrate breaks SU(2) spin-rotation symmetry as well as $C_2$ symmetry (180$^\circ$ rotations about the out-of-plane axis) and generates both Ising- and Rashba-type SOC in TBG with respective strengths $\lambda_I$ and $\lambda_R$. 
Consequently, the wire preserves only electronic time reversal $\mathcal{T}$ and a U$_{\rm v}(1)$ valley symmetry associated with conservation of $K$ and $K'$ valley quantum numbers (see Refs.~\onlinecite{lopesdossantosGrapheneBilayerTwist2007,bistritzerMoireBandsTwisted2011,po_origin_2018}). 
In terms of momentum-space operators $\psi_k$ for the wire and Pauli matrices $\tau^{x,y,z}$ and $s^{x,y,z}$ that respectively act on the (implicit) valley and spin degrees of freedom, these symmetries transform the operators as 
\begin{align}
  \mathcal{T}: \psi_k \rightarrow i s^y \tau^x \psi_{-k},~~~~~
  {\rm U}_{\rm v}(1)&: \psi_k \rightarrow e^{i \phi \tau^z}\psi_k,
  \label{symmetries}
\end{align}
where $\phi$ is an arbitrary phase and $\tau^z = \pm 1$ correspond to valleys $K$ and $K'$.  

We consider the following $\mathcal{T}$- and ${\rm U}_{\rm v}(1)$-invariant wire Hamiltonian:
\begin{equation}
    H_0 = \int_k\psi_k^\dagger\left(\frac{k^2}{2m}-\mu +c_1k \tau^z + k{\bm \alpha}_1\cdot{\bf s} + \tau^z{\bm \alpha_2}\cdot {\bf s}\right)\psi_k.
    \label{H0}
\end{equation}
Here, $m$ is the effective mass, $\mu$ is the wire's chemical potential, $c_1$ is a `valley-orbit' coupling, and ${\bm \alpha}_{1,2}$ arise from SOC.  
Figure~\ref{fig:ExampleBands}(a) sketches the wire band structure obtained from $H_0$.  Without SOC (dashed lines), the bands for the two valleys are split by valley-orbit coupling $c_1$ but retain two-fold spin degeneracy.  Resurrecting SOC (solid lines) lifts the spin degeneracy; importantly, the remaining band crossings in the spectrum are protected so long as U$_{\rm v}(1)$ is preserved. 
To emphasize this point, Fig.~\ref{fig:ExampleBands}(b) plots the band structure in the presence of a Zeeman term $H_{\rm Z} = \frac{1}{2}g \mu_B\int_k \psi_k^\dagger({\bf B}\cdot{\bf s})\psi_k$ arising from an in-plane magnetic field ${\bf B}$ ($g$ is the electron $g$ factor and $\mu_B$ is the Bohr magneton). Broken time reversal merely shifts the crossings to finite momentum.

{\bf \emph{Wire immersed in $\nu = 0$ IVC order.}}~Suppose that the wire is instead surrounded by correlated insulators emerging at charge neutrality, i.e., $\nu = 0$ [Fig.~\ref{fig_architecture}(c)].  Consider first the case \emph{without} SOC.   There, non-interacting bulk TBG band structure exhibits massless Dirac cones that underpin semimetallicity at $\nu = 0$.  
Hartree-Fock treatments for pristine TBG, by contrast, predict that Coulomb interactions stabilize an insulating ground state at $\nu = 0$ with IVC order \cite{bultinckGroundStateHidden2020,PhysRevB.102.035136,lian2020tbg} (see also Ref.~\onlinecite{TBG_QMC}).  We will discuss spin-singlet and triplet IVC states---respectively denoted sIVC and tIVC in Fig.~\ref{fig:ExampleBands}(c-f)---which are energetically competitive and differentiated by the short-range part of the Coulomb interaction and/or electron-phonon coupling \cite{bultinckGroundStateHidden2020}; both also appear compatible with existing measurements \cite{luSuperconductorsOrbitalMagnets2019}.

Continuing with the spin-orbit-free problem, spin-singlet IVC order spontaneously breaks time-reversal symmetry $\mathcal{T}$ and U$_{\rm v}(1)$ but preserves SU(2) spin rotations as well as an antiunitary operation $\tilde{\mathcal{T}}$ that flips the valley degree of freedom \cite{bultinckGroundStateHidden2020}.  The last symmetry satisfies $\tilde{\mathcal{T}}^2 = -1$ and thus, when present, guarantees Kramers degeneracy.
When acting on our wire fermions $\tilde{\mathcal{T}}$ sends $\psi_k \rightarrow i \tau^y \psi_{-k}$.  Resurrecting SOC generically breaks $\tilde{\mathcal{T}}$ symmetry, as can be seen by its nontrivial action on the ${\bm\alpha}_{1,2}$ terms in Eq.~\eqref{H0}.  An alternative antiunitary symmetry nevertheless persists,
\begin{equation}
    \mathcal{T}_{\rm sIVC}: \psi_k \rightarrow i s^y \tau^y \psi_{-k},
    \label{TsIVC}
\end{equation}
corresponding to $\tilde{\mathcal{T}}$ followed by a spin rotation, which indeed leaves Eq.~\eqref{H0} (and the singlet IVC order parameter characterizing the insulating regions) invariant.  Notice that $\mathcal{T}_{\rm s IVC}^2 = +1$---\emph{implying the demise of Kramers degeneracy with SOC}. Accordingly, the wire band structure in the presence of proximate singlet IVC order [Fig.~\ref{fig:ExampleBands}(c)] maintains $k\leftrightarrow -k$ symmetry but generically features no band crossings.  In-plane magnetic fields modify the band gaps and inject $k\leftrightarrow -k$ asymmetry as Fig.~\ref{fig:ExampleBands}(d) illustrates.

Without SOC, spin-triplet IVC order spontaneously breaks SU(2) spin symmetry and U$_{\rm v}(1)$ yet preserves both  $\mathcal{T}$ and  $\tilde{\mathcal{T}}$.  Reviving SOC once again breaks $\tilde{\mathcal{T}}$, but unlike the singlet IVC case we cannot append a spin rotation to obtain a proper symmetry because triplet IVC order breaks spin SU(2).  The system then preserves only the familiar electronic time-reversal symmetry $\mathcal{T}$---which satisfies $\mathcal{T}^2 = -1$ and underpins Kramers degeneracy---implying that proximity to triplet IVC order preserves the crossings in the band structure at $k = 0$ [Fig.~\ref{fig:ExampleBands}(e)], similar to the trivial wire case.  
Contrary to the latter problem, however, the loss of valley conservation from triplet IVC order allows in-plane magnetic fields to eliminate these band crossings; see Fig.~\ref{fig:ExampleBands}(f). 

{\bf \emph{Wire immersed in $\nu = \pm 2$ IVC order.}}~Next we immerse the wire within a $\nu = \pm 2$ correlated insulator [Fig.~\ref{fig_architecture}(d)].  The commonly observed insulating states at these fillings have also been predicted to display IVC order \cite{bultinckGroundStateHidden2020,PhysRevB.102.035136,lian2020tbg}.  Insulating IVC states at $\nu = -2$ ($+2$) can arise upon completely depleting (filling) two of the fourfold-degenerate flat bands, and then gapping the remaining `active' carriers via spontaneous inter-valley hybridization.  
We consider in detail two candidate phases that, without SOC, correspond to $(i)$ a ferromagnetic (FM) IVC state with 
active carriers spin-polarized in the out-of-plane $(s^z)$ direction
and 
$(ii)$ an `antiferromagnetic' (AFM) state with  
active carriers consisting of $s^z = +1$ electrons from one valley and $s^z = -1$ electrons from the other. 
The former state may be relevant to Ref.~\onlinecite{lin2021proximityinduced}---which reported ferromagnetism and an anomalous Hall effect at $\nu = 2$ in TBG on WSe$_2$---whereas $\nu = \pm 2$ insulators observed elsewhere appear compatible with the latter as argued in Ref.~\onlinecite{lake2021reentrant}.  

In the spin-orbit-free problem, both IVC orders  spontaneously violate SU(2) spin symmetry and U$_{\rm v}$(1).  The FM IVC state also breaks $\mathcal{T}$ but preserves $\tilde{\mathcal{T}}$; conversely, the AFM IVC state preserves $\mathcal{T}$ but violates $\tilde{\mathcal{T}}$.  Turning on SOC breaks $\tilde{\mathcal{T}}$ for the FM IVC state, and (just like the $\nu = 0$ triplet IVC) one cannot append a spin rotation to obtain a modified symmetry because the order parameter breaks spin SU(2). Hence the wire preserves no symmetries when proximate to ferromagnetic IVC order and only $\mathcal{T}$ in the antiferromagnetic IVC case.  Figures~\ref{fig:ExampleBands}(g)-(j) present the wire band structures resulting from proximate FM and AFM IVC order, both with zero (g,i) and non-zero (h,j) in-plane magnetic fields.  The band structures resemble those generated by $\nu = 0$ singlet and triplet IVC order, respectively, though ferromagnetic IVC order breaks $k\leftrightarrow -k$ symmetry in the band structure even at zero field due to the absence of $\mathcal{T}_{\rm sIVC}$ symmetry.

\newlength\mylenA 
\setlength\mylenA{0.175\textwidth}
\addtolength\mylenA{-2\tabcolsep}
\newlength\mylenB 
\setlength\mylenB{0.65\textwidth}
\addtolength\mylenB{-2\tabcolsep}
{\renewcommand{\arraystretch}{1.7}
\begin{table*}
  \centering
\begin{tabular}{p{\mylenA} p{\mylenA} p{\mylenB}}
    \multicolumn{1}{l}{Proximate Order} 
    & \multicolumn{1}{l}{Wire Symmetries} 
    & \multicolumn{1}{l}{Wire Perturbations} 
	\\ \hline\hline
	trivial wire
    & $\mathcal{T}^2=-1$, $\mathrm{U}_\mathrm{v}(1)$
    & 
	$\frac{k^2}{2m}-\mu+ck\tau^z+k\boldsymbol{\alpha}_1\cdot\mathbf{s}+\tau^z\boldsymbol{\alpha}_2\cdot\mathbf{s}$
	\\\hline
    $\nu = 0$ singlet IVC 
	& $\mathcal{T}_{\rm sIVC}^2 = +1$ 
	& 
	$a_1 k \tau^x + 
    \tau^x\left(\beta_1^xs^z+\beta_1^ys^y\right)+
	\beta_2\tau^xs^z $ 
	\\
    $\nu = 0 $ triplet IVC & $\mathcal{T}^2 = -1$ 
	& 
	$a_1'k\tau^xs^z + k\tau^y\left(\beta_1'^x s^x+\beta_1'^ys^y\right) + \beta_2'\tau^x $ 
	\\ 
	$\nu = \pm 2$ FM IVC 
	& none 
	& 
	$a_1 k \tau^x + a_1'k\tau^xs^z+ a_1''k\tau^zs^z+ a_2''s^z 
	+\tau^x(\beta_1^xs^x+\beta_1^ys^y) + \beta_2\tau^xs^z$
	\newline
	$+ k\tau^y(\beta_1'^xs^x+\beta_1'^ys^y)+\beta_2'\tau^x
	+\beta_1''k + \beta_2''\tau^z$
	\\ 
	$\nu = \pm 2$ AFM IVC 
	& $\mathcal{T}^2 = -1$  
	& $a_1'''k(\tau^xs^x+\tau^ys^y) + a_2'''k s^z+a_3'''\tau^zs^z 
	+ k\tau^x\boldsymbol{\beta}'''_1\cdot\mathbf{s}+ k\tau^y\boldsymbol{\beta}'''_2\cdot\mathbf{s}+ \beta_3'''\tau^x+\beta_4'''\tau^y $
  \end{tabular}
  \caption{
   Wire Hamiltonian terms for the trivial wire (first row) and perturbations generated by proximate IVC orders (subsequent rows).  
    Couplings labelled by $a$'s are generated by IVC order in the absence of SOC, 
   whereas ${\beta}$ terms can be viewed as additional IVC order-parameter components generated due to SOC, akin to the spin-orbit-induced admixture of singlet and triplet pairing in inversion-asymmetric superconductors \cite{PhysRevLett.87.037004}. 
   For $\nu = 0$ triplet and $\nu = \pm 2$ IVC orders, we assumed that, without SOC, the spins orient in the out-of-plane ($\pm s^z$) direction. 
  }
  \label{WireTable}
\end{table*}
}

Appendices~\ref{sec:ContMod_Int_NoSOC}$-$\ref{sec:IVCwire} complement the preceding symmetry-based analysis by deriving the dominant wire-Hamiltonian terms induced by proximate IVC states and SOC at first order in $\lambda_{I,R}$.  Table~\ref{WireTable} summarizes the results.  The band structures in Fig.~\ref{fig:ExampleBands} were obtained using the corresponding perturbations.  Appendix~\ref{sec:5bndmodel} also validates qualitative features of the wire band structures using  microscopic five-band model simulations.

{\bf \emph{Experimental IVC detection.}}~Electrical transport provides a straightforward diagnostic of the hallmark IVC-mediated wire band gaps. 
Blue curves in the lower panels of Fig.~\ref{fig:ExampleBands} sketch the zero-bias, zero-temperature conductance $G$ versus wire chemical potential $\mu$ assuming ballistic transport.  Most strikingly, proximate IVC order generates conductance dips (e.g., re-entrant $e^2/h$ plateaus) in Figs.~\ref{fig:ExampleBands}(c,d,f,g,h,j) associated with band-gap-induced reduction in the number of conducting channels; whether these dips appear at zero or finite magnetic fields additionally constrains the IVC spin structure.   Similar experiments have been conducted in semiconductor-based wires to detect odd-channel `helical' regimes driven by an interplay between SOC and magnetic fields \cite{SOgap,SOgap2}.  In our case the analogous odd-channel regimes additionally require inter-valley coherence but, interestingly, do not necessarily require a magnetic field [Figs.~\ref{fig:ExampleBands}(c,g)].  
Singlet IVC order at $\nu = 0$ can be identified even when the `dips' widen such that the conductance rises monotonically in $e^2/h$ increments, like the trivial case with $B\neq 0$ in Fig.~\ref{fig:ExampleBands}(b).  Indeed the complete lifting of band degeneracy at zero field combined with a vanishing bulk Hall conductance dictated by $\mathcal{T}_{\rm sIVC}$ symmetry distinguishes singlet IVC order from a state that breaks $\mathcal{T}$ but preserves U$_{\rm v}(1)$.
Ballistic conduction is inessential provided the conductance features highlighted above remain visible. 
The IVC-mediated gaps also qualitatively modify the density of states [lower panels of Figs.~\ref{fig:ExampleBands}(c,d,f,g,h,i), yellow curves] and can be  detected using scanning tunneling microscopy.

{\bf \emph{Internally engineered Majorana modes.}}~Shaded rectangles in the  Fig.~\ref{fig:ExampleBands} band structures indicate IVC-mediated odd-channel regimes that can be harvested for Majorana modes. Imagine now gating one side of the wire into a superconductor (Fig.~\ref{fig:SCphases} insets), which we assume is gapped \cite{rodanlegrain2020highly} and  pairs time-reversed partners.  
For accessing topological superconductivity it suffices to consider the proximity-induced wire pairing perturbation
\begin{equation}\label{eqn:ProximateSC}
    \delta H_{\rm SC} =  \frac{1}{2}\int_k [\psi_k^T(\Delta_1\tau^x s^y +\Delta_2 \tau^y s^x)\psi_{-k} + h.c.]
\end{equation}
with $\Delta_{1,2}\in \mathbb{R}$. 
The first term is a spin singlet, valley triplet, while the second is a spin triplet,  valley singlet; both preserve U$_{\rm v}(1)$ and $\mathcal{T}$.

Figure~\ref{fig:SCphases} illustrates the phase diagram versus $\mu$ and in-plane magnetic field $B$ at fixed $\Delta_2 = -0.4\Delta_1$ for a wire bordered on the other end by (a) $\nu = 0$ singlet IVC order and (b) $\nu = \pm2$ AFM IVC order.  Band structure parameters are the same as for the corresponding panels in Fig.~\ref{fig:ExampleBands}. 
The topological phases (labeled `topo') descending from odd-channel regimes host unpaired Majorana zero modes---which we confirm by simulating the wire model on a lattice with open boundaries.  Non-zero $\Delta_2$ enables the upper topological phase in Fig.~\ref{fig:SCphases}(a) and reduces somewhat the critical field for topological superconductivity in Fig.~\ref{fig:SCphases}(b).  Extended gapless regions arise due to suppression of pairing by field-induced $k\leftrightarrow-k$ band asymmetry and, for parameters chosen here, prevent a topological phase from emerging in the upper odd-channel regime in Fig.~\ref{fig:ExampleBands}(j). Strikingly, in Fig.~\ref{fig:SCphases}(a) topological superconductivity extends down to \emph{zero magnetic field} due to internal $\mathcal{T}$-breaking by the proximate singlet IVC order.  Similar behavior is expected from proximate $\nu = \pm2$ FM IVC order.  In (b),  topological superconductivity appears only at $B \neq 0$ since $\nu = \pm 2$ AFM IVC order preserves $\mathcal{T}$; $\nu = 0$ triplet IVC order shares this property and yields a similar phase diagram.  The formation of Majorana modes in the latter cases may be assisted by interaction-enhancement of graphene's nominally small $g$ factor \cite{Stoudenmire} as well as SOC-induced broadening of the field interval over which superconductivity survives in TBG.  Moreover, the field orientation comprises a practical tuning knob that can be used to optimize topological superconductivity: the optimal orientation depends on a non-universal interplay between the IVC order, SOC parameters, and wire geometry.

\begin{figure}[t]
\centering
\includegraphics[width=0.99\columnwidth]{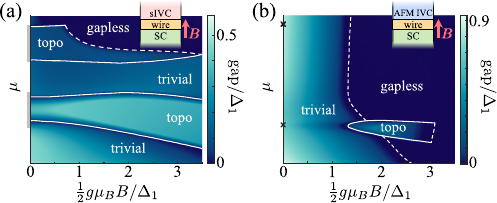}
\caption{Phase diagram for a wire bordered on one side by a superconducting TBG region and on the other by (a) $\nu = 0$ singlet IVC order and (b) $\nu = \pm 2$ AFM IVC order.  The color scale shows the wire's excitation gap.  `Topo' indicates topological superconductivity hosting unpaired Majorana zero modes.  Grey bars on the $\mu$ axis of (a) correspond to odd-channel regimes highlighted in  Fig.~\ref{fig:ExampleBands}(c), which give way to Majorana modes even at $B = 0$. 
The crosses on the $\mu$ axis of (b) label the energies at which the Kramers-enforced band crossings occur at $B=0$ in Fig.~\ref{fig:ExampleBands}(i).
Appendix~\ref{sec:WireProxCoupledSC} presents additional phase diagrams illustrating the dependence on $\Delta_{1,2}$.}
\label{fig:SCphases}
\end{figure}

{\bf \emph{Outlook.}}~Electrical detection of IVC order as envisioned here would not only provide a critical test for skyrmion-mediated superconductivity \cite{khalaf2020charged,chatterjee2020skyrmion}, but also lay the groundwork for topological qubit applications.  We stress that our proposed experiments extend to other types of IVC states beyond those examined above.  
Notably, recent Hartree-Fock simulations \cite{kwan_kekule_2021} predict that physically plausible strain levels stabilize a different IVC phase---the intervalley Kekulé spiral (IKS) state---at $\nu=\pm2$.
Like the AFM IVC state, IKS order preserves $\mathcal{T}$ but violates $\tilde{\mathcal{T}}$ and $\mathrm{U_v}(1)$.
The band structure and effective Hamiltonian of a wire immersed within IKS order (supplemented by SOC) thus takes the same generic form as with proximate AFM IVC order.
Furthermore, spin-polarized IKS states are proposed at $\nu=\pm1,\pm3$ and appear to be compatible with the experiments of Refs.~\onlinecite{yankowitzTuningSuperconductivityTwisted2019,stepanov2020competing}.
These states accordingly break $\mathcal{T}$ and spin $\mathrm{SU}(2)$ in addition to $\tilde{\mathcal{T}}$ and $\mathrm{U_v}(1)$; the band structure and effective Hamiltonian of a wire proximitized by these orders in turn  mimics the FM IVC case.  
Thus the IVC diagnostics outlined earlier extend straightforwardly to these cases.

Our proposed gate-defined wire platform offers numerous virtues for Majorana engineering: ease of gate-tunability, internal proximity effects that circumvent interface issues accompanying the merger of disparate materials, real-time control over the arrangement of phases in the device, and amenability to transport and various local probes.  Extensions to twisted trilayer graphene \cite{TTGtheory} are particularly interesting to pursue in future work given that superconductivity persists to higher temperatures \cite{TTGexpt1,TTGexpt2} and withstands O(10T) in-plane magnetic fields \cite{TTGexpt3}.  More generally, we anticipate that gate-defined wires in twisted heterostructures can be broadly employed to diagnose symmetry-breaking order and for quantum devices.

{\bf \emph{Acknowledgements.}}~We are grateful to Cory Dean, Ethan Lake, Cyprian Lewandowski, T. Senthil, and Andrea Young for
illuminating discussions. This work was supported by the
Army Research Office under Grant Award W911NF17-
1-0323; the U.S. Department of
Energy, Office of Science, National Quantum Information Science Research Centers, Quantum Science Center; the National Science Foundation through grant
DMR-1723367; an Aker Scholarship; the Caltech Institute for Quantum Information and Matter, an NSF
Physics Frontiers Center with support of the Gordon
and Betty Moore Foundation through Grant GBMF1250; and the Walter Burke Institute for Theoretical Physics at Caltech.

\hbadness=10000

\let\oldaddcontentsline\addcontentsline
\renewcommand{\addcontentsline}[3]{}
\bibliography{references}
\let\addcontentsline\oldaddcontentsline

\newpage
\appendix 
\fontfamily{lmr}\selectfont
\renewcommand{\rmdefault}{lmr}

\title{\Large Supplemental material}

\maketitle

\onecolumngrid

\tableofcontents
\linespread{1.3}
\fontsize{10pt}{11pt}
\selectfont

\section{Continuum model without spin orbit coupling}\label{sec:ContMod_Int_NoSOC}

\subsection{Continuum model definition}\label{sec:ModelDef}

We begin by giving a short summary of the continuum model \cite{lopesdossantosGrapheneBilayerTwist2007,bistritzerMoireBandsTwisted2011} of twisted bilayer graphene (TBG) 
in the absence spin orbit coupling (SOC).
We let the operators $d_{t/b}(\vk)$ denote electron annihilation operators in the momentum basis residing on the top ($t$) and bottom ($b$) graphene sheets, which we take to be rotated by an angle $\theta\ll1$ from one another.
In defining $d_{t/b}$, both sublattice and spin indices have been suppressed.
The regime of interest occurs at very low energies, and it is therefore sufficient to restrict our study to the low-energy excitations of each of the graphene monolayers.
These states take the form of Dirac fermions at the Brillouin zone (BZ) corners $K$ and $K'=-K$, and
we therefore define $\Psi_{K^{(\prime)},t/b}(\vk)=d_{t/b}(\vk+\vK^{(\prime)}_{t/b})$, where $K^{(\prime)}_t$ and $K^{(\prime)}_b$ differ by a small amount as a result of the twist angle offset between the two layers.

The continuum model Hamiltonian may be expressed as
\begin{align}\label{eqn:ContModDef}
H_\mathrm{cont}
&=
H_t+H_b+H_\mathrm{tun}.
\end{align}
The first two terms on the right-hand side respectively denote the Dirac Hamiltonian of the top and bottom layers in the absence of tunnelling:
\begin{align}
H_{t/b}&=\int_\vk \Psi^\dagger_{t/b}(\vk)\h_{t/b}(\vk)\Psi_{t/b}(\vk),
\end{align}
where
\begin{align}\label{eqn:DiracHamDefs}
\h_t(\vk)
&=-v_0e^{i\theta\tau^z\sigma^z/4}(k_x\tau^z\sigma^x+k_y\sigma^y)e^{-i\theta\tau^z\sigma^z/4},
&
\h_b(\vk)
&=-v_0e^{-i\theta\tau^z\sigma^z/4}(k_x\tau^z\sigma^x+k_y\sigma^y)e^{i\theta\tau^z\sigma^z/4}.
\end{align}
Here, $\tau^{x,y,z}$ and $\sigma^{x,y,z}$ act on the (suppressed) valley and sublattice indices, respectively.
The Fermi velocity is approximately $v_0\sim\unit[10^6]{\text{m/s}}$ \cite{CastroNeto07}.
The layers tunnel through
\begin{align}\label{eqn:HtunDefs}
H_\mathrm{tun}
&=\sum_{\ell=1,2,3}\sum_{v=K,K'}\int_\vk \Psi^\dagger_{v,t}(\vk)T^{(v)}_\ell\Psi_{v,t}(\vk+\eta_v\vq_\ell)+h.c.,
\end{align}
where $\eta_K=+1$, $\eta_{K'}=-1$. 
All other indices are suppressed.
The momenta exchanged are defined as
\begin{align}
\vq_\ell&=k_\theta\left( -\sin\left[\frac{2\pi}{3}\left(\ell-1\right)\right]\hat{\boldsymbol{x}} + \cos\left[\frac{2\pi}{3}\left(\ell-1\right)\right]\hat{\boldsymbol{y}}\right),
&
k_\theta
&=\frac{4\pi}{3a}2\sin(\theta/2),
\end{align}
while the tunnelling matrices themselves take the form
\begin{align}\label{eqn:InterlayerTunnellingDef}
T^{(\pm)}_\ell
&=
w_0
+
w_1\left( e^{\mp 2\pi(\ell-1) i/3}\sigma^+ + e^{\pm 2\pi i (\ell-1)/3} \sigma^- \right).
\end{align}
The tunnelling parameters are typically taken to be $(w_0,w_1)=\unit[(85,108)]{\text{meV}}$ \cite{bistritzerMoireBandsTwisted2011,NamN17,po_origin_2018,Carr19}.

Note that this model allows no tunnelling between valleys $K$, $K'$. 
This assumption is valid when the twist angle is very small, $\theta\ll1$, as considered here, since it then follows that the momentum $\vq_\ell$ is much smaller than the magnitude of the momentum separating the two valleys.
Below, this absence of inter-valley tunnelling is formulated in terms of an emergent U(1) symmetry.

In order to diagonalize $H_\mathrm{cont}$, we first observe that because the momentum arguments of $\Psi_{K^{(\prime)},t/b}(\vk)$ are measured relative to distinct momenta $K_{t}^{(\prime)}\neq K_b^{(\prime)}$, the interlayer tunnelling term in Eq.~\eqref{eqn:HtunDefs} does not in fact involve a momentum transfer of $\pm \vq_\ell$ between layers.
It's useful to move to a representation expressed in terms of operators  $\tilde{\Psi}_{K^{(\prime)},t/b}(\v{k})$ identical to the previous notation save that the momentum arguments of these new operators are now measured relative to a common point. 
In term of the $\tilde{\Psi}$ operators, the Hamiltonian $H_\mathrm{cont}$ may be written in a form where it's manifestly apparent that it only involves momentum exchanges by a set of Bravais lattice vectors $\v{G}$ defined by basis vectors
$\vG_1=\vq_1-\vq_2$ and $\vG_2=\vq_2-\vq_3$.
As an example, we might define, for valley $K$,
\begin{align}\label{eqn:UVElectronOperatorReDef}
	\tilde\Psi_{K,t}(\vk)
	&=
	\Psi_{K,t}(\vk),
	&
	\tilde{\Psi}_{K,b}(\vk)
	&=
	\Psi_{K,b}(\vk-\vq_1).
\end{align}
Regardless of the specific conventions chosen,
the Hamiltonian may then be written as
\begin{align}\label{eqn:HcontBravaisDef}
	H_\mathrm{cont}&=
	\sum_{v=K,K'}\sum_{\vG,\vG'}
	\int_{\vk\in\mBZ} \tilde\Psi^\dag_{v,i}(\vk+\vG)\h^{(v)}_{i,\vG;j,\vG'}(\vk)\tilde\Psi_{v,j}(\vk+\vG'),
\end{align}
where $v$ labels the $K$-valley, $i,j$ are combined indices including layer and sublattice, and the
spin index is suppressed. 
The momentum integration over $\vk$ only includes values within the moiré Brillouin zone (denoted `$\mBZ$') defined by the Bravais lattice vectors $\vG$.

Diagonalizing $H_\mathrm{cont}$ one obtains an infinite set of bands both above and below charge neutrality.
Our focus will be the bands closest to charge neutrality.
It can be shown that these bands will also possess Dirac cones located at the $\mBZ$ corners, which we denote $\kappa$ and $\kappa'$ to distinguish them from the Brillouin zone corners of the graphene monolayers.
By considering the limit of infinitesimally small interlayer tunnelling ($w_{0,1}\to0$ in Eq.~\eqref{eqn:InterlayerTunnellingDef}), one can show that for valley $K$, the Dirac cone at $\kappa$ descends from the top layer, $K_t$, while the Dirac cone at $\kappa'$ descends from the bottom layer, $K_b$. 
Similarly, for valley $K'$, the Dirac cone at $\kappa$ descends from the bottom layer, $K_b'$, while the Dirac cone at $\kappa'$ descends from the top layer, $K_t'$.

At the magic angle, $\theta\sim1.1^\circ$, the bands above and below charge neutrality become nearly completely flat, allowing interactions to dominate.
Further, provided $w_0<w_1$, as is believed to be the case, the flat bands are also isolated from the `dispersive' bands at higher or lower energy by a gap $\Delta_\mathrm{disp}$.
The filling of the flat bands is expressed via the filling factor $\nu$: all flat bands are empty at $\nu=-4$, all flat band are filled at $\nu=+4$, and $\nu=0$ corresponds to charge neutrality.

\subsection{Interactions}\label{sec:InteractionTerms}

The primary source of interactions believed to be relevant to TBG is the Coulomb interaction:
\begin{align}
    H_\mathrm{int}
    &=
    \frac{1}{2}
    \int d^2\vq\,\rho^\dag(\vq)V(\vq)\rho(\vq),
\end{align}
where the Coulomb potential satisfies $V(\vq)\propto 1/\abs{\vq}$.
In terms of the microscopic graphene operators introduced at the beginning of the previous section, the density operator takes the form $\rho(\vq)=\int_\vk d^\dag(\vk)d(\vk+\vq)$.
It is sufficient to focus on the low-energy states of the graphene monolayers, which is equivalent to restricting the momentum arguments of $d(\vk)$ to values close to $K$, $K'$, or, equivalently, writing everything in terms of the $\Psi(\vk)$ operators defined above, with $\abs{\vk}$ smaller than the momentum different $\abs{K-K'}=K$.
As a result only density operators whose arguments are either very small or are very close to $K$, $K'$ are important:
\begin{align}
    \rho(\vq)
    &\cong
    \sum_{v=K,K'}\int_\vk 
    \Psi^\dag_v(\vk)\Psi_v(\vk+\vq),
    &
    \rho(\vq+\vK)
    &\cong
    \int_\vk
    \Psi^\dag_K(\vk)\Psi_{K'}(\vk+\vq),
\end{align}
where $\abs{\vq}\ll K$.
In line with the reasoning above, the integration over $\vk$ is restricted to momenta that are small compared to $K$.
These expressions imply that the interaction term $H_\mathrm{int}$ may be separated into two pieces: $H_\mathrm{int}\cong H_C+H_J$ with
\begin{align}\label{eqn:CoulombIntExpansion}
    H_C
    &=
    \frac{1}{2}\int_\vq \rho^\dag(\vq)V(\vq)\rho(\vq),
    &
    H_J
    &=
    \int_\vq \rho^\dag(\vq+\vK)V(\vq+\vK)\rho(\vq+\vK).
\end{align}
Again, $\vq$ is restricted to small momenta. 
Given the form of $V(\vq)$ quoted above, it is easy to demonstrate that the magnitude of the second term, which we will denote the Hund's term, is suppressed by a factor of $\sim\theta\ll1$.

\subsection{Symmetry action on microscopic operators}\label{sec:SymmetryAction}

We now summarize the symmetries that are present in the system. 
We work with the operators $\Psi(\vk)$, but note that the momentum-shifted operators $\tilde{\Psi}(\vk)$ defined in the previous section transform in an identical fashion.

In addition to the usual $\mathrm{U_c}(1)$ symmetry associated with charge conservation, TBG without SOC possesses two continuous symmeties, a $\mathrm{U_v}(1)$ valley symmetry and the SU(2) spin symmetry:
\begin{align}\label{eqn:ContinuousSymmetryAction}
	\mathrm{U}_\mathrm{v}(1)&:\qquad\Psi(\vk)\to e^{i\phi\tau^z}\Psi(\vk),
	\nt
	\mathrm{SU}(2)&:\qquad\Psi(\vk)\to e^{i\theta(\hat{\v{n}}\cdot\v{s})/2}\Psi(\vk),
\end{align}
where $\tau^{x,y,z}$ ($s^{x,y,z}$) Pauli matrices act on the valley $K$ indices (spin indices) and $\hat{\v{n}}$ is a unit vector.
The valley symmetry is a direct consequence of the (physically correct) omission of tunnelling terms between fermions originating in valley $K$ and those of valley $K'$ in Sec.~\ref{sec:ModelDef}.
In fact, both $H_\mathrm{cont}$ and $H_C$ are invariant under a much larger continuous symmetry group, $\mathrm{U(2)\times U(2)}\cong \mathrm{U_c}(1)\times\mathrm{U_v}(1)\times\mathrm{SU(2)_K\times\mathrm{SU}(2)_{K'}}$, where $\mathrm{SU}(2)_{K/K'}$ correspond to \emph{independent} spin rotations in valleys $K/K'$:
\begin{align}\label{eqn:EnlargeContSym}
\mathrm{SU}(2)_K\times\mathrm{SU}(2)_{K'}&:\qquad
\Psi(\vk)
\to
\left( \mathcal{P}_K e^{i\phi_+\hat{\v{n}}_{K}\cdot\v{s}/2}
+
\mathcal{P}_{K'} e^{i\phi_-\hat{\v{n}}_{K'}\cdot\v{s}/2}
\right)\Psi(\vk).
\end{align}
Here, $\hat{\v{n}}_{K^{(\prime)}}\in S^2$ are arbitrary unit vectors and $\mathcal{P}_K=(\id+\tau^z)/2$ and $\mathcal{P}_{K'}=(\id-\tau^z)/2$ project onto the $K$ and $K'$ valleys, respectively.
This symmetry is only broken once the effects of the Hund's interaction in Eq.~\eqref{eqn:CoulombIntExpansion} is taken into account.

There are a number of additional discrete symmetries, as well as time reversal.
While all of these symmetry transformations should be composed with spin rotations, it is convenient in this appendix to separate the internal degrees of freedom deriving from spin from those of the discrete symmetry operations.
We therefore consider the action of ``spinless" versions of the discrete symmetries:
\begin{align}\label{eqn:NoSpinSym}
\Tns&:\qquad\Psi(\vk)\to \tau^x\Psi(-\vk),\qquad i\to-i,
\nt
\C_2&:\qquad \Psi(\vk) \to \tau^x\sigma^x \Psi(-\vk),
\nt
\C_2\Tns&:\qquad\Psi(\vk)\to \sigma^x\Psi(\vk),\qquad i\to-i,
\nt
\C_3&:\qquad\Psi(\vk)\to e^{-2\pi i \tau^z \sigma^z /3}\Psi(R_3\vk),
\nt
\M_y&:\qquad\Psi(\vk)\to \mu^x\sigma^x \Psi(R_y\vk),
\end{align}
where $\Psi(\vk)$ are the 16 component electron annihilation operators of the two graphene layers and the Pauli matrices $\tau^{x,y,z}$, $\sigma^{x,y,z}$, and $\mu^{x,y,z}$ act on $K$-valley, sublattice, and layer indices respectively.
The matrices $R_3$ and $R_y$ are given by
\begin{align}\label{eqn:R3RyDef}
R_3&=
\begin{pmatrix}
  -1/2  & -\sqrt{3}/2 \\  \sqrt{3}/2  & -1/2
\end{pmatrix},
&
R_y
&=
\begin{pmatrix} 
	1	&	0	\\
	0	&	-1	
\end{pmatrix}.
\end{align}
We have redundantly included the composite symmetry $\C_2\Tns$ as it commutes with the $\mathrm{U_v}(1)$ symmetry, making it especially useful when considering only a single $K$-valley.

The symmetries listed above are not the true physical symmetries of the problem, and they will clearly no longer be preserved when SOC is included below. 
The physical symmetries may be expressed as
\begin{align}	\label{eqn:PhysicalSymDef}
	\mathcal{T}&=is^y \Tns,
	&
	C_2&=is^z\C_2,
	&
	C_2\mathcal{T}&=is^x \C_2\Tns,
	\nt 
	C_3&=e^{-2\pi is^z/3}\C_3,
	&
	M_y&=is^x\M_y.
\end{align}
In reality, it is the `spinful' symmetries above that should be viewed as fundamental. 
The `spinless' symmetries of Eq.~\eqref{eqn:NoSpinSym} are then more appropriately obtained by appending an additional spin rotations using the SU(2) spin degree of freedom.
For bookkeeping purposes we nevertheless largely discuss symmetries in terms of the `spinless' symmetries.

\section{Spin orbit coupling in twisted bilayer graphene}\label{sec:ContMod_SOC}

In this section, we outline how spin-orbit coupling (SOC) is introduced to the continuum model of twisted bilayer graphene.
We begin by discussing a monolayer of graphene coupled to a transition metal dichalcogenide (TMD), before considering what happens in twisted bilayer graphene.

\subsection{Monolayer graphene with induced spin orbit coupling}\label{sec:MLG+SOC}

We begin by describing the induced spin-orbit felt by monolayer graphene adjacent to a TMD.
In the absence of SOC, the low-energy Hamiltonian for the monolayer is
\begin{align}\label{eqn:H0Dirac}
H_{\mathrm{MLG},0}&=
-v_0\int_\vk
\psi^\dagger(\vk) \Big( k_x \tau^z\sigma^x + k_y \sigma^y \Big) \psi(\vk),
\end{align}
where $\psi(\vk)$ is an eight component spinor with sublattice, valley, and spin indices.
The Pauli matrices $\sigma^{x,y,z}$ act on sublattice indices of the spinor, while $\tau^{x,y,z}$ act on the valley indices.
The proximate TMD induces both Ising and Rashba terms, which may be included by taking $H_{\mathrm{MLG},0}\to H_{\mathrm{MLG},0}+H_\mathrm{MLG,SO}$, where \cite{Wang2015,Gmitra15,Zaletel19}
\begin{align}
	\label{eqn:OneLayerSO}
H_\mathrm{MLG,SOC}
&=
\int_\vk
\psi^\dagger (\vk)
\Bigg(
\frac{\lambda_I}{2}\tau^z s^z
+
\frac{\lambda_R}{2}e^{-i\phi_Rs^z/2}\left(\tau^z \sigma^x s^y-\sigma^y s^x\right)e^{i\phi_Rs^z/2}
\Bigg)\psi(\vk).
\end{align}
Here, $s^{x,y,z}$ act on the spin indices.
The parameters $\lambda_I$ and $\lambda_R$ quantify the strength of the Ising and Rashba terms respectively.
 We further note that the Rashba term may be rotated in-plane by an angle $\phi_R$.

 Only Refs.~\citenum{li_twist-angle_2019} and~\citenum{david_induced_2019} considered the effects of the relative twist angle between the graphene sheet and the TMD monolayer, $\theta_\mathrm{TMD}$.
Their works implies that $\lambda_I$, $\lambda_R$ and $\phi_R$ are all dependent on the relative graphene-TMD twist angle $\theta_\mathrm{TMD}$.
In particular, Ref.~\citenum{li_twist-angle_2019} also shows that when $\theta_\mathrm{TMD}=0^\circ$ and $\theta_\mathrm{TMD}=30^\circ$ (or any $60^\circ$ rotation of these values), the system possess one of two possible reflection symmetries: $\mathcald{R}_x:(x,y,z)\to(-x,y,z)$ or $\mathcald{R}_y:(x,y,z)\to(x,-y,z)$.
If either symmetry is preserved, $\lambda_I$ must vanish and $e^{i\phi_R}$ must be real, \emph{i.e.}, $\phi_R=0,\pi\;\mathrm{mod}2\pi$.
Without loss of generality, we assume $\phi_R(\theta_\mathrm{TMD}=0^\circ)=0$.
There are two potential scenarios for how $\phi_R$ varies as $\theta_\mathrm{TMD}$ is tuned from $0^\circ$ to $30^\circ$.
The first possibility is that $\phi_R$ takes some nonzero values, but ultimately returns back to $0$ at $\theta_\mathrm{TMD}=30^\circ$.
The second option is that it instead equals $\pi$ when $\theta_\mathrm{TMD}=30^\circ$.
Further increasing $\theta_\mathrm{TMD}$ to $60^\circ$ sees $e^{i\phi_R}$ wrap around the unit circle in the complex plane.
Since Rashba SOC is ultimately a consequence of a net \emph{out-of-plane} electric field, we view this second option as extremely unlikely. 
It is more plausible that $\phi_R$ takes small values for all $\theta_\mathrm{TMD}$.

The numerically estimated values of the SOC coupling strengths vary substantially depending on the study, as well as the TMD under consideration:
 $\lambda_I\sim\unit[1-5]{\text{meV}}$, and $\lambda_R\sim\unit[1-15]{\text{meV}}$ \cite{Yang_2016,Wang2015,Gmitra15,Gmitra16,li_twist-angle_2019,david_induced_2019}. 
Calculations that included the effects of $\theta_\mathrm{TMD}$, however, predict substantially smaller values of the Rashba coupling strength: $\lambda_R\lesssim\unit[4]{\text{meV}}$.
The presence of SOC has also been confirmed experimentally  \cite{islandSpinOrbitdrivenBand2019,PhysRevB.97.075434,wang_quantum_2019,wakamuraSpinorbitInteractionInduced2019}, but extracting the magnitudes of $\lambda_I$ and $\lambda_R$ is difficult.

\subsection{Twisted bilayer graphene with induced spin orbit coupling}\label{sec:TBGwithSOC}

We now consider what occurs when a TMD is placed adjacent to one or both of the graphene monolayers that compose twisted bilayer graphene.
We assume for the moment the most physically relevant scenario in which a TMD monolayer is adjacent to a single layer of graphene, as shown in the Fig.~\ref{fig_architecture} of the main text.
In contrast to that figure, we first assume the TMD is alongside the top layer.
The quadratic portion of the Hamiltonian is thus modified to 
\begin{align}\label{eqn:ContModWithSOC}
	H_\mathrm{cont,SOC}
	&=
	H_\mathrm{cont}+H_{\mathrm{SOC}}
\end{align}
where
\begin{align}\label{eqn:ContModSOCterm}
	H_{\mathrm{SOC}}&=\int_\vk \Psi^\dag(\vk) 
	\big(\h_{I}+\h_{R}\big)\Psi(\vk).
\end{align}
Here, $\h_{I}$ and $\h_{R}$ are simply the terms from Eq.~\eqref{eqn:OneLayerSO}:
\begin{align}\label{eqn:SpinOrbitContributions}
\h_{I}
&=
\frac{\lambda_{I}}{2}\mathcal{P}_t\tau^zs^z,
&
\h_{R}
&=
\frac{\lambda_{R}}{2}\mathcal{P}_te^{-i\phi_Rs^z/2}\left(\tau^z\sigma^x s^y-\sigma^y s^x\right)e^{i\phi_Rs^z/2},
\end{align}
where $\mathcal{P}_t=(\id+\mu^z)/2$ projects onto the top layer: only the operators $\Psi_t$ are present in Eq.~\eqref{eqn:ContModSOCterm}.
We note that both $\h_R$ and $\h_I$ are rotationally invariant, which explains the absence of the rotational matrices present in the Dirac parts of the Hamiltonian in Eq.~\eqref{eqn:DiracHamDefs}.
We note that the influence of SOC in TBG has been observed experimentally in Ref.~\citenum{aroraSuperconductivityMetallicTwisted2020}.

The addition of $\h_{I}$ and $\h_{R}$ to the Hamiltonian breaks the SU(2) spin as well as a number of other symmetries.
In Table~\ref{tab:OrdParamSyms}, the symmetries preserved by the introduction of Rashba and Ising are separately listed.
Equivalently, the table may be interpreted as showing the symmetries preserved by the Hamiltonian $H_\mathrm{cont,SO}$ when $\lambda_R\neq0$, $\lambda_I=0$ and when $\lambda_R=0$, $\lambda_I\neq0$, respectively.

Table~\ref{tab:OrdParamSyms} will serve as the basis for the body of this appendix and so we describe the information it contains in detail. 
The top row of the table provides both the continuous subgroups and the generators of the discrete symmetry operations that are preserved by the interacting continuum model $H_\mathrm{cont}+H_C$ without SOC and in the absence of the Hund's coupling (see Eqs.~\eqref{eqn:HcontBravaisDef} and~\eqref{eqn:CoulombIntExpansion}).
The symmetry group represented by the top row of Table~\ref{tab:OrdParamSyms} is the largest group our theory can realize---all of the symmetry groups represented in the rows below are contained within the group defined by the top row.
In the columns labelled by the two continuous symmetries, SU(2) and $\mathrm{U_v}(1)$, the corresponding entry of the Rashba and Ising shows the subgroup of the symmetry still preserved when Rashba or Ising SOC is present or, if no subgroup survives, an `\xmark' is written instead.
(The table does not explicitly reference the emergent $\mathrm{SU}(2)_K\times\mathrm{SU}(2)_{K'}$ symmetry of Eq.~\eqref{eqn:EnlargeContSym}.
When the terms of interest break this symmetry in a way that the existence of the $\mathrm{SU}(2)_K\times\mathrm{SU}(2)_{K'}$ parent symmetry matters, the resulting preserved symmetries are expressed across both columns.)
The SOC terms also break a number of the discrete symmetries of Eq.~\eqref{eqn:NoSpinSym}.
When one or more of the continuous symmetries has also been broken, a residual composite symmetry in which a discrete symmetry operation is followed by a continuous symmetry operation may survive. 
In this case, that composite symmetry is listed instead.
For instance, for Rashba SOC, none of the `spinless' versions of the symmetries survive---$\h_R$ is only invariant under their action when they are composed with additional spin rotation transformations, as seen in the first row of Table~\ref{tab:OrdParamSyms}. 
We see  that the discrete symmetries preserved by Rashba are in fact the physical space group operations and electronic time reversal provided in  Eq.~\eqref{eqn:PhysicalSymDef} (see Sec.~\ref{sec:MirrorSymScenario} for a discussion of the mirror symmetry).
We note that the generators chosen in each instance are not unique, \emph{e.g.}, we could equally well have added a $\mathrm{U_v}(1)$ transformation to each of the operations for both Rashba and Ising SOC.

When both Rashba and Ising SOC are simultaneously present (\emph{i.e.}, $H_\mathrm{cont,SOC}$ with $\lambda_R,\lambda_I\neq0$), an even smaller set of symmetries remains. 
As an example, the Rashba term is preserved under the physical inversion operation, $is^z\C_2=C_2$, while the Ising term instead requires a spin flipped version, $is^x\C_2$. 
These symmetries are not compatible and thus we conclude that in the presence of both Ising and Rashba SOC, inversion is no longer present.
As we argue below, we are able to neglect such effects.

Since Rashba SOC will be the only term we consider to include the in-plane spins $s^x$ and $s^y$, the phase $\phi_R$ may be ``undone'' through the appropriate rotation about the $z$-spin axis. 
Such a transformation of the internal spin directions does not alter the symmetry transformations of Table~\ref{tab:OrdParamSyms} except for the column labelled $\M_y$, which we discuss in the next section. 
Despite this degree of freedom, redefining $s^x$ and $s^y$ does of course modify the relation between the internal SOC parameters and the physical spatial directions (some consequences are discussed in Sec.~\ref{sec:InternalSOCDirection}).

\subsection{\texorpdfstring{Multiple TMDs and mirror symmetry $\M_y$}{Multiple TMDs and mirror symmetry My}}\label{sec:MirrorSymScenario}

Inclusion of $H_{\mathrm{SOC}}$ in Eqs.~\eqref{eqn:ContModWithSOC} and~\eqref{eqn:ContModSOCterm} breaks the mirror symmetry $\M_y$ since the latter interchanges the layers, only one of which possesses proximity-induced SOC.
It is useful below to consider the most physically relevant scenario in which both graphene monolayers ``symmetrically'' possess SOC. 
More precisely, we will construct Rashba and Ising terms that preserve a spin rotated mirror symmetry and the spinless mirror, respectively.
To do so we need to couple \emph{both} graphene monolayers to a TMD, so that $H_\mathrm{cont,SOC}=H_\mathrm{cont}+H_{\mathrm{SOC}}+H_{\mathrm{SOC}}'$ with $H_{\mathrm{SOC}}'$ defined in direct analogy to $H_{\mathrm{SOC}}$ except using operators originating from the bottom graphene sheet with corresponding SOC strengths $\lambda_R'$ and $\lambda_I'$.
Here, we assume that $\lambda_{I/R}'$ possess the same magnitudes as $\lambda_{I/R}$, but allow them to take different signs. The choice most relevant to the physical scenario in which a single monolayer of TMD is present is determined below.

\subsubsection{Ising SOC}\label{sec:MirrorSymIsing}
We begin by outlining what occurs to first order when Ising SOC of strength $\lambda_{I}$ is added to the Dirac Hamiltonian of the top layer only (in the absence of Rashba).
We focus on the physics occurring close to charge neutrality at the Dirac cones at $\kappa$, $\kappa'$, for the moment specifying to the $K$ valley ($\tau^z=+1$). 
The analysis for the $K'$ valley follows directly.
Using first order perturbation theory similar to the analysis used to `derive' the magic angle in Ref.~\onlinecite{bistritzerMoireBandsTwisted2011}, one can show that the presence of $\lambda_{I}\neq0$  
induces \emph{effective} Ising SOC terms for the Dirac cones at both $\kappa=K_t$ 
and $\kappa'=K_b$.
More precisely, in the low energy theory of the Dirac cones of the moiré system, to first order in $\lambda_{I}$, the Dirac cone at $\kappa$ has an effective SOC parameter $\tilde{\lambda}_{I,\kappa}= c_{1}\lambda_{I}$ while the Dirac cone at $\kappa'$ has SOC parameter $\tilde{\lambda}_{I,\kappa'}= c_{2}\lambda_{I}$, where $c_{1}$ and $c_{2}$ are real numbers.
Importantly, one finds that $c_{1}$ and $c_{2}$ have the same sign, meaning that $\tilde{\lambda}_{I,\kappa}$ and $\tilde{\lambda}_{I,\kappa'}$ do as well.
These results are further supported by numerics.
The preceding scenario is what we want the setup with two TMD monolayers to resemble.

When Ising SOC is also present in the Dirac Hamiltonian of the bottom layer, it will similarly induce effective Ising SOC into the Dirac cones at $\kappa$, $\kappa'$: $\tilde{\lambda}_{I,\kappa}'=c_2\lambda'_{I}$ and $\tilde{\lambda}'_{I,\kappa'}=c_1\lambda_{I}'$, where $c_1$ and $c_2$ are the same constants appearing above and their assignment is determined by symmetry.
Within the first order analysis we consider here, the total effective Ising parameters are given by the sum, \emph{i.e.} $\tilde{\lambda}_{I,\kappa}^\mathrm{tot}=\tilde{\lambda}_{I,\kappa}+\tilde{\lambda}_{I,\kappa}'$ and $\tilde{\lambda}_{I,\kappa'}^\mathrm{tot}=\tilde{\lambda}_{I,\kappa'}+\tilde{\lambda}_{I,\kappa'}'$. 
It's clear that in order for $\tilde{\lambda}^\mathrm{tot}_{I,\kappa}$ to have the same sign as $\tilde{\lambda}^\mathrm{tot}_{I,\kappa'}$, we must have $\mathrm{sgn}(\lambda_{I})=\mathrm{sgn}(\lambda_{I}')$. 
As mentioned, we further specify to the situation in which they have the same magnitude.
We conclude that the appropriate symmetry-enhanced version of $\h_I$ is
\begin{align}\label{eqn:2layerIsing}
	\h_I^*
	&=
	\frac{1}{2}(\lambda_{I}\mathcal{P}_t  + \lambda_{I} \mathcal{P}_b)\tau^z s^z
	=
	\frac{\lambda_I}{2}\tau^zs^z,
\end{align}
where $\mathcal{P}_{t/b}=(\id\pm\mu^z)/2$ project onto the top or bottom graphene monolayer.
In Table~\ref{tab:OrdParamSyms}, in the column labelled  `$\M_y$,' we 
write $\M_y^*$
in the Ising column to indicate that the mirror symmetry $\M_y$ is only truly preserved when the first-quantized Hamiltonian $\h_I^*$ is used in place of $\h_I$. 
The Hamiltonian $\h_I^*$ is otherwise invariant under the same symmetries as $\h_I$.

\subsubsection{Rashba SOC}

An identical analysis to the one described above for the Ising SOC indicates that the presence of Rashba SOC in the Dirac cone descending from one of the graphene layers (say, the top at $\kappa=K_t$) should be accompanied by \emph{opposite sign} Rashba SOC strength $\lambda_R'$ and \emph{same sign} in-plane rotation $\phi_R'$ in the Dirac cone descending from the other monolayer (bottom, say, at $\kappa'=K_b$).
It follows that the two-layer extension of the Rashba SOC-coupled theory that most resembles the single layer case (\emph{i.e.}, the SOC contribution is not cancelled at linear order in $\lambda_R$), is obtained through
\begin{align}\label{eqn:MirrorSymRashbaDef}
	\h_R^*
	&=
	\frac{1}{2}(\lambda_R \mathcal{P}_t-\lambda_R\mathcal{P}_b)e^{-i\phi_Rs^z/2}(\tau^z\sigma^xs^y-\sigma^ys^x)e^{i\phi_Rs^z/2}
	=
	\frac{\lambda_R}{2}\mu^ze^{-i\phi_Rs^z/2}(\tau^z\sigma^xs^y-\sigma^ys^x)e^{i\phi_Rs^z/2}.
\end{align}
The action of all the symmetries is the same for $\h_R^*$ as for $\h_R$ save for the mirror symmetry $\M_y$.
It of course requires a spin flip operation, but in a way that depends on the angle $\phi_R$.
Namely, the composite operation $e^{-i\phi_Rs^z/2}is^xe^{i\phi_Rs^z/2}\M_y$ is preserved by $\h_R^*$.
As discussed at the end of Sec.~\ref{sec:TBGwithSOC}, we have the freedom to redefine $s^x$ and $s^y$ so that $\phi_R=0$.
In what follows, we assume that either $\phi_R=0$ or that such a transformation has been made---we emphasize, however, that this choice is only relevant when the mirror symmetry is discussed.
In such a limit, $\h_R^*$ preserves the $M_y=is^x\M_y$, where $M_y$ is the physical version of the symmetry operations of Eq.~\eqref{eqn:PhysicalSymDef}.
An asterisk is added in Table~\ref{tab:OrdParamSyms}  to indicate that $is^x\M_y$ is only preserved  by the modified Rashba term $\h^*_R$.
That is, we write $is^x\M_y^*$ in the appropriate column.

{\renewcommand{\arraystretch}{1.7}
\begin{table}[t]
	\centering
	\begin{tabular*}{\textwidth}{@{\extracolsep{\fill}}lc | ccc | rrrrr r| r}
		Order 
		&
		&   SU(2)	&	U$_{\rm v}$(1)
		&
		&	\multicolumn{1}{c}{$\Tns=\tau^x\K$}	
		&	\multicolumn{1}{c}{$\C_2$}	
		&	\multicolumn{1}{c}{$\C_2\Tns$}	
		&	\multicolumn{1}{c}{$\C_3$}	
		&	\multicolumn{1}{c}{$\M_y$} 
		&
		&	\multicolumn{1}{c}{Mixed}
		\\
		\hline\hline
		Rashba	(R)
		&
		&
		\xmark	&	U$_{\rm v}$(1) 	
		&
		& 
		$is^y\Tns$	&	$is^z\C_2$	&	$is^x\C_2\Tns$	&	$e^{-2\pi is^z/3}\C_3$	&	$is^x\M_y^*$    &
		\\
		Ising (I)
		&
		&
		$\mathrm{U}_z(1)$	&	U$_{\rm v}$(1) 
		&
		&	$is^y\Tns$	&	$is^x\C_2$	&	$\C_2\Tns$	&	$\C_3$	&	$\M_y^{*}$  &
		\\\hline
		sIVC 
		&  
		&   SU(2)	&	\xmark
		&
		&	$i\tau^y\K$	&	$\tau^z\C_2$	&	$\C_2\Tns$	&	$\C_3$	&	$i\tau^z\M_y$   &
		\\
		sIVC + R
		&
		&	\xmark	& \xmark
		&
		&	$\tau^ys^y\K$	&	$\tau^zs^z\C_2$	&	$is^x\C_2\Tns$	&	$e^{-2\pi is^z/3}\C_3$	&	$\tau^zs^x\M_y^*$   &
		\\
		sIVC + I
		&
		&	$\mathrm{U}_z(1)$	&	\xmark
		&
		&	$\tau^ys^y\K$	&	$\tau^zs^y\C_2$	&	$\C_2\Tns$	&	$\C_3$	&	$i\tau^z\M_y^*$   &
		\\\hline
		tIVC
		&
		&   \multicolumn{2}{c}{$\mathrm{SU(2)_{tIVC}}$} 
		&
		&   $is^y\Tns$
		&	$is^{x}\C_2$	
		&	$\C_2\Tns$	&	$\C_3$	
		&	$i\tau^z\M_y$ 
		&
		\\
		tIVC + R
		&
		&	\xmark	&	\xmark
		&
		&	$is^y\Tns$	&	$\tau^zs^z\C_2$	&	$\tau^zs^x\C_2\Tns$	&	$e^{-2\pi is^z/3}\C_3$	&	$is^x\M_y^*$    &
		\\
		tIVC + I
		&
		&	$\mathrm{U}_z(1)$	&	\xmark
		&
		&	$is^y\Tns$	&	$is^x\C_2$	&	$\C_2\Tns$	&	$\C_3$	&	$i\tau^z\M_y^*$ &
		\\\hline
		FM IVC 
		&
		&   $\mathrm{U}_z(1)$   &   \xmark
		&
		& $i\tau^y\K$    &   $i\tau^z\C_2$    &   $\C_2\Tns$   &   $\C_3$  &   $i\tau^z\M_y$   &
		\\
		FM IVC + R
		&	
		&	\xmark	&	\xmark	
		&
		&	\xmark	&	$\tau^zs^z\C_2$	&	\xmark	&	$e^{-2\pi is^z/3}\C_3$	&	\xmark	
		&
		&	$is^z\M_y^*\Tns$
		\\
		FM IVC + I
		&
		&	$S^z$	&	\xmark
		&
		&	\xmark	&	\xmark	&	$\C_2\Tns$	&	$\C_3$	&	$i\tau^z\M_y^*$ &
		\\\hline
		AFM IVC
		&
		&  \multicolumn{2}{c}{$\tilde{\mathrm{U}}_z(1)$, $\mathrm{U}_{\mathrm{v}\cdot z}(1)$}
		&
		&	$is^y\Tns$	&	$is^y\C_2$	&	$\C_2\Tns$	&	$\C_3$	&	$i\tau^z\M_y$   &   
		\\
		AFM IVC + R
		&
		&	\xmark	&	\xmark
		&
		&	$is^y\Tns$	&	\xmark	&	\xmark	&	$e^{-2\pi i(s^z+\tau^z)/3}\C_3$	&	\xmark	
		&
		&	$\tau^zs^y\C_2\M_y^*$
	\end{tabular*}
	\caption{\AppCaptions
	Symmetries preserved in 2$d$ in the presence of the various order parameters and SOC terms listed.
	The title row provides a list the generators of the symmetry group preserved by the Hamiltonian $H_\mathrm{cont}+H_C$ in the absence of SOC or IVC order.	
	The entries in the columns corresponding to the continuous symmetries, SU(2) and ${\mathrm{U_v}(1)}$, show the subgroup preserved by the order labelling the row. 
	Notably, when the $z$ component of spin is conserved, the subgroup $\mathrm{U}_z(1):c(\vk)\to e^{i\theta s^z/2} c(\vk)$ remains a good symmetry.
	When no subgroup is preserved, an `\xmark' is written instead.
	While we do not explicitly indicate reference the $\mathrm{SU}(2)_K\times\mathrm{SU}(2)_{K'}$ symmetry in the top row, we do record the scenarios in which a preserved continuous symmetry descends from this larger group, \emph{i.e.} the preserved subgroup or subgroups ``mix'' the SU(2) spin and $\mathrm{U_v}(1)$ symmetries.
	In this case, the preserved subgroups are listed across both columns. 
	This scenario is relevant for the tIVC state, which preserves the $\mathrm{SU}(2)_\mathrm{tIVC}$ symmetry with generators $\{s^z,\tau^zs^x,\tau^zs^y\}$, and for the AFM IVC state, which preserves the two U(1) symmetries $\tilde{\mathrm{U}}_z(1):\;c(\vk)\to e^{i\theta(\tau^z+s^z)/2}c(\vk)$ and $\mathrm{U}_{\mathrm{v}\cdot z}(1):c(\vk)\to e^{i\theta\tau^zs^z/2}c(\vk)$.
	The entries in the columns corresponding to the discrete time reversal and lattice symmetries show the generator of a conserved symmetry either equal to the discrete symmetry or obtained by composing it with an element of a broken continuous symmetry group.
	When no such combination exists, `\xmark' is present instead.
	We note that $is^y\Tns=\mathcal{T}$ is the physical time reversal symmetry and that $i\tau^y\mathcal{K}=\tilde{\mathcal{T}}$ is the non-unitary symmetry preserved by the sIVC state and discussed in the main text. 
	The relation of the other generators to the physical symmetries of the system is provided in Eq.~\eqref{eqn:PhysicalSymDef}. 
	When relevant, the final column, `Mixed,' lists any conserved generators 
	arising out of the composition of multiple discrete symmetries in addition to possible continuous symmetry operations.
	}
	\label{tab:OrdParamSyms}
\end{table}
}

\section{\texorpdfstring{Proximity-coupled wire at $\nu=\pm4$}{Proximity-coupled wire at nu=+-4}}\label{sec:ProxCoupWire_BandIns}

In this section, we illustrate our derivation of the ``trivial'' wire Hamiltonian.
We derive the effective Hamiltonians in two dimensions for the flat bands about the moiré BZ centre, denoted the $\gamma$ point (Secs.~\ref{sec:2dHeff-NoSOC} and~\ref{sec:2dHeff-WithSOC}), which allows us to extract useful information on the scaling of the parameters, as we describe in Sec.~\ref{sec:TrivialWireHam}.
We finish with a discussion in Sec.~\ref{sec:TrivialWireC3MyBreaking} of how our analysis would differ in the absence of the $\C_3$ and/or mirror symmetries.

\subsection{\texorpdfstring{Projection to flat bands in 2$d$ without SOC}{Projection to flat bands in 2d without SOC}}
\label{sec:2dHeff-NoSOC}

The foundation of our analysis is the flat band Hamiltonian without SOC or IVC order---these pieces will be added in a perturbative fashion in subsequent sections using the basis and symmetry action described here.

As mentioned at the end of Sec.~\ref{sec:ModelDef}, in order to diagonalize the continuum model, it is convenient to write it in terms of moiré lattice vectors $\vG$ and momenta $\vk$ restricted to the moiré BZ. 
We reproduce here Eq.~\eqref{eqn:HcontBravaisDef}, the continuum model Hamiltonian in the absence of SOC:
\begin{align}\label{eqn:HcontBravaisDef-2}
	H_\mathrm{cont}&=
	\sum_{v=K,K'}
	\int_{\vk\in\mBZ} \tilde\Psi^\dag_{v,i}(\vk+\vG)
	\h^{(v)}_{i,\vG;j,\vG'}(\vk)
	\tilde\Psi_{v,j}(\vk+\vG'),
\end{align}
where $i,j$ are combined indices including layer and sublattice, and $\vG$, $\vG'$ are moiré reciprocal lattice vectors. The spin index is suppressed. 
Recall that the $\tilde{\Psi}$ operators are related to the $\Psi$ operators   through a simple momentum shift (\emph{e.g.}, Eq.~\eqref{eqn:UVElectronOperatorReDef}), and that they therefore transform in the same way under the symmetry action detailed in Sec.~\ref{sec:SymmetryAction}.
The Hamiltonian $H_\mathrm{cont}$ is diagonalized through a unitary transformation
\begin{align}\label{eqn:TrivialCaseDiag-1}
	\tilde\Psi_{v,i}(\vk+\vG)&=\sum_\alpha \tilde{U}^{(v)}_{i,\vG;\alpha}(\vk) \tilde{c}_{v,\alpha}(\vk),
	&
	\tilde{c}_{v,\alpha}(\vk)
	&=
	\sum_{i,\vG}\tilde{U}^{(v)\dag}_{\alpha;i,\vG}(\vk)\tilde\Psi_{v,i}(\vk+\vG),
\end{align}
satisfying 
\begin{align}
	\label{eqn:TrivialCaseDiag}
	\sum_{i,\vG,j,\vG'}\tilde{U}^{(v)\dag}_{\alpha;i,\vG}\h_{i,\vG;j,\vG'}^{(v)}(\vk) \tilde{U}_{j,\vG';\beta}^{(v)}(\vk)
	=
	\delta_{\alpha\beta}\epsilon^{(v)}_\alpha(\vk).
\end{align}
Here, $\epsilon_\alpha^{(v)}(\vk)$ represents the energy of band $\alpha$ at momentum $\vk$ for valley $v$.
Time reversal requires that $\epsilon^{(K)}_\alpha(\vk)=\epsilon^{(K')}_\alpha(-\vk)$.

As mentioned at the end of Sec.~\ref{sec:ModelDef}, the flat bands, $\alpha\in\mathit{fl}$, comprise the band above and below charge neutrality for each spin and valley, meaning that the Fermi energy intersects these states for fillings $-4<\nu<+4$.
It turns out that the basis defined in Eq.~\eqref{eqn:TrivialCaseDiag} is not the most convenient for describing the flat bands.
We instead choose a basis in which the operators we work with transform in a certain way under the symmetries of the Hamiltonian. 
This basis change is accomplished simply through a rotation 
\begin{align}\label{eqn:FlatBandBasis}
	\tilde{c}_{v,\alpha}(\vk)
	&=
	V^{(v)}_{\alpha\beta}(\vk){c}_{v,\beta}(\vk),
	&
	c_{v,\alpha}(\vk)
	&=
	V^{(v)\dagger}_{\alpha\beta}(\vk)
	\tilde{c}_{v,\beta}(\vk),
\end{align}
where $V_{\alpha\beta}^{(v)}(\vk)$ is a $2\times2$ unitary matrix.
It was demonstrated in Ref.~\onlinecite{zou_band_2018,bultinckGroundStateHidden2020} that there exists a basis (\emph{i.e.}, a set of matrices $V_{\alpha\beta}^{(v)}(\vk)$) in which the flat band operators $c(\vk)$ transform as
\begin{align}\label{eqn:FlBandSyms-1}
	\Tns&:\quad
	c(\vk)\to \tau^x c(-\vk),\; i\to-i,
	\nt
	\C_2&:\quad
	c(\vk)\to e^{i\theta(\vk)} \tau^x \sigma^x c(-\vk),
	\nt
	\C_2\Tns&:\quad
	c(\vk)\to e^{i\theta(\vk)} \sigma^x c(\vk),
	\;i\to-i.
\end{align}
Here, $\tau^{x,y,z}$ Pauli matrices continue to represent transformations acting on the valley indices of the operators. 
Conversely, the $\sigma^{x,y,z}$ Pauli matrix is no longer acting on the sublattice indices (since sublattice is typically not a good quantum number), but are instead acting on an additional `band index' in our basis\footnote{%
In certain approximations, these flat band indices acted on by $\sigma^{x,y,z}$ in fact coincide with the actual sublattice index of the UV theory---that is not quite the case here, but explains the nomenclature.
}.
Close to the ${\gamma}$ point, $\theta(\vk)$ may be set to zero.
Further, we find that, when applicable, the transformation of $c(\vk)$ under $\C_3$ and $\M_y$ can be written
\begin{align}\label{eqn:FlBandSyms-2}
	\C_3&:\quad
	c(\vk)\to c(R_3\vk),
	&
	\M_y&:\quad
	c(\vk)\to \sigma^x c(R_y\vk),
\end{align}
within an open region containing the $\gamma$ point.
The matrices $R_3$ and $R_y$ are provided in Eq.~\eqref{eqn:R3RyDef}.
(See Sec.~\ref{sec:TrivialWireC3MyBreaking} for a discussion of what in the absence of $\M_y$ and/or $\C_3$.)
The SU(2) spin and U$_{\rm v}$(1) symmetries act on the operators $c(\vk)$ in exactly the same fashion as they act on the operators $\Psi$ in Eq.~\eqref{eqn:NoSpinSym}.

The Hamiltonian itself is generically no longer diagonal in this basis. 
It takes the form
\begin{align}\label{eqn:FlBandContMod}
	H^{(\fl)}_\mathrm{cont}
	&=
	\int_{\vk\in\mBZ}c^\dag(\vk)h^{(0)}(\vk)c(\vk),
\end{align}
where we have suppressed all indices, including the valley index $v$. 
That is, $c(\vk)$ is an 8-component vector, while $h^{(0)}(\vk)$ is an $8\times8$ dimensional matrix.
We will keep with the convention that flat band effective Hamiltonians are written in normal, serifed script, whereas calligraphic script will continue to be used for first-quantized Hamiltonians in the basis of the monolayers operators $\Psi$, $\tilde{\Psi}$.
Although determining $h^{(0)}(\vk)$ analytically is a completely intractable task, there is nevertheless a great deal that can be said only using the continuous symmetries $\mathrm{U_v}(1)$ and $\mathrm{SU}(2)$ of Eq.~\eqref{eqn:ContinuousSymmetryAction} as well as the discrete symmetries of Eqs.~\eqref{eqn:FlBandSyms-2} and~\eqref{eqn:FlBandSyms-1}.

\subsubsection{\texorpdfstring{2$d$ flat band Hamiltonian without SOC}{2d flat band Hamiltonian without SOC}}\label{sec:FlatBandHeff-NoSOC}

We start by considering the form the Hamiltonian in the gauge just described takes in the absence of spin orbit coupling, \emph{i.e.} no explicit spin terms in the Hamiltonian. 
The SU(2) spin symmetry thus prohibits the presence of spin Pauli matrices $s^{x,y,z}$, while the U$_{\rm v}$(1) symmetry limits the valley Pauli matrices to $\tau^{0,z}$.
Imposing
$\C_2\Tns$ requires that terms proportional to $\sigma^z$ or $\tau^z\sigma^z$ vanish.
Finally, of the terms remaining, time reversal indicates that they are either even or odd under $\vk\to-\vk$ according to
\begin{align}
	h_{e}^{(0)} (\vk)
	&=
	h^{(0)}_{0,0,0}(\vk) +{ h^{(0)}_{0,x,0}(\vk)\sigma^x} 
	+ 
	{h^{(0)}_{z,y,0}(\vk)\tau^z \sigma^y},
	\nt
	h_{o}^{(0)}(\vk)
	&=
	{h^{(0)}_{0,y,0}(\vk)\sigma^y }
	+ 
	{h^{(0)}_{z,0,0}(\vk)\tau^z}
	+
	{h^{(0)}_{z,x,0}(\vk)\tau^z\sigma^x},
\end{align}
where the subscripts `$e$' and `$o$' indicate the parity (even and odd) of the two terms.
The total effective Hamiltonian is $h^{(0)}(\vk)=h^{(0)}_e(\vk)+h^{(0)}_o(\vk)$.

As explained below, we are specifically interested in the physics of the lower, hole-doped flat band close to the $\gamma=(0,0)$ point at the $\mBZ$ centre. 
The action of the mirror symmetry $\M_y$ implies that at the $\gamma$ point, $h^{(0)}_{z,y,0}$ must vanish.
As a result, $h^{(0)}(\vk=0)=h^{(0)}_{0,0,0}(0)+h_{0,x,0}^{(0)}(0)\sigma^x$. 
Since SOC and the proximity-induced IVC order will both be considered in a perturbative limit below, it is appropriate to further project the Hamiltonians onto the $\sigma^x=\pm1$ basis (the actual sign of $\sigma^x$ will not matter).
Doing so yields
\begin{align}\label{eqn:Ham-1bnd-NoSOC-2d}
	\bar{h}^{(0)}_e(\vk)
	&=
	t_{0,0}^{(0)}(\vk),
	&
	\bar{h}_o^{(0)}(\vk)
	&=
	t_{z,0}^{(0)}(\vk)\tau^z.
\end{align}
In the left equation $t_{0,0}$ can be expressed as
\begin{align}\label{eqn:Ham-1bnd-NoSOC-2d-kexpansion-Even}
	t_{0,0}^{(0)}(\vk)
	=
	\frac{\vk^2}{2m}
	-\mu.
\end{align}
On the other hand, the $\C_3$ symmetry requires that 
\begin{align}\label{eqn:Ham-1bnd-NoSOC-2d-kexpansion-Odd}
	{t}_{z,0}^{(0)}(\vk)
	=
	\tilde{t}_{o}^{(0)}k_x(k_x^2-3k_y^2)
	+
	\tilde{t}_{o}^{(0)\prime}k_y(3k_x^2-k_y^2).
\end{align}
The terms we derived are recorded in Table~\ref{tab:kexpansion}.
Mirror symmetry $\M_y$ further sets $\tilde{t}_{o}^{(0)\prime}=0$.
Nevertheless, since $\M_y$ is technically not a good symmetry of the problem, $\M_y$-breaking perturbations may be generated. 
We therefore have included the $\M_y$-breaking couplings in Table~\ref{tab:kexpansion}, delineated by curly braces `$\{\}$' to distinguish them from $\M_y$-breaking terms allowed within our perturbative expansion.
We emphasize that the applicability of mirror symmetry in this context is independent of whether SOC is ultimately included through $\h_{I/R}$ or the mirror-symmetrized $\h_{I/R}^*$.

\subsection{\texorpdfstring{Effective flat band Hamiltonian in 2$d$ with SOC}{Effective flat band Hamiltonian in 2d with SOC}}
\label{sec:2dHeff-WithSOC}

We treat the addition of SOC perturbatively, most notably in the sense that we assume that it induces minimal mixing between the flat and non-flat/dispersive bands at $\nu<-4$, $+4<\nu$. 
Our goal is therefore to derive the effective Hamiltonian in terms of the operators $c(\vk)$ defined in Eq.~\eqref{eqn:FlatBandBasis}. 
Such a perturbative expansion is well-defined
provided $\lambda_{R/I}\ll\Delta_\mathrm{disp}$, where $\Delta_\mathrm{disp}$ is the gap separating the flat and dispersive bands.  
Since $\Delta_\mathrm{disp}\sim\unit[30]{\text{meV}}$ is the typically obtained experimentally close to the magic angle \cite{choiElectronicCorrelationsTwisted2019,polshynLargeLinearInTemperatureResistivityInTBG} while $\lambda_{I/R}\lesssim\unit[5]{\text{meV}}$ \cite{Yang_2016,Wang2015,Gmitra15,Gmitra16,li_twist-angle_2019,david_induced_2019,aroraSuperconductivityMetallicTwisted2020}, this assumption is reasonable. 

We will specifically restrict our analysis to a \emph{first order} approximation.
We envision obtaining our effective Hamiltonian by projecting $H_\mathrm{SOC}$ onto the flat bands through the identification
\begin{align}
	\Psi_{v,i}(\vk+\vG)
	&\sim
	\sum_{\beta\in\mathit{fl}}{U}^{(v)}_{i,\vG;\beta}(\vk)c_{v,\beta}(\vk),
	&
	{U}^{(v)}_{i,\vG;\beta}(\vk)&=
	\sum_{\alpha\in\mathit{fl}}\tilde{U}^{(v)}_{i,\vG;\alpha}V_{\alpha\beta}^{(v)}(\vk),
\end{align}
where $\tilde{U}^{(v)}_{i,\vG;\beta}(\vk)$ and $V_{\alpha\beta}^{(v)}(\vk)$ are defined in Eqs.~\eqref{eqn:TrivialCaseDiag-1} and~\eqref{eqn:TrivialCaseDiag}.
It therefore follows that
\begin{align}\label{eqn:FlatBandProject}
	H_{\mathrm{SOC}}
	&=
	\sum_{\vG}\int_{\vk\in\mBZ}\Psi^\dag(\vk)\h_{\mathrm{SOC}}
	\Psi(\vk+\vG)
	\to
	H_\mathrm{SOC}^{(\fl)}
	\int_{\vk\in\mBZ} 
	c^\dag(\vk) 
	\underbrace{U^\dag(\vk) \h_\mathrm{SOC} U(\vk)}_{h_\mathrm{SOC}}
\end{align}
Consistent with the conventions above, the effective flat band Hamiltonian, $h_\mathrm{SOC}(\vk)$, is written in normal font, while the full first-quantized Hamiltonian in the basis of the microscopic graphene operators, $\h_\mathrm{SOC}$, is expressed in calligraphic font.
Note that although $\h_\mathrm{SOC}$ is momentum independent, the effective flat band Hamiltonian $h_{\mathrm{SOC}}$ depends on $\vk$.

A straightforward consequence of the approximation of Eq.~\eqref{eqn:FlatBandProject} is that, like $\h_\mathrm{SOC}$, the effective Hamiltonian $h_\mathrm{SOC}(\vk)$ may be divided into a strictly Rashba and a strictly Ising part:
\begin{align}\label{eqn:EffFlatBandSOCHam}
	h_\mathrm{SOC}(\vk)
	&=
	h_R(\vk)
	+
	h_I(\vk),
\end{align}
where $h_R(\vk)\propto \lambda_R$ and $h_I(\vk)\propto\lambda_I$.
Further, because $\h_R$ and $\h_I$ are both proportional to Pauli matrices acting on the spin indices, $s^{x,y,z}$, and the unitary transformations $U^{(v)}_{i,\vG;\alpha}(\vk)$ do not act on the spin, the effective spin-orbit contributions must also be proportional to spin Pauli matrices.

As in the previous section, it is not necessary to derive $h_{\mathrm{SOC}}(\vk)$ directly---we instead acquire its form by imposing the symmetries listed in Table~\ref{tab:OrdParamSyms}.
Further, given our restriction to first order, terms prohibited by a symmetry that is only broken when \emph{both} Rashba and Ising SOC are present will not be present in $h_\mathrm{SOC}(\vk)$ in Eq.~\eqref{eqn:EffFlatBandSOCHam} (\emph{e.g.}, $\C_2$).
This simplification accounts for the absence of a row in Table~\ref{tab:OrdParamSyms} listing the symmetries preserved when both Rashba and Ising SOC are present.

\subsubsection{\texorpdfstring{2$d$ flat band Rashba Hamiltonian}{2d flat band Rashba Hamiltonian}}\label{sec:2d-Rashba-Heff}

We now use the symmetries outlined in Table~\ref{tab:OrdParamSyms} to restrict the form of the Rashba effective Hamiltonian.
From $is^x\C_2\Tns$, we find that
\begin{align}
	h^{(R)}(\vk)
	&=
	\sum_{\ell=0,z}\bigg(\sum_{i=0,x,y}\sum_{a=0,x,y}h^{(R)}_{\ell,i,a}(\vk)\tau^\ell\sigma^i s^a
	+
	h^{(R)}_{\ell,z,z}(\vk) \tau^\ell\sigma^zs^z\bigg).
\end{align}
From the action of $is^y{\Tns}$, we separate these terms into even and odd components, \emph{i.e.} $h^{(R)}_e(-\vk)=h^{(R)}_e(\vk)$, $h^{(R)}_{o}(-\vk)=-h^{(R)}_o(\vk)$:
\begin{align}
	h^{(R)}_e(\vk)
	&=
	h^{(R)}_{0,0,0}(\vk)
	+
	h^{(R)}_{0,x,0}(\vk)\sigma^x
	+
	\sigma^y(h^{(R)}_{0,y,x}(\vk)s^x + h^{(R)}_{0,y,y}(\vk)s^y)
	\nt&\quad
	+
	\tau^z \Big[
		\beta_{z,0,x}(\vk)s^x+h^{(R)}_{z,0,y}(\vk)s^y
		+
		\sigma^x(h^{(R)}_{z,x,x}(\vk)s^x+h^{(R)}_{z,x,y}(\vk)s^y)
		+h^{(R)}_{z,y,0}(\vk)\sigma^y
		+h^{(R)}_{z,z,z}(\vk)\sigma^zs^z
	\Big],
	\nt
	h^{(R)}_o(\vk)
	&=
	h^{(R)}_{0,0,x}(\vk)s^x+h^{(R)}_{0,0,y}(\vk)s^y
	+
	\sigma^x(\beta_{0,x,x}(\vk)s^x+h^{(R)}_{0,x,y}(\vk)s^y)
	+h^{(R)}_{0,y,0}(\vk)\sigma^y
	+h^{(R)}_{0,z,z}(\vk)\sigma^zs^z
	\nt&\quad
	+
	\tau^z\big[
	\beta_{z,0,0}(\vk)
	+
	\beta_{z,x,0}(\vk)\sigma^x
	+
	\sigma^y(\beta_{z,y,x}(\vk)s^x + \beta_{z,y,y}(\vk)s^y)\big].
\end{align}
To derive the effective one-band Hamiltonian close to the $\gamma$ point, we start by projecting onto states with either $\sigma^x=+1$ or $\sigma^x=-1$, as outlined in Sec.~\ref{sec:FlatBandHeff-NoSOC}.
Doing so, we obtain
\begin{align}\label{eqn:Rashba-2dHam}
	\bar{h}^{(R)}_{e}(\vk)
	&=
	t^{(R)}_{0,0}(\vk)
	+
	\big( t^{(R)}_{z,x}(\vk)\tau^zs^x+ t^{(R)}_{z,y}(\vk)\tau^zs^y\big),
	\nt
	\bar{h}^{(R)}_{o}(\vk)
	&=
	t^{(R)}_{0,x}(\vk)s^x+t^{(R)}_{0,y}(\vk)s^y + t^{(R)}_{z,0}(\vk)\tau^z.
\end{align}
We next expand the functions $t^{(R)}_{a,n}(\vk)$ in powers of $\vk$ with the $e^{-2\pi is^z/3}\C_3$ symmetry as a restriction.
Both $t_{0,0}^{(R)}$ and $t_{z,0}^{(R)}$ will take identical forms as in the case without SOC, and so these terms need not be considered separately from the others.
Moreover, as indicated in the discussion below Eq.~\eqref{eqn:EffFlatBandSOCHam}, the Rashba SOC term is only able to generate such spin-diagonal terms at higher orders in perturbation theory; in the first order approximation used here, $t_{0,0}^{(R)}$ and $t_{z,0}^{(R)}$ thus vanish.
It follows that the addition of mirror-symmetry-breaking Rashba SOC does \emph{not} generate the valley-orbit term $\tilde{t}^{(0)\prime}_ok_y(3k_x^2-k_y^2)\tau^z$ (Eq.~\eqref{eqn:Ham-1bnd-NoSOC-2d-kexpansion-Odd}) that we argued was forbidden by mirror symmetry in the spin symmetric portion of the Hamiltonian. 
For the remaining, non-zero terms, rotation symmetry implies that they take the form
\begin{align}\label{eqn:Rashba-2dHam-kexpansion}
	t^{(R)}_{z,x}(\vk)\tau^zs^x+ t^{(R)}_{z,y}(\vk)
	&=
	\tilde{t}^{(R)}_e
	\tau^z\left( \frac{1}{2}(k_x^2-k_y^2)s^x - k_xk_y s^y \right)
	+
	\tilde{t}^{(R)\prime}_e
	\tau^z\left( k_xk_ys^x + \frac{1}{2}(k_x^2-k_y^2) s^y \right),
	\nt
	t^{(R)}_{0,x}(\vk)s^x + t^{(R)}_{0,y}(\vk)s^y
	&=
	\tilde{t}^{(R)}_o\big( k_x s^x + k_y s^y \big)
	+
	\tilde{t}^{(R)\prime}_o\big( k_y s^x - k_x s^y \big),
\end{align}
where $\tilde{t}^{(R)}_e$, $\tilde{t}^{(R)\prime}_e$, $\tilde{t}^{(R)}_o$, and $\tilde{t}^{(R)\prime}_o$ are real numbers whose values would have to be obtained numerically.
Since the correction is linear in $\lambda_R$, we trivially conclude that $\tilde{t}^{(R)}_{e/o},\tilde{t}^{(R)\prime}_{e/o}\propto\lambda_R$.
Each independent term in Eq.~\eqref{eqn:Rashba-2dHam-kexpansion} is recorded in Table~\ref{tab:kexpansion}.

If we further impose mirror symmetry, we see that both $\tilde{t}^{(R)\prime}_e=\tilde{t}^{(R)\prime}_0=0$ as well.
However, unlike the situation without SOC, there is no reason for these terms to vanish in our scheme when the (physically relevant) Hamiltonian $\h_R$ is used instead of $\h_R^*$---the expressions that $\tilde{t}^{(R)\prime}_e$ and $\tilde{t}^{(R)\prime}_0$ multiply are thus included in Table~\ref{tab:kexpansion} without curly braces.

\subsubsection{\texorpdfstring{2$d$ flat band Ising Hamiltonian}{2d flat band Ising Hamiltonian}}\label{sec:IsingOnly}

Again, we use the symmetry action provided in Table~\ref{tab:OrdParamSyms} to obtain an effective Hamiltonian for the Ising SOC.
The preservation of both $\C_2\Tns$ and $\mathrm{U}_z(1)$, the symmetry responsible for spin rotations about the $z$-axis, implies
\begin{align}
	h_I(\vk)
	&=
	\sum_{\ell=0,z}\sum_{i=0,x,y}\tau^\ell \sigma^i\Big(
		h^{(I)}_{\ell,i,0}(\vk) + h^{(I)}_{\ell,i,z}(\vk)s^z
	\Big).
\end{align}
From the action of time-reversal, $\Tns$, we separate these expressions into even, `$e$,' and odd, `$o$,' components like in the previous sections:
\begin{align}
	h_{I,e}(\vk)
	&=
	h^{(I)}_{0,0,0}(\vk)+h^{(I)}_{0,x,0}(\vk)\sigma^x + h^{(I)}_{0,y,z}(\vk)\sigma^ys^z
	+
	\tau^z \Big[
		h^{(I)}_{z,0,z}(\vk)s^z + h^{(I)}_{z,x,z}(\vk)\sigma^xs^z+h^{(I)}_{z,y,0}(\vk)\sigma^y
	\Big],
	\nt
	h_{I,o}(\vk)
	&=
	h^{(I)}_{0,0,z}(\vk)s^z+h^{(I)}_{0,x,z}(\vk)\sigma^xs^z+h^{(I)}_{0,y,0}(\vk)\sigma^y
	+
	\tau^z\Big[
		h^{(I)}_{z,0,0}(\vk) + h^{(I)}_{z,x,0}(\vk)\sigma^x+h^{(I)}_{z,y,z}(\vk)\sigma^ys^z
	\Big].
\end{align}
As before, we project onto a single flat band close to $\gamma$, which is equivalent to setting $\sigma^x=\pm1$:
\begin{align}\label{eqn:Ham-1bnd-Ising-2d}
	\bar{h}^{(I)}_e(\vk)
	&=
	t^{(I)}_{0,0}(\vk) + t^{(I)}_{z,z}(\vk)\tau^zs^z,
	&
	\bar{h}^{(I)}_o(\vk)
	&=
	t^{(I)}_{0,z}(\vk)s^z + t^{(I)}_{z,0}(\vk) \tau^z.
\end{align}
Both $t^{(I)}_{0,0}$ and $t^{(I)}_{z,0}$ take an identical form to the SU(2) symmetric Hamiltonian of Eq.~\eqref{eqn:Ham-1bnd-NoSOC-2d}, and, further, as these terms are independent of spin, they are only generated by the Ising SOC at higher orders in $\lambda_I$.
We now use the $\C_3$ symmetry respected by Ising SOC to expand the remaining terms generated by the Ising SOC at linear order in $\lambda_I$, $t^{(I)}_{z,z}$ and $t^{(I)}_{0,z}$, in powers of $\vk$ about the $\gamma$ point:
\begin{align}\label{eqn:Ham-1bnd-Ising-2d-kexpansion}
	t^{(I)}_{z,z}(\vk)
	&=
	\tilde{t}^{(I)}_e,
	&
	t^{(I)}_{0,z}(\vk)
	&=
	\tilde{t}^{(I)}_{o} k_x(k_x^2-3k_y^2)
	+
	\tilde{t}^{(I)\prime}_{o} k_y(3k_x^2-k_y^2),
\end{align}
where $\tilde{t}^{(I)}_e$, $\tilde{t}^{(I)}_o$ and $\tilde{t}^{(I)\prime}_o$ are constants whose values depend on the details of the theory.
Analogous to the Rashba case, by explicit construction these constants are proportional to the Ising SOC coupling constant, $\lambda_I$: $\tilde{t}^{(I)}_e,\tilde{t}^{(I)}_o,\tilde{t}^{(I)\prime}_o\propto\lambda_I$.
In the above, only the lowest non-zero power of $\vk$ is kept.
When we choose to work with $\h_I^*$ in place of the (physically relevant) $\h_I$, $t^{(I)\prime}_o=0$.

\subsection{Trivial wire Hamiltonian}\label{sec:TrivialWireHam}

{\renewcommand{\arraystretch}{1.7}
\begin{table}
  \centering
  \begin{tabular*}{\textwidth}{@{\extracolsep{\fill}}p{20ex} l p{42ex} p{31ex}}
    \multicolumn{1}{l}{Proximate Order} 
    & \multicolumn{1}{l}{Wire Symmetries} 
    & \multicolumn{1}{l}{Wire Perturbations} 
	&	\multicolumn{1}{l}{Parameter Scaling}
	\\ \hline\hline
	trivial wire
    & $\mathcal{T}^2=-1$, $\mathrm{U}_\mathrm{v}(1)$
    & 
	$\frac{k^2}{2m}-\mu+ck\tau^z+k\boldsymbol{\alpha}_1\cdot\mathbf{s}+\tau^z\boldsymbol{\alpha}_2\cdot\mathbf{s}$
	&	$c\propto1/W^2$, 
	\newline
	$\alpha_1^{x,y}\propto\lambda_R$,
	$\alpha_2^{x,y}\propto\lambda_R/W^2$,
	\newline
	$\alpha_1^z\propto\lambda_I/W^2$,
	$\alpha_2^z\propto\lambda_I$
	\\\hline
    $\nu = 0$ singlet IVC 
	& $\mathcal{T}_{\rm sIVC}^2 = +1$ 
	& 
	$a_1 k \tau^x + 
    \tau^x\left(\beta_1^xs^x+\beta_1^ys^y\right)+
	\beta_2\tau^xs^z $ 
	& $a_1\propto1/W^2$, $\beta_1^{x,y} \propto \lambda_R/W^2$, 
	\newline
	$\beta_2 \propto \lambda_I$
	\\
    $\nu = 0 $ triplet IVC & $\mathcal{T}^2 = -1$ 
	& 
	$a_1'k\tau^xs^z + k\tau^y\left(\beta_1'^x s^x+\beta_1'^ys^y\right) + \beta_2'\tau^x $ 
	& $a_1'\propto 1/W^2$, 
	$\beta_1'^{x,y} \propto \lambda_R$, 
	\newline
	$\beta_2' \propto \lambda_I$
	\\ 
	$\nu = \pm 2$ FM IVC 
	& none 
	& 
	$a_1 k \tau^x + a_1'k\tau^xs^z+ a_1''k\tau^zs^z+ a_2''s^z $
	\newline 
	$+\tau^x(\beta_1^xs^x+\beta_1^ys^y) + \beta_2\tau^xs^z$
	\newline
	$+ k\tau^y(\beta_1'^xs^x+\beta_1'^ys^y)+\beta_2'\tau^x$
	\newline
	$+\beta_1''k + \beta_2''\tau^z$
	& 	
	$a_1,a_1',a_1''\propto1/W^2$,
	\newline
	$\beta_1^{x,y}\propto\lambda_R/W^2$, 
	$\beta_1'^{x,y}\propto\lambda_R$,
	\newline
	$\beta_2,\beta_2',\beta_2''\propto\lambda_I$,
	$\beta_1''\propto\lambda_I/W^2$
	\\ 
	$\nu = \pm 2$ AFM IVC 
	& $\mathcal{T}^2 = -1$  
	& $a_1'''k(\tau^xs^x+\tau^ys^y) + a_2'''k s^z+a_3'''\tau^zs^z$ 
	\newline
	$+ k\tau^x\boldsymbol{\beta}'''_1\cdot\mathbf{s}+ k\tau^y\boldsymbol{\beta}'''_2\cdot\mathbf{s}+ \beta_3'''\tau^x+\beta_4'''\tau^y $
	&
	$a_{1,2}'''\propto1/W^2$, 
	$|\boldsymbol{\beta}_{1,2}'''|\propto\lambda_R$,
	\newline
	$\beta_{3,4}'''\propto\lambda_R/W^2$
  \end{tabular*}
  \caption{\AppCaptions
  Reproduction of Table~\ref{WireTable} from the main text with an additional column providing the scaling of the parameters listed.
  Wire Hamiltonian terms for the trivial wire (first row) and perturbations generated by proximate IVC orders (subsequent rows) are shown.  
  The valley-orbit coupling $c$ persists in the absence of both IVC order and SOC;
  ${\bm \alpha}_{1,2}$ denote SOC terms that appear already for the trivial wire;
  couplings labelled by $a$'s survive with IVC order in the absence of SOC;
  and ${\beta}$ terms can be viewed as additional IVC order-parameter components generated due to SOC, akin to the spin-orbit-induced admixture of singlet and triplet pairing in inversion-asymmetric superconductors \cite{PhysRevLett.87.037004}. 
  For $\nu = 0$ triplet and $\nu = \pm 2$ IVC orders, we assumed that, without SOC, the spins orient in the out-of-plane ($\pm s^z$) direction. 
  The rightmost column lists the parameter scaling as a function of the SOC strengths, $\lambda_{I/R}$, and the wire width $W$.
  The $1/W^2$ dependence signals that $\C_3$ symmetry must be broken for the relevant terms to take nonzero values, which we assume in the majority of this appendix only occurs through the wire's presence.
  However, in the presence of strain or interaction-induced nematic order, the $1/W^2$ scaling should be replaced by functions quantifying such effects. 
  }
  \label{tab:WireTable}
\end{table}
}

We are finally in a position to discuss the effective Hamiltonian for the trivial wire presented in Table~1 of the main text and reproduced here in Table~\ref{tab:WireTable} (with an additional column).
As depicted in Fig.~1(b) of the main text, the wire is constructed by electrostic confinement of the 2$d$ system, which allows us to extract estimates of the wire Hamiltonian from the effective expressions we just derived.
Specifically, we assume that within a region around $\sim~10$ moiré unit cells wide, the chemical potential intersects the lower flat band band bottom, which is situated near the $\gamma$ point in momentum space; the filling in this region is $\nu_\mathrm{wire}=-4+\delta$, for some small $\delta$.
Outside the wire, in the bulk region, the chemical potential is tuned so that the flat bands are completely empty with $\nu_\mathrm{bulk}=-4$.
(Our analysis would proceed in a nearly identical fashion for $\nu_\mathrm{wire}=+4-\delta$ and
$\nu_\mathrm{bulk}=+4$).

We remark that within this approximation scheme, the Hamiltonian we derive  is equivalent to what would be obtained for a strip of finite width  (provided one ignores details related to the edges of the `wire').
The primary difference is that instead of tuning the chemical potential in the bulk region to lie within the gap separating the flat and dispersive bands, as done in our setup, in a finite-width strip the existence of any other bands is not considered and the `outer' regions are taken to be vacuum.
In terms of the symmetry-obtained Hamiltonians, however, the result is identical.

\subsubsection{Wire Hamiltonian without SOC}\label{sec:WireHamNoSOC}

We begin by considering the wire Hamiltonian when SOC is absent.
It is also convenient for the moment to ignore the `valley-orbit' coupling defined in Eq.~\eqref{eqn:Ham-1bnd-NoSOC-2d-kexpansion-Odd}: $\tilde{t}^{(0)}_{0,z},\tilde{t}^{(0)\prime}_{z,0}\to0$.
It is re-introduced below.
This simplification leaves us with the rotationally invariant quadratic Hamiltonian of Eq.~\eqref{eqn:Ham-1bnd-NoSOC-2d-kexpansion-Even}.
We assume for the moment that the wire extends in the $x$ direction. 
It follows that $k_x$ remains a good quantum number, allowing us to write the wavefunction as $\Phi_{k_x}(x,y)=e^{ik_xx}\phi_{k_x}(y)$.
The $y$-dependent portion of the wavefunction is obtained through the standard quantum mechanical particle in a potential well problem.
In the simplest case, one considers a box potential of width $W$.
The solution proceeds by first solving the Schrodinger equation in the three regions, $-W/2<y$, $-W/2<y<W/2$, and $W/2<y$ and subsequently implementing the appropriate boundary conditions.
Regardless of the wire profile details, the negative energy bound states are confined to the wire region, $-W/2\lesssim y\lesssim W/2$, and, further, assuming an even wire profile, the lowest energy states will possess even parity: $\phi_{k_x}(y)=\phi_{k_x}(-y)$.
These wavefunctions form the foundation of the analysis that follows here and in subsequent sections.

We now re-introduce the valley-orbit terms by projecting them to the space spanned by the confined wavefunctions, $\phi_{k_x}(y)$.
In particular, the odd portion of the wire Hamiltonian is obtained via
\begin{align}\label{eqn:TrivialWire-OddProj}
    {h}^{(0)}_{\mathrm{wire},o}(k_x=k)
    &=
    \tau^z
    \int dy\,\phi_k^\dag(y)\left[	
    \tilde{t}_{z,0}^{(0)}k\big(k^2-3(-i\partial_y)^2\big)
	+
	\tilde{t}_{z,0}^{(0)\prime}(-i\partial_y)\big(3k^2-(-i\partial_y)^2\big)\right]
	\phi_k(y)
	\nt
	&=
	\left[
	\tilde{t}_{z,0}^{(0)}k\big(
	    k^2 - 3 \braket{k_y^2}
	    \big)
	    +
	   \tilde{t}_{z,0}^{(0)\prime}\big(
	   3\braket{k_y}k^2-\braket{k_y^3}\big)
	\right]\tau^z,
\end{align}
where
\begin{align}\label{eqn:WireProjDef}
    \Braket{k_y^n}
    &=
    \int dy\,\phi_k^\dag(y)(-i\partial_y)^n \phi_k(y).
\end{align}
The $k$-dependence of the correlation function is left implicit.
A consequence of the even parity of $\phi_k(y)$ is that only even powers of $n$ return non-zero values.
The total trivial wire Hamiltonian we find is
\begin{align}  \label{eqn:TrivialWireHam}
    h_\mathrm{wire}^{(0)}(k)
    &=
    \frac{k^2}{2m}-\mu
    -
	3\tilde{t}_o^{(0)}\Braket{k_y^2}k\tau^z,
\end{align}
where a constant has been absorbed into the chemical potential and
higher powers of $k$ are ignored.
Up to shifts that can be easily accounted for by the back gate voltage, the mass term and chemical potential are simply the 1$d$ version of the flat band Hamiltonian, \emph{i.e.}, what one would get by ignoring all $k_y$ dependence. 
By contrast, we see that the breaking of $\C_3$ symmetry allows a linearly dependent valley-orbit term proportional to $\tau^z$, whose coefficient $c$ in the first row of Table~\ref{tab:WireTable} is equal to $-3\tilde{t}_o^{(0)}\Braket{k_y^2}$.
One generally expects $\Braket{k_y^2}\sim (2\pi/W)^2$, and it follows that $c\propto1/W^2$, as indicated in the fourth column of Table~\ref{tab:WireTable}.
Importantly, the scaling of $c$ is itself independent of the wire direction: we always find $c\propto\braket{k^2_\perp}\propto1/W^2$ ($k_\perp$ is the transverse momentum).
Note, however, that had we instead chosen our wire to lie along the $y$-direction, we would have found $c=3\tilde{t}_o^{(0)\prime}\braket{k_x^2}$.
This relation satisfies the scaling $c\propto1/W^2$, but  within our perturbative expansion, the mirror symmetry should be imposed upon the spin symmetric portion of the trivial wire Hamiltonian---in which case $\tilde{t}^{(0)\prime}_o=0$.
In formulating Table~\ref{tab:WireTable}, we assume that the wire is oriented in an arbitrary direction that does not possess these cancellations.

As we discuss in the next section, explicit breaking of the $\C_3$ symmetry either through strain or interaction-induced nematicity would also contribute to the coefficient $c$.
This observation follows from noting that it was the $\C_3$ symmetry that prohibited the linear terms from appearing in the expression for $t_{0,z}^{(0)}(\vk)$ in Eq.~\eqref{eqn:Ham-1bnd-NoSOC-2d-kexpansion-Odd}.

Finally, one can ask what would occur had we derived the confined wavefunctions $\phi_k$ starting with the valley-orbit-coupled Hamiltonian.
The result is analogous save for an additional phase: $\tilde{\phi}_{k_x}(y)\sim e^{i\theta y}\phi_{k_x}(y)$ with $\theta\sim 2m\cdot3\tilde{t}^{(0)}_o\braket{k_y^2}$.
Our assumption that this term is small is supported by the factor of $1/W^2$---any errors resulting from the fact that we study the problem using the functions $\phi_{k_x}$ instead of $\tilde{\phi}_{k_x}$ will only manifest at higher orders in $1/W^2$, which is equivalent to our truncation in powers of $\vk$.

\subsubsection{Wire Hamiltonian for Rashba SOC}

As in the case without SOC, we extract the wire Hamiltonian directly from the results of Sec.~\ref{sec:2d-Rashba-Heff} (or, alternatively, from Table~\ref{tab:kexpansion}) under the assumption that the wire points along the $x$-direction.
We find
\begin{align}
	h^{(R)}_\mathrm{wire}(k)
	&=
	-\frac{1}{2}\braket{k_y^2}
	\tau^z\big(
	    \tilde{t}^{(R)}_e s^x
	    +
	    \tilde{t}^{(R)\prime}_es^y
	    \big)
	  +
	  k\big(
	  \tilde{t}^{(R)}_o s^x
	  -
	  \tilde{t}^{(R)\prime}_o s^y
	  \big)
\end{align}
where $k\equiv k_x$ and the correlations are defined via Eq.~\eqref{eqn:WireProjDef}, with correlation functions for all odd powers of $k_y$ assumed to vanish.
Comparing with the notation of the main text reproduced in Table~\ref{tab:WireTable}, we conclude that
\begin{align}
	\alpha_1^x&=\tilde{t}_0^{(R)},
	&
	\alpha_1^y&=-\tilde{t}_0^{(R)\prime},
	&
	\alpha_2^x&=-\frac{1}{2}\Braket{k_y^2}\tilde{t}^{(R)}_e,
	&
	\alpha_2^y&=-\frac{1}{2}\Braket{k_y^2}\tilde{t}^{(R)\prime}_e.
\end{align}
While it's clear that all four parameters above are proportional at leading order to the UV Rashba coupling constant, $\lambda_R$ (per Eq.~\eqref{eqn:SpinOrbitContributions}), both $\alpha_2^x$ and $\alpha_2^y$ are also proportional to $\Braket{k_y^2}\sim1/W^2$.
These scalings are noted in Table~\ref{tab:WireTable}.

\subsubsection{Wire Hamiltonian for Ising SOC}

We follow the procedure of the previous two sections, now with Eqs.~\eqref{eqn:Ham-1bnd-Ising-2d} and~\eqref{eqn:Ham-1bnd-Ising-2d-kexpansion} (or the row labelled `Ising' in Table~\ref{tab:kexpansion}) as our starting point. 
We find
\begin{align}
	h^{(I)}_\mathrm{wire}(k)
	&=
	\tilde{t}^{(I)}_e\tau^zs^z
	3\tilde{t}^{(I)}_{o}\Braket{k_y^2} k,
\end{align}
implying that
\begin{align}
    \alpha_1^z&=-3\tilde{t}^{(I)}_o\Braket{k_y^2},
	&
	\alpha_2^z&=\tilde{t}^{(I)}_e.
\end{align}
Once more, we conclude that $\alpha_1^z\propto\lambda_I$ and $\alpha_2^z\propto\lambda_I/W^2$ as shown in Table~\ref{tab:WireTable}.

\subsection{\texorpdfstring{Analysis in the absence of $\M_y$ or $\C_3$}{Analysis in the absence of My or C3}}\label{sec:TrivialWireC3MyBreaking}

It is worth emphasizing which aspects of the analysis above hold when $\M_y$ and/or $\C_3$ are not present in even the  continuum model $H_\mathrm{cont}$ without SOC, especially in light of the recently proposed intervalley Kekulé spiral phase \cite{kwan_kekule_2021}.
In particular, we explain that provided $\Tns$, $\C_2$, and the relevant continuous symmetries are preserved, the expressions derived in Eqs.~\eqref{eqn:Ham-1bnd-NoSOC-2d}, \eqref{eqn:Rashba-2dHam}, and~\eqref{eqn:Ham-1bnd-Ising-2d} remain valid, including the parity assignments.

We first consider the case where $\M_y$ is not present in the SOC-free Hamiltonian.
The basis chosen in Eq.~\eqref{eqn:FlatBandBasis} is no longer relevant, and it is more convenient to keep to the operators $\tilde{c}(\vk)$ of Eq.~\eqref{eqn:TrivialCaseDiag-1}.
We restrict this operator to a single band, $\tilde{c}_{v,n,\alpha=1}(\vk)\to \tilde{c}_{v,n}(\vk)$, where $v$ and $n$ are valley and spin indices respectively, and $\alpha=1$ specifies the band of interest.
Importantly, because the Hamiltonian $H_\mathrm{cont}$ preserves $\Tns$ and $\C_2\Tns$, these symmetries may be chosen to act as 
\begin{align}
    \Tns&:\quad
    \tilde{c}(\vk)
    \to\tilde{c}(-\vk),
    \quad
    i\to-i,
    &
    \C_2\Tns&:\quad
    \tilde{c}(\vk)
    \to 
    e^{i\tilde{\theta}(\vk)}c(\vk),
    \quad
    i\to -i,
\end{align}
which is identical to what you obtain by projecting onto $\sigma^x=\pm1$ in Eq.~\eqref{eqn:FlBandSyms-1}, as argued in Sec.~\ref{sec:FlatBandHeff-NoSOC}.
Although it is now clear that the projection procedure in the sections above could have been avoided, the basis Eq.~\eqref{eqn:FlatBandBasis} will prove useful in the following section. 

When $\C_3$ is preserved, its action on the $\tilde{c}(\vk)$ operators also remains unchanged, and the analysis of the previous sections follows as described above with the minor difference that certain terms previously prohibited by the mirror symmetry are now present, \emph{i.e.}, the curly braced terms in Table~\ref{tab:kexpansion} are now equally valid.
Conversely, when $\C_3$ is broken, the effective Hamiltonians of Eqs.~\eqref{eqn:Ham-1bnd-NoSOC-2d}, \eqref{eqn:Rashba-2dHam}, and~\eqref{eqn:Ham-1bnd-Ising-2d} may be directly translated to the one dimensional wire limit by simply taking $\vk\to k$ and expanding the functions $t_{\mu,a}(\vk)$ in even or odd powers of $k$.
Returning to Table~\ref{tab:kexpansion}, this step is equivalent to noting only which matrices $\tau^\mu s^a$ are allowed in each scenario as well as the parity of the term $t_{\mu,a}(\vk)$ that multiplies it.
While no information is gained on the scaling of the parameters in the theory except for the $\lambda_{R/I}$ dependence, by comparison with the $\C_3$ symmetric analysis, it is clear that all terms described as scaling with $1/W^2$ in Sec.~\ref{sec:TrivialWireHam} and Table~\ref{tab:WireTable} must now depend on the strain or nematic order parameter responsible for the $\C_3$ breaking.

\section{\texorpdfstring{Wire adjacent to IVC order at $\nu=0$ and $\nu=\pm2$}{IVC and IVC proximity-coupled wire at nu=0 and nu=+-2}}\label{sec:IVCwire}

Here we detail our derivation of the $1d$ Hamiltonians describing a wire proximity-coupled to various IVC insulating states at charge neutrality and $\nu=\pm2$. 
In Sec.~\ref{sec:GroundStateSelection}, we begin by sketching the ideas of Ref.~\citenum{bultinckGroundStateHidden2020} that lead to the proposed IVC ground states and discussing the effect SOC may have on their analysis.
We continue in Sec.~\ref{sec:Methodology} by outlining the methodology we use to obtain the effective Hamiltonians of IVC-proximity-coupled wires.
The details of the specific IVC orders themselves are described in Secs.~\ref{sec:CNPordParamDefs} and~\ref{sec:nu2_ordParamDefs}; the results of these sections appear in Table~\ref{tab:WireTable} and are based on the information provided in Tables~\ref{tab:OrdParamSyms} and~\ref{tab:kexpansion}.
In the final two subsections, we address some subtleties of the analysis---namely the relation between the internal SOC direction and the physical wire direction (Sec.~\ref{sec:InternalSOCDirection}) and topological aspects of the various IVC states (Sec.~\ref{sec:IVCTopoAspects}).

\subsection{Selection of IVC ground states}\label{sec:GroundStateSelection}

Our focus in the main text on IVC states is motivated in part by the perturbative expansion outlined in Ref.~\citenum{bultinckGroundStateHidden2020}.
Before delving into the wire setup itself, we briefly summarize the authors' reasoning and pertinent conclusions, but we stress that this discussion is not intended to capture all of the details or physics of that article.
The authors of Ref.~\citenum{bultinckGroundStateHidden2020} did not consider the effect of SOC on the ground state, so we restrict our discussion for the moment to the spin symmetric case.

The Hamiltonian of interest is
\begin{align}\label{eqn:Htot_NoSOC}
	H
	&=
	H_\mathrm{cont}
	+ 
	H_C
	+
	H_J,
\end{align}
where $H_\mathrm{cont}$ is the continuum model Hamiltonian (see Eq.~\eqref{eqn:HcontBravaisDef}), 
and $H_C$ and $H_J$ together represent the Coulomb interaction (see Eq.~\eqref{eqn:CoulombIntExpansion}).
The first step taken in Ref.~\onlinecite{bultinckGroundStateHidden2020} is to project the full Hamiltonian onto the flat bands, which are defined by diagonalizing $H_\mathrm{cont}$ as done in Sec.~\ref{sec:2dHeff-NoSOC}:
\begin{align}\label{eqn:Htotfl_NoSOC}
	H
	&\to
	H^{(\fl)}
	=	
	H_\mathrm{cont}^{(\fl)}
	+ 
	H_C^{(\fl)}
	+
	H_J^{(\fl)}.
\end{align}
Here, $H_\mathrm{cont}^{(\fl)}$ has appeared already in Eq.~\eqref{eqn:FlBandContMod}.
This simplification is valid provided the gap separating the flat and remote bands is larger than the other scales of the theory. 
In experiments, it is measured to be approximately $\Delta_\mathrm{disp}\sim\unit[30]{\text{meV}}$ \cite{choiElectronicCorrelationsTwisted2019,polshynLargeLinearInTemperatureResistivityInTBG}---larger than the scales discussed below, though not by orders of magnitude.

Within this flat band subspace, a number of approximate particle-hole-like symmetries are identified, which are then employed to separate $H_C^{(\mathit{fl})}$ into symmetric and antisymmetric contributions, respectively denoted $H_{C,S}^{(\mathit{fl})}$ and $H_{C,A}^{(\mathit{fl})}$.
Close to the magic angle, $\theta\sim1.1^\circ$, the corresponding energy scales are estimated to be $U_S\sim\unit[15-20]{\text{meV}}$ for $H_{C,S}^{(\mathit{fl})}$ and $U_A\sim\unit[4-6]{\text{meV}}$ for $H_{C,A}^{(\mathit{fl})}$. 
The band energies implied by the quadratic term $H_\mathrm{cont}^{(\fl)}$---the eigenvalues $\epsilon_\alpha(\vk)$, $\alpha\in\mathit{fl}$ of Eq.~\eqref{eqn:TrivialCaseDiag}---are also small compared to $U_S$ close to the magic angle (note, however, that the quadratic part of the \emph{full} Hamiltonian, $H_\mathrm{cont}$, provides the leading energy scale $\Delta_\mathrm{disp}$ and is responsible for the projector used to obtain $H_C^{(\mathit{fl})}$ and thus $H_{C,S}^{(\mathit{fl})}$ and $H_{C,A}^{(\mathit{fl})}$).
In Ref.~\onlinecite{bultinckGroundStateHidden2020}, 
the typical scale set by $H_\mathrm{cont}^{(\fl)}$ (and thus the energies $\epsilon_\alpha(\vk)$) is estimated as $t_\mathit{fl}\sim \unit[4-6]{\text{meV}}$ in the absence of strain close to the magic angle.
This figure takes into account the specific Hartree-Fock scheme the authors consider, and we refer the reader to the appendix of Ref.~\citenum{bultinckGroundStateHidden2020} for details.
Finally, the Hund's term $H_J^{(\mathit{fl})}$ is characterized by a much smaller scale $J_H\sim\unit[0.2-0.5]{\text{meV}}$ (which is approximately $\theta\cdot U_{S}$ for $\theta\sim1.1^\circ$).
We therefore roughly arrive at the following hierarchy of scales:
\begin{align}
    \label{eqn:ScaleHierarchy_NoSOC}
    \Delta_\mathrm{disp}\gg U_S\gg U_A\sim t_\mathit{fl} \gg J_H.
\end{align}
While some of these inequalities may be questioned, the conclusions reached in Ref.~\citenum{bultinckGroundStateHidden2020} are supported by numerics.

This separation of scales is useful since since it further implies a hierarchy of symmetries and thus of symmetry breaking.
The symmetric part of the long-range Coulomb interaction, $H_{C,S}^{(\mathit{fl})}$, is invariant under an enlarged symmetry group $\mathrm{U}(4)\times\mathrm{U}(4)$.
The inclusion of $H_{C,A}^{(\mathit{fl})}$ and $H_\mathrm{cont}^{(\fl)}$ break this symmetry to the $\mathrm{U}(2)_K\times \mathrm{U}(2)_{K'}\cong \mathrm{U_c}(1)\times\mathrm{U_v}(1)\times\mathrm{SU}(2)_K\times\mathrm{SU}(2)_{K'}$ symmetry described above (see Eq.~\eqref{eqn:EnlargeContSym}).
Only at scales given by the Hund's term is this enlarged symmetry reduced to the `physical' $\mathrm{U_c}(1)\times\mathrm{U_v}(1)\times\mathrm{SU}(2)$.

Following the above arguments, we first focus solely on $H_{C,S}^{(\mathit{fl})}$, whose ground states are determined to spontaneously break the $\mathrm{U}(4)\times\mathrm{U}(4)$ effective symmetry---hence, action on one ground state by this large symmetry group returns an equally valid ground state.
This degeneracy is subsequently broken by the inclusion of the subleading terms $H_{C,A}^{(\mathit{fl})}$ and $H_\mathrm{cont}^{(\fl)}$.
Within this $\mathrm{U}(4)\times\mathrm{U}(4)$ manifold of $H_{C,S}^{(\mathit{fl})}$ ground states, states possessing IVC order have the lowest energy when both $H_\mathrm{cont}^{(\fl)}$ and $H_{C,A}^{(\mathit{fl})}$ are present.
Like the terms so far considered, the IVC ground states also possesses a $\mathrm{U}(2)_K\times\mathrm{U}(2)_{K'}$ degeneracy.
We consider the inclusion of
these three terms---$H_{C,S}^{(\fl)}$, $H_{C,A}^{(\fl)}$, and $H_\mathrm{cont}^{(\fl)}$---as the ``zeroth order'' and ``first order'' contributions. 
Below the scales set by $U_A$ and $t_\fl$, we make no definitive conclusions regarding the energetically preferred ground state.
For this reason, it is convenient to define the ``leading order Hamiltonian,''
\begin{align}\label{eqn:LeadingOrderHam}
	H^{(\fl)}_{0,1}
	&=
	H_{C,S}^{(\fl)}
	+
	H_{C,A}^{(\fl)}
	+
	H_\mathrm{cont}^{(\fl)},
\end{align}
whose ground states are  $\mathrm{U}(2)_K\times\mathrm{U}(2)_{K'}$-degenerate IVC states.
The details of the relevant states for $\nu=0$ and $\nu=\pm2$ are given below in Secs.~\ref{sec:CNPordParamDefs} and~\ref{sec:nu2_ordParamDefs}, respectively.

The Hund's term $H_J^{(\mathit{fl})}$ is only relevant in selecting among ground states that are otherwise degenerate at the level of $H_{0,1}^{(\mathit{fl})}$.
However, given its small magnitude, $H_J^{(\mathit{fl})}$ is susceptible to substantial renormalization by effects so far not considered; in particular, phonon interactions could alter the sign of the coupling in Eq.~\eqref{eqn:CoulombIntExpansion}: $V_\mathrm{eff}(\vK+\vk)\cong V_\mathrm{eff}(\vK)$ \cite{chatterjee_skyrmion_20} (where $J_H\sim\abs{V_\mathrm{eff}(\vK)}$).
Given the inherent uncertainty in such calculations, we take the sign of the Hund's coupling as an  unknown, which leads us to consider two distinct (but related) many-body ground states for the both $\nu=0$ and $\nu=\pm$2.
Note that because $H_J^{(\fl)}$ preserves the spin SU(2), spin rotations of whichever ground state is chosen return an energetically equivalent ground state.

We now ask how these considerations are altered by the inclusion of SOC.
Equations~\ref{eqn:Htot_NoSOC} and~\ref{eqn:Htotfl_NoSOC} are modified to 
\begin{align}
	H'
	&=
	H_\mathrm{cont}
	+
	H_C
	+
	H_J
	+
	H_\mathrm{SOC}
	\to
	H^{(\fl)\prime}
	=
	H_\mathrm{cont}^{(\fl)}
	+
	H_C^{(\fl)}
	+
	H_J^{(\fl)}
	+
	H_\mathrm{SOC}^{(\fl)},
\end{align}
where $H_\mathrm{SOC}$ and $H_\mathrm{SOC}^{(\fl)}$ are give in Eqs.~\eqref{eqn:ContModSOCterm} and~\eqref{eqn:FlatBandProject}, respectively. 
Guided by TBG experiments, we assume that the energy scale of $H_\mathrm{SOC}^{(\fl)}$ is $\lambda_\mathrm{SOC}\sim\lambda_{I,R}\sim \unit[2-3]{\text{meV}}$,\footnote{We note that this estimate is lower than the bound provided above in Sec.~\ref{sec:2dHeff-WithSOC}. 
There, $\lambda_{I,R}$ represented the parameters that appear directly in the full continuum model. 
By contrast, $\lambda_\mathrm{SOC}$ is the energy scale of the flat-band-projected Hamiltonian $H_\mathrm{SOC}^{(\fl)}$, which we naturally expect to be smaller than $\lambda_{I,R}$. 
Based on the perturbative calculation alluded to in Sec.~\ref{sec:MirrorSymIsing}, we could identify $\lambda_\mathrm{SOC}\sim\tilde\lambda_{I,R}$ (where $\tilde{\lambda}_R$ is defined through the same procedure as $\tilde\lambda_I$).
This perspective is also supported by experiments.
While transport is able to resolve the SOC gap \cite{aroraSuperconductivityMetallicTwisted2020},
STM studies, with resolutions of $\sim\unit[1-2]{\text{meV}}$ cannot \cite{choiCorrelationdrivenTopologicalPhases2021,choi_interaction-driven_2021}.
}
which is intermediate between $t_S$, $U_A$ and the Hund's scale, $J_H$.
That is, we modify Eq.~\eqref{eqn:ScaleHierarchy_NoSOC} to give 
\begin{align}\label{eqn:ScaleHierarchy_SOC}
    \Delta_\mathrm{disp}\gg U_S\gg U_A\sim t_\fl \gtrsim
    \lambda_\mathrm{SOC}\gtrsim J_H.
\end{align}
Hence, it is plausible that $H_\mathrm{SOC}^{(\fl)}$---as opposed to the Hund's term---selects the true ground state among the $\mathrm{U}(2)_K\times\mathrm{U}(2)_{K'}$ symmetric manifold of IVC states degenerate at the level of $H_{0,1}^{(\fl)}$.,
In this case, since SOC violates the spin SU(2), the resulting ground state may possess a preferred spin direction, in contrast to the ground states selected by the spin-symmetric Hund's term.
When restricting to uniform IVC order, we do not expect this distinction to affect the results, as we address below.

We acknowledge that Eq.~\eqref{eqn:ScaleHierarchy_NoSOC} suggests that since $t_\fl$ and $U_S$ are only two or three times larger than $\lambda_\mathrm{SOC}$ it may not be entirely valid to consider on the effects of $H_{C,A}^{(\fl)}$ and $H_\mathrm{cont}^{(\fl)}$ in the absence of $H_\mathrm{SOC}^{(\fl)}$.
We note, however, that the scale associated with $H_\mathrm{cont}^{(\fl)}$ can be substantially increased through strain \cite{biDesigningFlatBands2019}, so that the effects of SOC may again be treated a subleading.
As addressed in the outlook section of the main text, while the presence of strain may ultimately favour a different type of IVC order, the intervalley Kekulé spiral \cite{kwan_kekule_2021}, the primary conclusions of our analysis remain unchanged (see Sec.~\ref{sec:TrivialWireC3MyBreaking}).

\subsection{Outline of methodology}\label{sec:Methodology}

Having presented the reasoning behind our focus on IVC states, we now describe our derivation of the effective Hamiltonian of an IVC proximity-coupled wire.
The details of the specific IVC insulators alluded to in the previous section are given in subsequent sections (see Secs.~\ref{sec:CNPordParamDefs} and~\ref{sec:nu2_ordParamDefs}).

As discussed, the effective Hamiltonian derived in Sec.~\ref{sec:TrivialWireHam} could equally well describe a `wire' in the form of a nanoribbon with physical (though assumed unimportant) boundaries in addition to the electrostatically defined `wires' considered here.
The derivation of the proximity-coupled wire could then proceed by first coupling the degrees of freedom of the trivial wire Hamiltonian to the gapped bulk degrees of freedom of the adjoining IVC phase under consideration.
Since the IVC phases are gapped, an effective Hamiltonian for the proximity-coupled wire could be obtained by systematically integrating out the bulk fermions.
The chemical potential necessary to obtain the adjoining IVC phases of interest and
whether or not the wire was obtained via electrostic gating or in some nanoribbon scenario are not expected to alter the universal, symmetry-constrained physics of interest here.
However, even with the assumptions already in place, this integrating-out procedure remains impractical.
We instead consider a closely related companion problem whose qualitative features are expected to be the same.

In the scenario we consider, we start with two 2$d$ TBG systems stacked atop one another and allowed to tunnel \emph{very} weakly (much more weakly than the graphene sheets of each TBG system are coupled). 
We denote these the primary system and the auxiliary system. 
In the primary system, the chemical potential is tuned to lie close to the flat band bottom around the $\gamma$ point.
By constrast, the chemical potential of the auxiliary system is chosen to lie either at $\nu=0$ (charge neutrality) or at $\nu=\pm2$.
In both situations, we assume that as a result of interactions, the auxiliary system spontaneously breaks $\mathrm{U_v}(1)$ and possibly other symmetries, resulting in a intervalley coherent (IVC) insulator whose nature is described in the sections below.
Because the auxiliary system is gapped, it only has a perturbative effect on the gapless degrees of freedom in the primary system.
In particular, in second order perturbation theory, it alters the energy of the primary system's Hamiltonian at $\sim\mathcald{O}(t^2/\Delta_\mathrm{IVC})$ where $t$ is the tunnelling strength between the primary and auxiliary systems and $\Delta_{\mathrm{IVC}}$ is the gap of the auxiliary system---set, of course, by the IVC order.
We can then derive the effective Hamiltonian of the primary system in the 2$d$ limit using symmetry arguments in a manner directly analogous to the analysis of Secs.~\ref{sec:FlatBandHeff-NoSOC}, \ref{sec:2d-Rashba-Heff}, and~\ref{sec:IsingOnly}. 
Importantly, even though the order parameter may fundamentally alter the structure of the flat bands within the auxiliary system, it does not do so in the primary system in the sense that the basis and electron operators defined in Eq.~\eqref{eqn:FlatBandBasis} continue to provide a good description of the system.
Finally, we imagine obtaining our wire Hamiltonian by reducing the primary system to a narrow strip, allowing the methodology of Sec.~\ref{sec:TrivialWireHam} to be applied.
That is, we envision projecting the IVC corrections onto the wire wavefunctions $\phi_k(y)$ defined through Eqs.~\eqref{eqn:TrivialWire-OddProj} and~\eqref{eqn:WireProjDef}. 
The primary and auxiliary systems are now more correctly denoted the wire and bulk systems, respectively. 
Similary, the `interlayer' tunnelling constant $t$ is identified as the tunnelling strength of the narrow wire region to the adjoining bulk states.

We consider two possible IVC order parameters for the auxiliary/bulk system at both $\nu=0$ and $\nu=\pm2$, the details of which are provided below.
For each IVC order, we break our study into three scenarios, which we consider in order:
\begin{enumerate}
	\item IVC order without SOC
	\item IVC with Rashba SOC
	\item IVC with Ising SOC
\end{enumerate}
(Following the arguments of Sec.~\ref{sec:2dHeff-WithSOC}, terms arising from the simultaneous presence of Rashba and Ising SOC are neglected.)
For each of these scenarios, we perform the following analysis:
\begin{itemize}
	\item We note which symmetries remain and record the results in Table~\ref{tab:OrdParamSyms}.
	\item For these symmetries, we determine all terms allowed in the 2$d$ effective Hamiltonian of the primary system in an expansion about the $\gamma$ point.
	\item Terms that were prohibited in the effective Hamiltonian without IVC order (first three rows of Table~\ref{tab:kexpansion}) or in a previous scenario (\emph{i.e.}, terms allowed with IVC but without SOC) are typically allowed for each new scenario. 
	These new terms are recorded in Table~\ref{tab:OrdParamSyms}.
	Note that this step implies a built-in hierarchy, in which more symmetric scenarios are considered first.
	\item Following Sec.~\ref{sec:TrivialWireHam}, we use the 2$d$ Hamiltonian we just derived to obtain the corresponding 1$d$ wire Hamiltonian along with the scaling of the parameters within the theory. 
	These results are shown in Table~\ref{tab:WireTable}.
\end{itemize}
While the terms generated in scenarios 2 and 3 will be proportional to $\lambda_R$ and $\lambda_I$, we cannot conclude that any of the additional terms derived in this fashion are proportional to $\Delta_\mathrm{IVC}$ or even $t^2/\Delta_\mathrm{IVC}$.
The confounding factor is that the second order perturbative expansion necessarily alters the quasiparticle weight of the effective flat-band operators, $Z\sim[1+t^2/\Delta_\mathrm{IVC}]^{-1}$, which in turn requires that the effective Hamiltonian pick up an additional factor, $h_\mathrm{eff}\to Z h_\mathrm{eff}$.
The result is an effective IVC coupling $\tilde{\Delta}_\mathrm{IVC}\sim Z\cdot t^2/\Delta_\mathrm{IVC}=-(1-Z)\Delta_\mathrm{IVC}$. 
The terms generated accordingly depend on $\Delta_\mathrm{IVC}$ and $t$ in a complicated and nonuniversal fashion that may not prove relevant to the physical situation of interest.

Although the steps outlined above all assume $\C_3$ and mirror symmetry at the level of the trivial Hamiltonian, the discussion in Sec.~\ref{sec:TrivialWireC3MyBreaking} is equally valid here.
Without mirror, the terms included in Table~\ref{tab:kexpansion} are allowed. 
When $\C_3$ is absent, the information contained in Table~\ref{tab:kexpansion} should be interpreted in a drastically simpler fashion: only the matrices allowed and the parity of the terms they multiple are relevant to the wire problem.

\subsection{\texorpdfstring{$\nu=0$ IVC order parameters}{nu=0 IVC order parameters}}\label{sec:CNPordParamDefs}

In the previous section, we briefly sketched the arguments of Ref.~\citenum{bultinckGroundStateHidden2020} that the leading order contribution in TBG was $H_{0,1}^{(\fl)}$ in Eq.~\eqref{eqn:LeadingOrderHam} and that the ground state of this Hamiltonian was an IVC insulator at $\nu=0$.
Following the treatment given in the appendix of Ref.~\citenum{bultinckGroundStateHidden2020}, we now present these candidate $\nu=0$ IVC ground states in more detail.

The ground state may be defined through a projection operator $\mathcal{P}_{\nu=0}$: all occupied (empty) states in the IVC insulator satisfy $\mathcal{P}_{\nu=0}\Ket{\phi}=\Ket{\phi}$ ($\mathcal{P}_{\nu=0}\Ket{\phi}=0$).
In a given basis, the projector can generally be written in terms of what we will refer to as the \emph{order parameter} $Q_{\nu=0}$:
\begin{align}
	\mathcal{P}_{\nu=0}(\vk)&=\frac{1}{2}\big(\id+ Q_{\nu=0}(\vk)\big),
	&
	Q^2_{\nu=0}(\vk)&=\id,
\end{align}
where both $Q_{\nu=0}$ and $\mathcal{P}_{\nu=0}$ are $8\times8$  matrices.
In the basis satisfying Eq.~\eqref{eqn:FlBandSyms-1}, the matrix $Q_{\nu=0}(\vk)$ corresponding to an IVC state may be expressed in a momentum independent fashion as
\begin{align}\label{eqn:CNP_IVC_general}
	Q_{\nu=0}&=\sigma^y\otimes 
	\begin{pmatrix}
		0	&	V	\\
		V^\dagger&	0	
	\end{pmatrix}_{\!\tau},
	&
	V&=e^{i\phi}e^{i\frac{\theta}{2}\hat{\v{n}}\cdot\v{s}},
\end{align}
where the $\sigma^a$ Pauli matrices act on the ``band'' indices, $\tau^a$ Paulis act on the valley indices, and $s^a$ Paulis act on the spin. 
As discussed, although not true across the full moiré Brillouin zone, close to the $\gamma$ point, we can set $\theta(\vk)=0$ in Eq.~\eqref{eqn:FlBandSyms-1} and describe the action of $\C_3$ and $\M_y$ via Eq.~\eqref{eqn:FlBandSyms-2}.

The Hamiltonian $H_{0,1}^{(\fl)}$ possesses an effective symmetry under which the spins and phases of states originating in each valley can be rotated separately, $\mathrm{U}(2)_K\times\mathrm{U}(2)_{K'}\cong\mathrm{U_v(1)}\times\mathrm{U_c(1)}\times\mathrm{SU}(2)_K\times \mathrm{SU}(2)_{K'}$, where $\mathrm{U_v}(1)$ ($\mathrm{U_c}(1)$) is the U(1) valley (charge) symmetry and $\mathrm{SU}(2)_K$ ($\mathrm{SU}(2)_{K'}$) correspond to spin rotations in the $K$ ($K'$) valley (see Eq.~\eqref{eqn:ContinuousSymmetryAction}).
The order parameter $Q_{\nu=0}$ breaks this symmetry down to the usual charge symmetry $\mathrm{U_c}(1)$ as well as a residual $\mathrm{SU}(2)$.
That is, for all choices of $V$, there exists a subgroup $\mathrm{SU}(2)\subset \mathrm{U_v}(1)\times\mathrm{SU}(2)_K\times\mathrm{SU}(2)_{K'}$ whose action leaves $Q_{\nu=0}$ invariant.
A Chern number can be assigned to the bands through $C=\tau^z\sigma^z$ \cite{bultinckGroundStateHidden2020}, and one finds that within each (residual) SU(2) sector, the corresponding IVC insulator has a filled band with Chern number $C=+1$ and $C=-1$ that are related by an antiunitary ``time reversal'' symmetry.
As the residual SU(2) is in reality broken, the $\nu=0$ IVC states are not true topological states, as we discuss below for the specific cases under consideration.

At the level of $H_{0,1}^{(\fl)}$, ground states defined by $Q_{\nu=0}$ for all functions $V$ are energetically equivalent.
In Sec.~\ref{sec:GroundStateSelection}, we discussed how states lying within this SU(2) manifold are distinguished by the valley Hund's term when SOC is not present.
Depending on the sign of $J_H$, $H_J^{(\fl)}$ either prefers a singlet state in which $V=e^{i\phi}$ or a triplet state in which $V=e^{i\phi}e^{i\pi\hat{\v{n}}\cdot\v{s}/2}=e^{i\phi}i\hat{\v{n}}\cdot\v{s}$.
Although SOC may in fact constitute a larger energy scale than the Hund's term $J_H$, these two states remain the natural candidate ground states.
The only alternative is a state in which singlet and triplet orders are mixed, but since these two states are distinguished by the action of time reversal (the singlet state breaks $\mathcal{T}=is^y\Tns$, while the triplet state preserves it), they will not be mixed by the addition of a time-reversal respecting perturbation like SOC.
To ensure that $\C_3$ symmetry is preserved in the triplet case, we restrict our study to $\hat{\v{n}}\propto\hat{\v{z}}$.

\subsubsection{{IVC singlet order}}\label{sec:sIVCWireHam}

We focus now on the singlet IVC (sIVC) state, defined by $V=e^{i\phi}$ in Eq.~\eqref{eqn:CNP_IVC_general}.
The order parameters are equivalent for all values of $\phi$ and so we select $\phi=0$.
This choice returns the order parameter
\begin{align}
	\label{eqn:QsIVCdef}
	Q_\mathrm{sIVC}&=\tau^x\sigma^y.
\end{align}
It is clear that the residual SU(2) symmetry preserved by this state is the usual spin rotation symmetry.
The discrete symmetries preserved by $Q_\mathrm{sIVC}$ without SOC and with Rashba and with Ising SOC (separately) in 2$d$ are listed in Table~\ref{tab:OrdParamSyms}.
{As detailed above, we use these symmetries to restrict the form of an effective 2$d$ Hamiltonian close to the $\gamma$ point. 
The additional terms generated by the presence of sIVC order and subsequently by the pairwise presence of sIVC order and Rashba/Ising are shown in Table~\ref{tab:kexpansion}.
These in turn directly lead to the wire Hamiltonian shown in Table~\ref{tab:WireTable}.}

{We note that one of the terms obtained in the manner just described is necessarily second order in the SOC coupling constants.
In particular, when the sIVC order above is present alongside Rashba SOC, a constant term proportional to $\tau^ys^z$ is allowed by symmetry. 
However, as explained in Sec.~\ref{sec:2dHeff-WithSOC}, the leading order contributions of the Rashba term are all proportional to the spin Pauli matrices present in the UV Rashba SOC induced from the TMD monolayer $\h_R$---namely, they must be proportional to $s^x$ or $s^y$. 
Since the sIVC order parameter can itself not contribute any additional powers of spin, we conclude that the $\tau^ys^z$ term necessarily arises at order $\lambda_R^2$ or possibly an even higher power.
For this reason, the $\tau^ys^z$ term is not included in the 1$d$ wire Hamiltonian of Table~\ref{tab:WireTable}.}

Above, we mentioned that the $\nu=0$ IVC states were ``topological'' when the corresponding residual SU(2) was preserved, which here is the spin SU(2).
We find that within each spin $\uparrow$ or $\downarrow$ sector, the system is a $\mathds{Z}_2$ topological insulator protected by the emergent time reversal $\tilde{\mathcal{T}}=i\tau^y\mathcal{K}$.
That is, for each spin, a band with Chern number $C=+1$ and $C=-1$ is filled, where the Chern number is given by the eigenvalue of $\tau^z\sigma^z$ \cite{bultinckGroundStateHidden2020}.
The presence of SOC, of course, breaks the SU(2) spin symmetry and would accordingly induce scattering between counterpropagating modes in either spin sector at the sample boundary. 
Spin-orbit-induced scattering is not, however, the only mechanism that spoils the topological nature of the sIVC insulator.
Even in a system with perfect spin symmetry, the effective time reversal symmetry, $\tilde{\mathcal{T}}$, descends from the spontaneously-broken $\mathrm{U_v}(1)$ symmetry.
This emergent symmetry is dependent on the absence of momentum exchange occurring at the scale of the microscopic graphene lattice scale, and while this assumption is reasonable within the bulk, the physical boundary of the TBG samples 
are 
expected to strongly break this valley symmetry.
We can no longer append the $\mathrm{U_v}(1)$ rotation to the physical time reversal symmetry and are instead left with the `spinless' antiunitary symmetry $\Tns=i\tau^x \mathcal{K}$ which satisfies $\Tns^2=+1$. 
The edge modes are thus able to backscatter even in the absence of SOC.

\subsubsection{IVC triplet order}\label{sec:tIVCWireHam}

The triplet IVC (tIVC) state is obtained by setting $V=e^{i\phi}i\hat{\v{n}}\cdot\v{s}$.
As mentioned, to preserve the $\C_3$ symmetry, the spin must point in the out-of-plane direction, $\hat{\v{n}}\propto\hat{\v{z}}$.
We use the $\mathrm{U_v}(1)$ degrees of freedom to set $\phi=-\pi/2$, resulting in the order parameter
\begin{align}
	\label{eqn:QtIVCdef}
	Q_\mathrm{tIVC}&=
	\tau^x\sigma^xs^z.
\end{align}
While clearly not invariant under the physical SU(2) symmetry, as discussed, this order parameter nonetheless preserves an SU(2) subgroup of $\mathrm{U_v}(1)\times\mathrm{SU}(2)_K\times\mathrm{SU}(2)_{K'}$.
In particular, the action of the group $\mathrm{SU(2)_{tIVC}}$ generated by the matrices $\{s^z,\tau^zs^x,\tau^zs^y\}$ leaves $Q_\mathrm{tIVC}$ unchanged.
Note that the $\mathrm{U}(1)$ subgroup of the spin symmetry responsible for $z$-axis spin rotations, which we denote $\mathrm{U}_z(1)$, is a subset of this conserved symmetry, \emph{i.e.}, $\mathrm{U}_z(1)\subset\mathrm{SU(2)_{tIVC}}$.
The lattice symmetries preserved by $Q_\mathrm{sIVC}$ without SOC are shown in the row labelled `tIVC' in Table~\ref{tab:OrdParamSyms}. 
{Below, the symmetries preserved when either Rashba SOC or Ising SOC is also present are given as well.
We follow the procedure outlined above to obtain the terms of Table~\ref{tab:kexpansion}, leading to the wire Hamiltonian of Table~\ref{tab:WireTable}.}

{As with the singlet IVC order, one of the terms allowed by symmetry when tIVC order is present alongside Rashba SOC necessarily only occurs at higher orders in $\lambda_R$ and is therefore not included in the wire Hamiltonian shown in Table~\ref{tab:WireTable}.
Specifically, the term $\tau^y$ contains no powers of spin. 
The Rashba contribution must therefore have contributed a power of $s^z$ or a power of the identity---neither of which occur at linear order in $\lambda_R$.}

As with the sIVC insulator, the tIVC insulator is also `topological' in a certain limit. 
Without SOC, the system is invariant under $\mathrm{SU(2)_{tIVC}}$ transformations generated by $\{s^z,\tau^zs^x,\tau^zs^y\}$.
Although it was more convenient to list the physical time reversal symmetry, $is^y\Tns=i\tau^xs^y\mathcal{K}$, in Table~\ref{tab:OrdParamSyms}, by composing this operation with an $\mathrm{SU(2)_{tIVC}}$ rotation, it is clear that the tIVC insulator also preserves $\tilde{\mathcal{T}}=i\tau^y\mathcal{K}$.
Dividing the bands according to their spin-$z$ value, $s^z=\,\uparrow$ or $s^z=\,\downarrow$, we find that each sector realizes a $\mathds{Z}_2$ topological insulator protected by the $\tilde{\mathcal{T}}$ antiunitary symmetry.
Just as for the sIVC case, this protection is negated by both the inclusion of SOC (the $\mathrm{SU(2)_{tIVC}}$ symmetry is broken) as well as through the explicit valley symmetry breaking present at a physical boundary (the symmetry $\tilde{\mathcal{T}}$ is broken).
Finally, in the main text we discussed how an in-plane magnetic field is required to reach the odd channel regime needed to realize Majorana zero modes.
This field would further induce scattering between any edge modes along a sample boundary.

\subsection{\texorpdfstring{$\nu=\pm2$ IVC order parameters}{nu=+-2 IVC order parameters}}
\label{sec:nu2_ordParamDefs}

The IVC states at $\nu=\pm2$ are defined analogously to the $\nu = 0$ order parameters of Sec.~\ref{sec:CNPordParamDefs} \cite{bultinckGroundStateHidden2020}.
For concreteness, we focus on the $\nu=-2$ scenario.
The primary difference between $\nu=-2$ and $\nu=0$ is that additional states must be projected away in the former case. 
That is, the projection operator satisfies $\mathrm{tr}\,\mathcal{P}_{\nu=-2}=2$, compared to the $\nu=0$ projection operator, which satisfied $\mathrm{tr}\,\mathcal{P}_{\nu=0}=4$.
One may therefore view the $\nu=-2$ order parameter as a combination of two commuting order parameters, $Q_{1,2}$:
\begin{align}\label{eqn:Nu2-OrdParamDef}
	Q_{\nu=-2}&=\mathcal{P}_+ +\mathcal{P}_- Q_1,
	&
	\mathcal{P}_\pm&=\frac{1}{2}(\id\pm Q_2),
	&
	Q_{1,2}^2&=\id,
	&
	[Q_1,Q_2]&=0.
\end{align}
The actual projection operator defining the $\nu=-2$ IVC insulator is then $\mathcal{P}_{\nu=-2}=(\id- Q_{\nu=-2})/2$.

The authors of Ref.~\citenum{bultinckGroundStateHidden2020} demonstrate that within the perturbative scheme they consider, the lowest energy insulators at $\nu=\pm2$ are the so-called spin-polarized IVC states. 
These states may be obtained starting from an order parameter defined by the choice $Q_1=\tau^x\sigma^y$ and $Q_2=s^z$.
As for $\nu=0$, to leading order---at the level of $H_{0,1}^{(\fl)}$ in Eq.~\eqref{eqn:LeadingOrderHam}---the system possesses a $\mathrm{U(2)\times U(2)}\cong \mathrm{U_c(1)}\times\mathrm{U_v}(1)\times\mathrm{SU}(2)_K\times\mathrm{SU}(2)_{K'}$ symmetry, and so acting on $Q_{\nu=-2}$ by any element of this group returns an energetically equivalent ground state.
Doing so, one finds that the generic form of the spin-polarized IVC state is
\begin{align}\label{eqn:Qnu2-generaldef}
	Q_{\nu=-2}&=\begin{pmatrix}
			\frac{1}{2}(\id+\hat{\v{n}}_+\cdot\v{s})	&	V\sigma^y	\\
			V^\dag \sigma^y		&	\frac{1}{2}(\id+\hat{\v{n}}_-\cdot\v{s})
	\end{pmatrix}_{\!\tau},
\end{align}
where 
\begin{align}\label{eqn:Qnu2-generaldef-2}
	\hat{\v{n}}_\pm 
	&=
	\frac{1}{2}\mathrm{tr}\big(U_\pm^\dag s^zU_\pm \v{s}\big),
	&
	V&=U_+^\dag \frac{1}{2}(\id-s^z)U_-,
\end{align}
with $U_\pm$ $2\times2$ unitary matrices acting on the spin indices.
It can be shown that $Q_{\nu=-2}$ is invariant under the action of a $\mathrm{U}(1)\times\mathrm{U}(1)\times\mathrm{U}(1)$ symmetry (one of which is $\mathrm{U_c}(1)$) \cite{bultinckGroundStateHidden2020}, implying that it takes values in the manifold $\mathrm{U}(2)\times \mathrm{U}(2)/[\mathrm{U}(1)\times\mathrm{U}(1)\times\mathrm{U}(1)]\cong\mathrm{U_v}(1)\times S^2\times S^2$. 
Here, the $\mathrm{U_v}(1)$ symmetry corresponds to the usual IVC phase we set to zero.
The two factors of the unit sphere, $S^2$, correspond to the spin directions in either valley, $\hat{\v{n}}_\pm$.

Just as for the IVC orders at $\nu=0$, the large degeneracy of $Q_{\nu=-2}$ is broken by both the SOC term, $H_\mathrm{SOC}^{(\fl)}$, and the valley Hund's term, $H_J^{(\fl)}$.
In the absence of SOC, either a ferromagnetic (FM) state given by $U_+=e^{i\phi}U_-$ or an antiferromagnetic (AFM) IVC state given by $U_+=s^xe^{i\phi}U_-$ is preferred depending on the sign of the Hund's coupling, $J_H$.
These choices accordingly select $\hat{\v{n}}_+=\hat{\v{n}}_-$ and $\hat{\v{n}}_+=-\hat{\v{n}}_-$ for the FM IVC and AFM IVC states, respectively.
These two orders represent the most natural choices of order parameters even in the presence of SOC.
In particular, they are the sole choices that are either even or odd under time reversal, and a time-reversal preserving perturbation like SOC is therefore unable to prefer a groundstate in which they are mixed.
As was the case at $\nu=0$, in order to preserve $\C_3$ symmetry, the spins must point in the out-of-plane direction, $\hat{\v{n}}_\pm\propto \hat{\v{z}}$, which we assume to be the case.

\subsubsection{Ferromagnetic IVC order}
\label{sec:FM-IVC}

The first state we consider is the ferromagnetic (FM) IVC state, which is obtained by choosing $\hat{\v{n}}_+=\hat{\v{n}}_-=\hat{\v{z}}$ in Eq.~\eqref{eqn:Qnu2-generaldef}.
An equivalent definition of this state follows by selecting $Q_1$ and $Q_2$ in Eq.~\eqref{eqn:Nu2-OrdParamDef} equal to two distinct elements of $\{s^z,\tau^x\sigma^y,\tau^x\sigma^ys^z\}$ up to some signs; the unassigned element is then equal to $\pm Q_1Q_2$.
Regardless, the resulting order parameter is
\begin{align}\label{eqn:FMIVC-ordparam}
	Q_\mathrm{FM\,IVC}&=\frac{1}{2}(\id + s^z)
	+
	\frac{1}{2}\tau^x \sigma^y (\id-s^z).
\end{align}
We can check that the three residual U(1) symmetries preserved by this order parameter are generated by $\id,s^z,\tau^z(\id-s^z)$, which respectively correspond to $\mathrm{U_c}(1)$, 
$\mathrm{U}_z(1)$, and a restricted version of $\mathrm{U_v}(1)$ that only acts on the $\downarrow$ spins.
While the first two are good symmetries, the second is an artifact of the simplistic, fine-tuned representation implied by Eq.~\eqref{eqn:FMIVC-ordparam}. 
Essentially, $Q_\mathrm{FM\,IVC}$ has degenerate eigenvalues equal to $-1$ for the two states with both $s^z=\,\downarrow$ and $\tau^x\sigma^y=-1$ and eigenvalues equal to $+1$ for the remaining six states; only the former two states survive the projector $(\id-Q_\mathrm{FM\,IVC})/2$.
However, although $Q_{\mathrm{FM\,IVC}}$ is a good description of the ground state, it is unlikely to describe the effective mean field Hamiltonian one would actually obtain in an interacting system. 
Instead, one would expect something like $H_\mathrm{FM\,IVC}\sim a s^z + b \tau^x\sigma^y - c\tau^x\sigma^ys^z$, with, say, $a,b,c>0$.
While $(\id-Q_\mathrm{FM\,IVC})/2$ indeed projects onto the filled states of $H_\mathrm{FM\,IVC}$ at $\nu=-2$, the Hamiltonian $H_\mathrm{FM\,IVC}$ itself only preserves the U(1) symmetry generated by $\tau^z(\id-s^z)$ when $b=c$, which is not required by any symmetry.
We therefore do not include this symmetry in our analysis.

All of the terms present in the effective Hamiltonian for a wire proximity-coupled to a $\nu=0$ sIVC insulator and for a wire proximity-coupled to a $\nu=0$ tIVC insulator (Secs.~\ref{sec:TrivialWireHam},~\ref{sec:sIVCWireHam}, and~\ref{sec:tIVCWireHam}) will also be generated here. 
In Table~\ref{tab:kexpansion}, these terms are summarized in rows 1-8.
Additional terms are obtained by considering the FM IVC order by itself and alongside the Rashba and Ising SOC individually. 
The symmetry group for each of these cases is provided in Table~\ref{tab:OrdParamSyms} and the additional terms present in each case are listed in Table~\ref{tab:kexpansion}.
From the results of this table, we follow the prescription outlined in Sec.~\ref{sec:TrivialWireHam} to obtain the effective 1$d$ Hamiltonian and the corresponding parameter scaling shown in Table~\ref{tab:WireTable}.

The FM IVC is also topological in some respects.
For the parameters chosen in Eq.~\eqref{eqn:FMIVC-ordparam}, the two filled bands have Chern numbers (given by $\tau^z\sigma^z$) $+1$ and $-1$ as well as $\tau^x\sigma^y=-1$ and $s^z=\,\downarrow$.
As for the $\nu=0$ IVC insulators, these two bands are mapped to one another under the emergent symmetry $\tilde{\mathcal{T}}=i\tau^y\mathcal{K}$.
Since this antiunitary symmetry descends from the $\mathrm{U_v}(1)$ symmetry, it is again broken by the sample boundary itself, similar to what occurs for the sIVC and tIVC insulators.
The topological protection of this state is also spoiled by the addition of SOC: as Table~\ref{tab:OrdParamSyms} indicates, when FM IVC order coexists with either Rashba or Ising SOC, no antiunitary symmetry survives.
Note, however, that unlike for the $\nu=0$ IVC states, the SOC terms do not directly mediate backscattering between counterpropagating modes. 
In particular, to first order, Rashba necessarily induces a spin slip, $\uparrow\,\to\,\downarrow$, while
Ising SOC flips the value of $\tau^x\sigma^y$. 
Higher order processes will nevertheless induce a gap.

\subsubsection{Antiferromagnetic IVC order}\label{sec:AFM-IVCorder}

The AFM IVC order parameter is given by selecting $U_+=\id$ and $U_-=s^x$ in Eqs.~\eqref{eqn:Qnu2-generaldef} and~\eqref{eqn:Qnu2-generaldef-2}:
\begin{align}\label{eqn:AFMIVCorderparam}
	Q_\mathrm{AFM\,IVC}&=\frac{1}{2}\big(
		\id+\tau^zs^z+\tau^x\sigma^ys^x+\tau^y\sigma^ys^y
	\big)
\end{align}
Comparing against Eq.~\eqref{eqn:Nu2-OrdParamDef}, we see that that $Q_\mathrm{AFM\,IVC}$ could equivalently have been obtained by choosing $Q_1$, $Q_2$ equal to two out of $\tau^zs^z$, $\tau^x\sigma^ys^x$, and $\tau^y\sigma^ys^y$ (as always, up to $\mathrm{U_v}(1)$ rotations).
We note that the mean field generation of a $\tau^zs^z$ term is equivalent to the spontaneous generation of Ising SOC; thus, when we consider the effect of adding SOC to the system, we will not need to consider Ising SOC as we did for the other cases. 
We recognize the other two order parameters, $\tau^x\sigma^ys^x$ and $\tau^y\sigma^ys^y$, as in-plane versions of the IVC triplet order parameter $Q_\mathrm{tIVC}=\tau^x\sigma^ys^z$ considered in Sec.~\ref{sec:tIVCWireHam}.

As written, $Q_{\mathrm{AFM\,IVC}}$ preserves the three U(1) symmetries generated by $\id$, $\tau^zs^z$, and $\tau^z+s^z$, and,
unlike in the previous case, there is a physical reason to expect that the AFM IVC may be invariant under all of these U(1) symmetries.
The primary issue is the symmetry generated by $\tau^z+s^z$. 
It appears to be similar to the U(1) symmetry that we argued in Sec.~\ref{sec:FM-IVC} was a non-physical artifact, the result of fine-tuning.
Here, however, the symmetry generated by $\tau^z+s^z$ can be interpreted as a residual version of $\mathrm{U_v}(1)\times \mathrm{U}_z(1)$, in which a spin rotation about the $z$ axis  
is followed by a valley rotation: $\tilde{\mathrm{U}}_z(1):c(\vk)\to e^{i\theta(\tau^z+s^z)/2}c(\vk)$. 
Although the introduction of SOC ultimately breaks both the $\mathrm{U}_z(1)$ and `residual' $\tilde{\mathrm{U}}_z(1)$ symmetry, the presence of the latter allows a residual version of the spinful 120$^\circ$ rotation symmetry to survive
even when SOC is included: $\tilde{C}_3=e^{-2\pi i\tau^z/3}C_3=e^{-2\pi i(s^z+\tau^z)/3}\C_3$.
Since the extra valley rotation that has been appended occurs only at the scale of the microscopic graphene lattice, this `residual' $\tilde{C}_3$ symmetry is largely indistinguishable from the `physical' $C_3$ symmetry we started with---all lattice symmetries are after all only well-defined in a low-energy limit, at length scales much larger than the microscopic lattice constant.
As a result, provided $\tau^z+s^z$ is preserved in the AFM IVC state without SOC, the AFM IVC state with SOC will \emph{not} be nematic.
While some experiments show evidence for nematicity around $\nu=\pm2$ \cite{choiElectronicCorrelationsTwisted2019,kerelskyMaximizedElectronInteractions2019,jiangChargeOrderBroken2019,cao2020nematicity}, for simplicity, we restrict ourselves to this fully symmetric case.
This assumption is equivalent to requiring that the two in-plane IVC order parameters occur with identical coefficients, \emph{i.e.}, in the form $\tau^x\sigma^ys^x + \tau^y\sigma^ys^y$.

The symmetries preserved by this order parameter are listed in Table~\ref{tab:OrdParamSyms} in the limit just discussed.
In turn, its presence induces the terms
listed in Table~\ref{tab:kexpansion} in the 2$d$ effective Hamiltonian.
Following the prescription outlined in Sec.~\ref{sec:TrivialWireHam}, we arrive at the effective wire Hamiltonian provided in Table~\ref{tab:WireTable} with the parameter scalings shown.

Unlike the other IVC insulators considered, the AFM IVC state is a true $\mathds{Z}_2$ topological insulator protected by the physical time reversal symmetry \cite{lake_re-entrant_2021}.
The order parameter of Eq.~\eqref{eqn:AFMIVCorderparam} implies that the two bands filled at $\nu=-2$ have definite $\tau^zs^z=-1$ and $\tau^x\sigma^ys^x=-1$, as well as Chern numbers $C=+1$ and $C=-1$ (where $C$ is given by the eigenvalue of $\tau^z\sigma^z$). 
At the phase boundary, backscattering between the modes of each of these two bands is prohibited by the true electronic time reversal symmetry $\mathcal{T}=i\tau^xs^y\mathcal{K}$ and so neither the addition of spin orbit nor edge disorder-induced intervalley scattering obviate the topological protection of the system.
Scattering and thus a gap will, however, result upon the application of an in-plane magnetic field (as needed to realize Majorana zero modes; see the main text for details). 

\subsection{Relation of internal SOC parameters and wire direction}\label{sec:InternalSOCDirection}

One aspect of the above analysis that may not be immediately clear is the relation it implies between the spin and wire direction.
This point is best illustrated by direct comparison with a more conventional spin-orbit coupled wire.
In a typical $2d$ SOC-coupled material, the Rashba term takes the form $\sim k_x s^y-k_ys^x$. 
Reducing to a wire scenario (directed along the $x$ direction) then returns a $1d$ Rashba contribution $\sim k_xs^y$. 
In the presence of an in-plane Zeeman field, different behaviour should be observed depending on whether the magnetic field points parallel to the wire, $\sim \frac{1}{2}g\mu_B Bs^x$, or perpendicular to the wire, $\sim\frac{1}{2}g\mu_BBs^y$.

By contrast, our symmetry analysis does not fix the relation between the spatial wire direction and the internal spin directions favored by SOC.
This uncertainty occurs at two (related) levels.
Firstly, the unknown in-plane rotation angle $\phi_R$ in the definition of $\h_R$ in Eq.~\eqref{eqn:SpinOrbitContributions} (as well as in the definition of $\h_R^*$ in Eq.~\eqref{eqn:MirrorSymRashbaDef}) signals our fundamental inability to relate spatial directions (encoded in the sublattice Pauli matrices $\sigma^{x,y}$) and spin directions (encoded in the Pauli matrices $s^{x,y}$).
Only when the twist angle between the \emph{TMD substrate} and neighbouring graphene monolayer, $\theta_\mathrm{TMD}$, is precisely $0^\circ$ or $30^\circ$ (mod~$60^\circ$) can we definitively say that $e^{i\phi_R}$ is real \cite{li_twist-angle_2019}.
This restriction follows from the presence of a reflection symmetry about either the $x$- or $y$-axes at those special angles.
As discussed in Sec.~\ref{sec:MLG+SOC}, we nevertheless expect the angle $\phi_R$ to deviation very little from $0$ or $\pi$ ($\mathrm{mod}\,2\pi$).

Even setting $\phi_R=0$ exactly does not resolve the issue of how to relate spatial and spin directions.
The more relevant issue is likely to be the absence of mirror symmetry.
In particular, we repeat here the second line of Eq.~\eqref{eqn:Rashba-2dHam-kexpansion}, which gives the odd parity Rashba SOC terms generated at first order:
\begin{align}
	\bar{h}^{(R)}_o(\vk)
	&=
	\tilde{t}^{(R)}_o\big( k_x s^x + k_y s^y \big)
	+
	\tilde{t}^{(R)\prime}_o\big( k_y s^x - k_x s^y \big).
\end{align}
Generically both $\tilde{t}^{(R)}_o$ and $\tilde{t}^{(R)\prime}_o$ are non-zero. 
In the $1d$ limit we therefore obtain a sum $\sim k_x\left(\tilde{t}^{(R)}_o s^x - \tilde{t}^{(R)\prime}_o s^y\right)$, indicating that the spin axis favored by Rashba SOC depends on non-universal details.
Since to a good approximation an applied magnetic field couples only to the wire's spin degrees of freedom, the non-universal relation between the wire orientation and spin directions implies a non-universal relation between the wire and magnetic field directions. 

This non-universality is the primary motivation for our introduction of the fine-tuned mirror-symmetric Hamiltonians of Sec.~\ref{sec:MirrorSymScenario}.
When $\h^*_R$ is present instead of $\h_R$, the $\M_y^*$ (or, more correctly, assuming $\phi_R=0$, the $is^x\M_y^*$) symmetry sets $\tilde{t}^{(R)\prime}_o$ to zero, in which case we can definitively `tie' the wire and internal SOC directions to one another, even at the level of the symmetry analysis.
Moving beyond this fine-tuned limit, although we cannot say that $\tilde{t}^{(R)\prime}_o$ vanishes in the  physical situation, we may nevertheless predict that because of the strong interlayer mixing that occurs between the two layers, the mirror symmetry is only weakly broken, implying that $\abs{\tilde{t}^{(R)\prime}_o}\ll\abs{\tilde{t}^{(R)}_o}$.
(A caveat to this reasoning is that mirror symmetry is most strongly broken in the region of interest, close to the $\gamma$ point.)

In selecting parameters for Figs.~\ref{fig:ExampleBands} and~\ref{fig:SCphases} of the main text, this weak breaking of the reflection and the mirror symmetries was imposed.
Within this set of assumptions, the magnetic field direction chosen is consistent with being perpendicular to the wire. 
We further assumed that the wire was directed in such a way that the 1$d$ projection of neither $k_x(k_x^2-3k_y^2)$ nor $k_y(3k_x^3-k_y^2)$ vanished.

\subsection{Topological aspects of IVC states}\label{sec:IVCTopoAspects}

In Secs.~\ref{sec:CNPordParamDefs} and~\ref{sec:nu2_ordParamDefs} we described the different IVC orders under consideration as well as the topological nature of these states, which we assessed by examining the fate of any edge modes at the physical sample boundary.
The wire profile interpolates between IVC and `trivial' regions of the phase diagram and may therefore be interpreted as two distinct boundaries separating the IVC and trivial insulators.
It's thus possible that topological (or pseudo-topological) edge modes may be present, and we address this possibility here.

Firstly, even in a true topological insulator, edge modes are only protected in the limit that they are infinitely separated spatially.
The narrowness of the wire, however, means that any edge modes arising out of the termination of either adjoining phase will overlap spatially, immediately encouraging the formation of a gap.
In addition to this admittedly trivial backscattering mechanism, to varying degrees, all of the means discussed in Secs.~\ref{sec:CNPordParamDefs} and~\ref{sec:nu2_ordParamDefs} by which the topological nature of the system may be spoiled are again relevant here.
Notably, with the exception of the AFM IVC insulator, inter-mode scattering mediated by SOC promotes a gap. 
An in-plane magnetic field should also induce scattering between modes; this mechanism is particularly relevant for the tIVC and AFM IVC insulators, which require the external breaking of time reversal to enter the odd channel regime.
Finally, explicit breaking of the $\mathrm{U_v(1)}$ symmetry may also be a relevant gap-forming mechanism, although this effect is not expected to be as important at a gate-mediated chemical potential shift as it would be at the physical sample boundary.
(Note that if such a chemical potential gradient were able to open substantial gaps via inter-valley scattering, proximate IVC order would \emph{not} be necessary to realize Majorana zero modes.)
Regardless of whether physical proximity, SOC, an applied field, or valley symmetry breaking is the primary gap-generating mechanism, the result may still be smaller than the bulk gap, resulting in pseudo-topological subgap states.

Although we have focused throughout this paper on wire states realized through a confining potential (Sec.~\ref{sec:TrivialWireHam}), the pseudo-topological states just described may also be used to obtain the odd channel regime necessary to realize Majorana zero modes. 
The body of the analysis used to obtain the effective wire Hamiltonians of Table~\ref{tab:WireTable} is equally valid for these topological states.
The primary distinction lies in the applications of the momentum expansions in Table~\ref{tab:kexpansion}, and thus the derivation of the fourth column of Table~\ref{tab:WireTable}, which relates the scaling of the parameters.
These observations follow from our use of the $2d$ $\vk$-dependent terms shown in Table~\ref{tab:kexpansion}, as well as the fact that these terms were derived using the $\C_3$ and $\M_y$ action expressed in Eq.~\eqref{eqn:FlBandSyms-2}, which only hold in the region close to the $\gamma$ point.
Conversely, the pseudo-topological states are fundamentally real-space objects and cannot be described as emerging from any location in the moiré BZ.
The only information in Table~\ref{tab:kexpansion} applicable to these modes is the matrix structure of the terms (which combinations of $\tau^\mu s^a$ are allowed) and the parity of the functions they multiply (the terms $t_{\mu,a}(k)$ found in Secs.~\ref{sec:FlatBandHeff-NoSOC}, \ref{sec:2d-Rashba-Heff}, and~\ref{sec:IsingOnly}); see Sec.~\ref{sec:TrivialWireC3MyBreaking} for more discussion.
(For wires directed specifically along high symmetry directions such as the $x$- or $y$-axes, mirror could yield additional constraints, but would require slightly different methods.)
In certain circumstances, the differences in the parameter scaling in Table~\ref{tab:WireTable}, in particular the dependence on the wire width, may make the pseudo-topological states better hosts for the Majorana physics proposed here. 
Given their topological provenance, one may further expect such states to possess additional stability against unwanted perturbations.

While we view these subgaps as a boon, one may be concerned that their presence will mar some of the conductance signatures of the IVC phase discussed in the main text.
Although these edge-like modes certainly increase the system's complexity, their presence may be discerned by the persistence of gapless wire states even when the chemical potential of the wire is well below the band bottom (see Sec.~\ref{sec:5bnd_wiresim} for further discussion).

The topological-like states traversing the gate-defined wire may coexist with topological states at the sample boundary.
The main text illustration of Fig.~\ref{fig_architecture}(a) as well as the insets in Fig.~\ref{fig:ExampleBands} all depict the ends of the wire as coinciding with the physical sample boundary, implying that pseudo-topological edge modes (if present) would reside in close proximity to the wire ends---potentially destroying the topological protection of any Majorana zero modes.
For the sIVC, tIVC, and FM IVC insulators, we expect any pseudo-topological state to acquire a significant gap as a result of the inter-valley scattering induced by the sharp physical boundary, and we are therefore not concerned about such a scenario.
The same considerations do not hold for AFM IVC state, which is a true topological insulator. 
Instead we must rely on the applied magnetic field to also open a gap along the sample boundary.
The chemical potential in this proximate region must further be tuned to lie within this energy window. 
Alternatively, judicious chemical potential variation along the wire direction could be used to push the Majorana zero modes away from any sample boundaries, far from possible gapless edge states.

\section{Wire proximity-coupled to IVC and superconducting states}\label{sec:WireProxCoupledSC}

\begin{figure}
    \centering
    \includegraphics[width=0.99\textwidth]{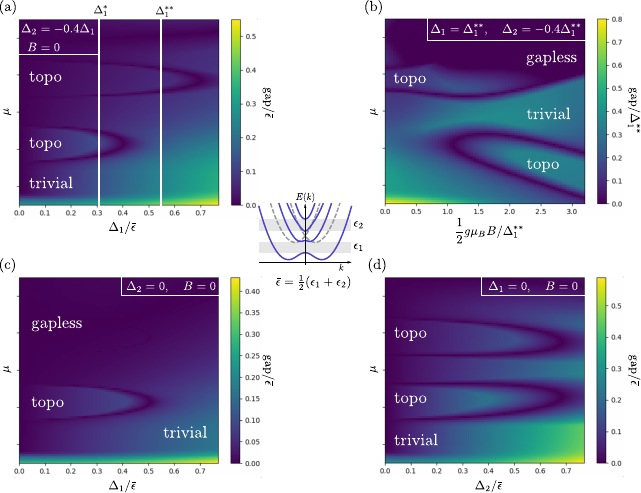}
    \caption{\AppCaptions
	Plots of excitation gap in a wire proximity-coupled to both an sIVC insulator and a superconductor that contributes pairing terms of the form provided in Eq.~\eqref{eqn:ProximateSC} of the main text.
    Topological (topo), trivial, and gapless phases are noted.
    The same parameters used in Figs.~\ref{fig:ExampleBands}(c) and~\ref{fig:SCphases}(a) are used above.
    The inset in the centre reproduces Fig.~\ref{fig:ExampleBands}(c) of the main text.
    The two odd-channel energy windows marked with grey boxes are labelled $\epsilon_{1,2}$ and $\bar{\epsilon}$ is defined as their average.
    (a) Excitation gap as a function of $\Delta_1/\bar{\epsilon}$ and $\mu$ for zero magnetic field.
    The vertical line on the left, marked $\Delta_1^*$, corresponds to the pairing used in Fig.~\ref{fig:SCphases}(a) of the main text.
    As in that figure, $\Delta_2=-0.4\Delta_1$.
    (b) Excitation gap as a function of magnetic field for $\Delta_1$ equal to $\Delta_1^{**}$, shown in (a) by the vertical line on the right.
    $\Delta_2$ is again equal to $-0.4\Delta_1$.
    Although the lower topological phase is absent at zero magnetic field, it re-emerges at finite field.
    (c) Excitation gap as a function of $\Delta_1/\bar{\epsilon}$ and $\mu$ at zero magnetic field with $\Delta_2=0$.
    The upper topological phase that arises out of the three-channel regime shown in the inset is now absent.
    (d)
    Excitation gap as a function of $\Delta_2/\bar{\epsilon}$ for zero magnetic field with $\Delta_1=0$.
    Both topological phases are clearly present.
    Furthermore, the one-channel phase extends to larger values of the externally induced pairing term $\Delta_2$.
    }
    \label{fig:sIVC_SC}
\end{figure}

\begin{figure}[t]
    \centering
    \includegraphics[width=0.99\textwidth]{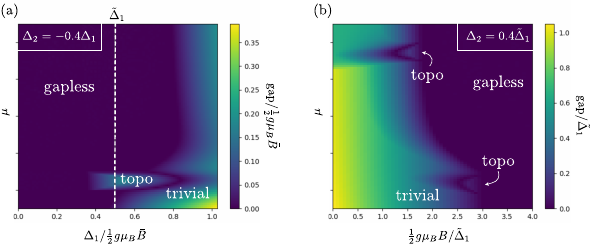}
    \caption{\AppCaptions
	Plots of the excitation gap in a wire proximity-coupled to both an AFM IVC insulator and a superconductor that contributes pairing terms of the form provided in Eq.~\eqref{eqn:ProximateSC} of the main text.
    Topological (topo), trivial, and gapless phases are noted.
    The same parameters used in Fig.~\ref{fig:ExampleBands}(i) as well as Fig.~\ref{fig:SCphases}(b) are used above.
    (a) Excitation gap as a function of $\Delta_1/(\frac{1}{2}g\mu_B \bar{B})$ and $\mu$ for fixed magnetic field $\bar{B}$ chosen such that $\frac{1}{2}g\mu_B\bar{B}/\tilde\Delta_1=2.0$, where $\tilde\Delta_1$ is the fixed pairing strength used to generate the Fig.~\ref{fig:SCphases}(b) of the main text and shown with the white dashed line.
    (b) Excitation gap in the wire as a function of magnetic field $\frac{1}{2}g\mu_BB/\tilde\Delta_1$ and $\mu$ for fixed singlet pairing strength $\tilde\Delta_1$---the same value used to obtain Fig.~\ref{fig:SCphases}(a) of the main text and plotted (a).
    Unlike Fig.~\ref{fig:SCphases}(b) of the main text, the triplet pairing strength is set to equal $\Delta_2=+0.4\tilde\Delta_1$. 
    There are now two distinct topological lobes, although they are quite a bit smaller in size.
    }
    \label{fig:AFMIVC_SC}
\end{figure}

Figures~\ref{fig:sIVC_SC} and~\ref{fig:AFMIVC_SC} build on the information in Fig.~\ref{fig:SCphases} of the main text.

Through the excitation gap, Fig.~\ref{fig:sIVC_SC} presents the phase diagram of a wire proximity coupled to both the sIVC insulator and the superconducting state in a number of scenarios.
In (a), the excitation gap is plotted as a function of $\Delta_1$ and $\mu$ at zero magnetic field. 
As in the main text, we assume $\Delta_2=-0.4\Delta_1$.
Two distinct topological states are observed at precisely the location of the odd channel regimes, as shown in Fig.~\ref{fig:ExampleBands}(c) of the main text and reproduced in the central inset of Fig.~\ref{fig:sIVC_SC}.
The white line on the left of (a), labelled $\Delta_1^*$,  indicates the pairing strength used to generate Fig.~\ref{fig:SCphases}(a) in the main text. 
Figure~\ref{fig:sIVC_SC}(b) plots the excitation gap as a function of magnetic field for the fixed pairing strength indicated by the line labelled $\Delta_1^{**}$ in (a).
Again, $\Delta_2=-0.4\Delta_1^{**}$.
Notably, since $\Delta_1^{**}$ does not intersect the lower topological lobe of (a) (corresponding to the one-channel regime), there is topological phase at $B=0$ in (b). 
At finite $B$, however, the topological phase re-emerges.
Figures~\ref{fig:sIVC_SC}(c) and~(d) are similar to (a) in that they also plot the excitation gap as a function of external pairing in zero magnetic field.
In (c), the triplet pairing is absent, $\Delta_2=0$, and we see that $\Delta_1$ is unable to open a gap in the three channel regime---only the lower topological lobe is present.
In (d), the singlet pairing is instead set to zero, $\Delta_1=0$, and topological phases are observed in both the one and three channel regimes as a function of the triplet pairing $\Delta_2$.

The plots in Fig.~\ref{fig:AFMIVC_SC} similarly expand upon the image shown Fig.~\ref{fig:SCphases}(b) of the main text.
In (a), the excitation gap is plotted as a function of the chemical potential $\mu$ and the pairing strength $\Delta_1$ at a fixed magnetic field, $\bar{B}$.
The value of $\bar{B}$ is chosen such that $\frac{1}{2}g\mu_B \bar{B}/\tilde{\Delta}_1=2.0$, where $\tilde{\Delta}_1$ is the singlet pairing strength used to generate Fig.~\ref{fig:SCphases}(b) of the main text and indicated in (a) with a vertical dashed white line.
Figure~\ref{fig:AFMIVC_SC}(b) is directly analogous to Fig.~\ref{fig:SCphases}(b) of the main text, save that $\Delta_2=+0.4\tilde{\Delta}_1$.
Two distinct lobes are now apparent at finite magnetic field; they are, however, much smaller than topological region shown in (a).

\section{Five band model}
\label{sec:5bndmodel}

As a proof of concept, we perform numerical simulations of a wire proximity-coupled to various IVC phases within the context of the five model for the flat bands of TBG described in Ref.~\onlinecite{poFaithfulTightbindingModels2019}.
In Secs.~\ref{sec:BasicDefs},~\ref{sec:5bndSyms}, and~\ref{sec:5bndNoSOC}, we describe the model setup, symmetry actions, and the inter-orbital hopping terms first provided in Ref.~\onlinecite{poFaithfulTightbindingModels2019}.
We subsequently use symmetries in conjunction with the information contained in Table~\ref{tab:OrdParamSyms} to incorporate SOC and IVC orders into the five-band model---focusing in particular on the sIVC insulator at $\nu=0$ and the AFM IVC insulator at $\nu=\pm2$.
We describe how the wire is realized and present out results in Sec.~\ref{sec:5bnd_wiresim}. 
We finish in Sec.~\ref{sec:5bnd_disc} by discussing some limitations of this study.

\subsection{Basic definitions}\label{sec:BasicDefs}

\begin{figure}
\centering
\includegraphics[width=0.99\textwidth]{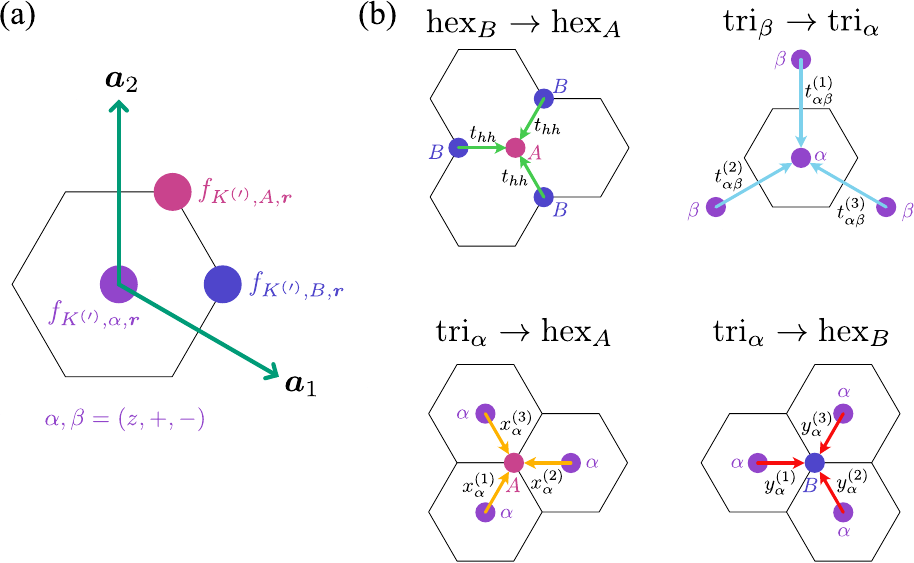}
\caption{\AppCaptions
	(a) Unit cell of the five band model with Bravais lattice vectors $\veo$ and $\vet$ shown in green. Note that the hexagon shown is the \emph{moiré}  unit cell, which itself comprises thousands of graphene unit cells. The location of the three $p$ orbitals, $p_{z,\pm}$, at the centre of the moiré hexagon is shown in purple, the $s$ orbital at the A sublattice of the hexagon corners is shown in magenta, and the $s$ orbital at the B sublattice of the hexagon corner is shown in blue.
	The corresponding electron annihilation operators are written alongside each orbital.
	(b) Inter-orbital hopping terms for $K$ valley orbitals as described in Sec.~\ref{sec:5bndNoSOC}.
	}
\label{fig:unitcell}
\end{figure}

While the two flat bands (per spin per valley) would ideally be describable using a microscopic two band model, the bands themselves possess a `fragile topology' \cite{ahn_fragiletopo_19,poFaithfulTightbindingModels2019,po_fragile_18} that prohibits such a description.
Instead, the minimal description one may formulate involves five orbitals centered at different locations within the moiré unit cell.
Here, we study a model with $p_z$, $p_+$, and $p_-$ orbitals on the triangular lattice sites at the centre of each moiré unit cell (where the graphene sheets have AA stacking) and $s$ orbitals on the A and B sublattices of the honeycomb lattice (where the graphene sheets have AB and BA stacking).
Below, the three $p$ orbitals on the triangular lattice are sometimes denoted $\mathrm{tri}_{z,\pm}$, while the two $s$ orbitals on the hexagon sites are sometimes denoted $\mathrm{hex}_{A/B}$.
See Fig.~\ref{fig:unitcell}(a) for details.
We arrange the corresponding electron annihilation operators as
\begin{align}\label{eqn:orbitalDef}
	f_{v,\vr}
	&=
	\Big(
	f_{v,z,\vr},f_{v,+,\vr},f_{v,-,\vr},f_{c,A,\vr},f_{v,B,\vr}
	\Big)^T,
	&
	v&=K,K'.
\end{align}
A spin index is suppressed in the above.
The position `$\vr$' is defined via the primitive vectors shown in Fig.~\ref{fig:unitcell}.
That is, $\vr=n_1\veo+n_2\vet$ where $n_{1,2}\in\mathds{Z}$ and
\begin{align}
	\veo
	&=
	\left( \frac{\sqrt{3}}{2}\cCom-\frac{1}{2}\right),
	&
	\vet
	&=
	\left( 0,1\right).
\end{align}
The moir\'{e} lattice constant has been set to unity.

\subsection{Symmetries}\label{sec:5bndSyms}
In this section, we outline the action of the symmetries on the orbitals of the five band model.
With some minor differences in presentation, we again use the `spinless' version of the symmetries, which are related to the physical symmetries via Eq.~\eqref{eqn:PhysicalSymDef}.
We opt to only explicitly describe $\C_2\Tns$ and $\Tns$, from which the $\C_2$ transformation may be derived. 
We further only explicitly relate the action of the symmetries on the operators originating from valley $K$; the action on valley $K'$ may be obtained using time reversal symmetry.
The continuous internal symmetries act on the spin and valley indices as they do in Eq.~\eqref{eqn:ContinuousSymmetryAction}.
\subsubsection{\texorpdfstring{120$^{\,\circ}$ rotation, $\C_3$}{120 degree rotation, C3}}

The $\C_3$ symmetry acts as
\begin{align}
	\v{a}_1
	&\to
	\vet,
	&
	\vet
	&\to
	-\veo-\vet,
\end{align}
and the electron operators transform as
\begin{align}
	\C_3:\quad
	f_{K,\vr}
	&\to
	\begin{pmatrix}
		f_{K,z,\vr'}	\\
		\omega^* f_{K,+,\vr'}	\\
		\omega f_{K,-,\vr}	\\
		f_{K,A,\vr'-\veo-\vet}	\\
		f_{K,B,\vr'-\veo}
	\end{pmatrix}
\end{align}
where $\vr=r_1\veo+r_2\vet$, $\vr'=-r_2\veo+(r_1-r_2)\vet$ for $r_{1,2}\in\mathds{Z}$.

\subsubsection{\texorpdfstring{Mirror, $\M_y$}{Mirror, My}}

Under $\M_y$, the primitive vectors transform as
\begin{align}
	\veo&\to\veo+\vet,
	&
	\vet&\to-\vet.
\end{align}
The real space electron operator transform as
\begin{align}
	\M_y:\quad
	f_{K,\vr}&\to
	\begin{pmatrix}
		-f_{K,z,\vr'}	\\
		f_{K,-,\vr'}	\\
		f_{K,+,\vr'}	\\
		f_{K,A,\vr'-\vet}	\\
		f_{K,B,\vr'}
	\end{pmatrix},
\end{align}
where $\vr=r_1\veo+r_2\vet$, $\vr'=r_1\veo+(r_1-r_2)\vet$ for $r_{1,2}\in\mathds{Z}$.

\subsubsection{\texorpdfstring{
180$^{\,\circ}$ rotation + time reversal, $\C_2\Tns$}
{180 degree rotation + time reversal, C2T}}

$\C_2\Tns$ acts on the primitive vectors as
\begin{align}
	\C_2\Tns&:
	&
	\veo&\to-\veo,
	&
	\vet&\to-\vet,
	&
	i\to-i.
\end{align}
Its action on the real space electron operators is
\begin{align}
	\C_2\Tns:\quad
	f_{K,\vr}
	\to
	\begin{pmatrix}
		f_{K,z,-\vr}	\\
		f_{K,-,-\vr}	\\
		f_{K,+,-\vr}	\\
		f_{K,B,-\vr-\veo-\vet}	\\
		f_{K,A,-\vr-\veo-\vet}
	\end{pmatrix},
	\qquad
	i\to-i.
\end{align}
\subsubsection{Time reversal}

Finally, time reversal acts as
\begin{align}
    \label{eqn:5bndTRS}
	\Tns:
	\quad
	f_{K,\vr}
	\to 
	\begin{pmatrix}
		f_{K',z,\vr}	\\
		f_{K',-,\vr}	\\
		f_{K',+,\vr}	\\
		f_{K',A,\vr}	\\
		f_{K',B,\vr}
	\end{pmatrix},
	\qquad 
	i\to -i.
\end{align}

\subsection{Five band model without SOC}\label{sec:5bndNoSOC}
We describe the model as provided in Ref.~\citenum{poFaithfulTightbindingModels2019}, focusing on valley $K$.
All terms corresponding to the $K'$ valley may be obtained via time reversal, Eq.~\eqref{eqn:5bndTRS}.
We divide the terms considered into the tunnelling types shown in Fig.~\ref{fig:unitcell}(b): hopping between the honeycomb lattice orbitals ($\mathrm{hex}_{A/B}\to \mathrm{hex}_{B/A}$), hopping between the three orbitals on the triangular lattice sites ($\mathrm{tri}_{z,\pm}\to\mathrm{tri}_{z,\pm}$), and hopping between the triangular lattice orbitals and the A and B sublattice of the honeycomb lattice ($\mathrm{tri}_{z,\pm}\to \mathrm{hex}_A$ and $\mathrm{tri}_{z,\pm}\to \mathrm{hex}_B$).
The equations below are parametrized by real numbers $a,b,c,d,$ and $\alpha$ (which should not to be confused with the effective wire parameters of Table~\ref{tab:WireTable}); an energy scale $t_0$; and a set of orbital potentials.

\subsubsection{\texorpdfstring{Hopping between $\mathrm{hex}_A$ and $\mathrm{hex}_B$ orbitals}{Hopping between hexA and hexB orbitals}}
Nearest-neighbour hopping between $s$ orbitals on the honeycomb lattice
is schematically shown on the top left of Fig.~\ref{fig:unitcell}(b) and may be expressed using the notation of Sec.~\ref{sec:BasicDefs} as
\begin{align}\label{eqn:5bnd-hh}
	H_{hh}^{(K)}
	&=
	-t_0d^2 e^{-2i\alpha}\sum_\vr
	f_{K,A,\vr}^\dag\Big(
	f_{K,B,\vr} + f_{K,B,\vr+\vet} + f_{K,B,\vr-\veo}
	\Big) + h.c.
\end{align}

\subsubsection{\texorpdfstring{Hopping between $\mathrm{tri}_\alpha$ orbitals}{Hopping between tri orbitals}}
The hopping between the $p_z$, $p_+$ and $p_-$ orbitals on triangular lattice sites is depicted on the top right of Fig.~\ref{fig:unitcell}(b).
The corresponding $K$-valley Hamiltonian is
\begin{align}
	H_{tt}^{(K)}
	&=
	-t_0\sum_{\alpha,\beta=z,\pm}
	\sum_\vr
	f_{\alpha,\vr}^\dagger \Big(
	t_{\alpha\beta}^{(1)}f_{\beta,\vr+\vet}
	+
	t_{\alpha\beta}^{(2)}f_{\beta,\vr-\veo-\vet}
	+
	t_{\alpha\beta}^{(3)}f_{\beta,\vr+\veo}
	\Big) +h.c.,
\end{align}
where 1, 2, 3 label the different tunnelling directions and
the inter-triangular lattice hopping parameters are $3\times3$ matrices given by
\begin{align}
	t^{(1)}
	&=
	\begin{pmatrix}
		-a^2	&	-iab	&	-iac	\\
		-iac	&	bc	&	c^2	\\
		-iab	&	b^2	&	bc
	\end{pmatrix},
	&
	t^{(2)}
	&=
	U_r^\dag t^{(1)} U_r,
	&
	t^{(3)}
	&=
	U_r t^{(1)} U_r^\dag.
\end{align}
Above we defined $U_{r}=\mathrm{diag}(1,\omega^*,\omega)$; here and below $\omega=e^{2\pi i/3}$.
The matrices $t^{(\ell)}$, $\ell=1,2,3$ and $U_r$ are expressed in a basis with the $p$ orbital ordering given in Eq.~\eqref{eqn:orbitalDef}.

\subsubsection{\texorpdfstring{Hopping between $\mathrm{tri}_\alpha$ and $\mathrm{hex}_A$ orbitals}{Hopping between tri and hexA orbitals}}

As shown to the bottom left of Fig.~\ref{fig:unitcell}(b), there are three types of nearest-neighbour tunnelling directions connecting orbitals on the A honeycomb sublattice to those on the triangular lattice.
We label these directions 1, 2, 3.
The Hamiltonian expressing these terms is
\begin{align}
	H_{t h_A}^{(K)}
	&=
	-t_0\sum_{\mu=\pm}\sum_\vr
	f_{K,A,\vr}^\dag \Big(
	x_\mu^{(1)} f_{K,\mu,\vr}
	+
	x_\mu^{(2)} f_{K,\mu,\vr+\veo+\vet}
	+
	x_\mu^{(3)} f_{K,\mu,\vr+\vet}
	\Big)+h.c.
\end{align}
The $x_\ell^{(\mu)}$ parameters are constrained by symmetry and can be decomposed according to
\begin{align}
	x_+^{(1)}
	&=
	d\,e^{-i\alpha}(\omega^* b+ c),
	&
	x_-^{(1)}
	&=
	d\,e^{-i\alpha}(b+\omega c),
	\nt
	x_+^{(2)}
	&=
	\omega^* x_+^{(1)},
	&
	x_-^{(2)}
	&=
	\omega x_-^{(1)},
	\nt
	x_+^{(3)}
	&=
	\omega x_+^{(1)},
	&
	x_-^{(3)}
	&=
	\omega^* x_-^{(1)}.
\end{align}

{\renewcommand{\arraystretch}{1.4}
\begin{table}
\centering
\begin{tabularx}{\textwidth}{BBBBB|BBBBB|B}
	\multicolumn{11}{c}{Tight-binding model parameters}
	\\\hline\hline
	$a$	&	$b$	&	$c$	&	$d$	&	$\alpha$    
	&
	$\bar\mu_z$	&	$\bar\mu_+$ &	$\bar\mu_-$	&	$\bar\mu_A$ & $\bar\mu_-$
	&
	$t_0$
	\\\hline
	$0.25$	&	$0.2$	&	$0.1$	&	$0.67$	&	$0$
	&
	$-0.043$	&	$0$	&	$0$	&	$0.05$	&	$0.05$
	&
	$80$ meV
\end{tabularx}
\caption{\AppCaptions
	Parameters defining five band model without SOC as per Ref.~\citenum{poFaithfulTightbindingModels2019}.}
\label{tab:TightBindingParam}
\end{table}}

\subsubsection{\texorpdfstring{Hopping between $\mathrm{tri}_\alpha$ and $\mathrm{hex}_B$ orbitals}{Hopping between tri and hexB orbitals}}

Finally, we consider the Hamiltonian describing the nearest-neighbour tunnelling between the B honeycomb orbitals and the orbitals on the triangular lattice sites.
As in the previous section, there are three directions and corresponding tunnelling constants labelled 1, 2, 3; see the image on the bottom right of Fig.~\ref{fig:unitcell}(b).
We find
\begin{align}
	H_{t h_B}^{(K)}
	&=
	-\sum_{\mu=\pm}\sum_\vr
	f_{K,B,\vr}^\dag
	\Big(
	y_\mu^{(1)} f_{K,\mu,\vr}
	+
	y_\mu^{(2)} f_{K,\mu,\vr+\veo}
	+
	y_\mu^{(3)} f_{K,\mu,\vr+\veo+\vet}
	\Big)+h.c.
\end{align}
with
\begin{align}
	y_+^{(1)}
	&=
	d\,e^{i\alpha}(\omega^* b+ \omega c),
	&
	y_-^{(1)}
	&=
	d\,e^{i\alpha}(\omega^* b+\omega c),
	\nt
	y_+^{(2)}
	&=
	\omega^* y_+^{(1)},
	&
	y_-^{(2)}
	&=
	\omega y_-^{(1)},
	\nt
	y_+^{(3)}
	&=
	\omega y_+^{(1)},
	&
	y_-^{(3)}
	&=
	\omega^* y_-^{(1)}.
\end{align}

\subsubsection{Complete Hamiltonian}

\begin{figure}
	\centering
	\includegraphics[width=0.99\textwidth]{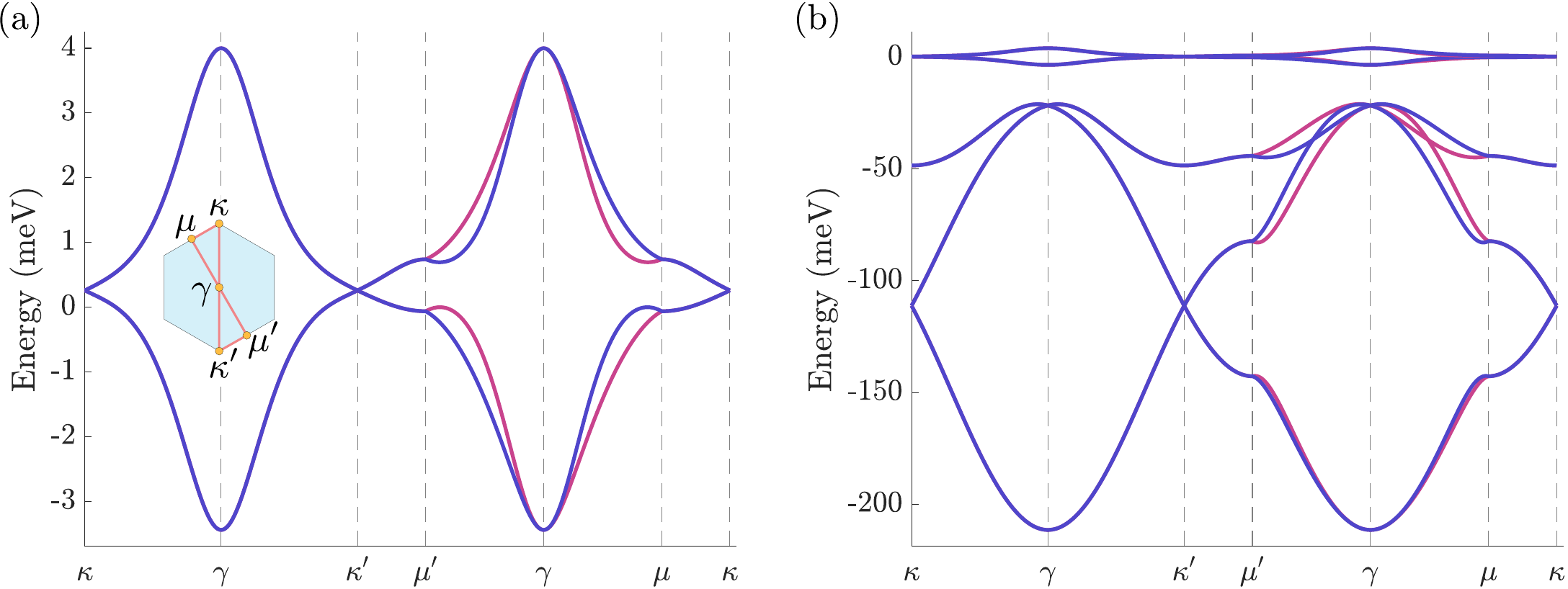}
	\caption{\AppCaptions
	Electronic structure of five band model without SOC defined in Sec.~\ref{sec:5bndSOC} along the linecut shown in the inset of (a).
	The parameters used are given in Table~\ref{tab:TightBindingParam}. 
	Blue and magenta lines respectively plot the energies of valley $K$ and $K'$.
	(a) The flat bands.
	(b) All bands represented within the five band band model.}
	\label{fig:5bnd_linecut}
\end{figure}

The full Hamiltonian is obtained by assembling the terms of the previous sections and including some orbital dependent chemical potential shifts:
\begin{align}
	H_0
	&=
	H_0^{(K)}+H_0^{(K')},
	\nt
	H_0^{(v)}
	&=
	H_{hh}^{(v)}+H_{t h_A}^{(v)}+H_{t h_B}^{(v)}+H_{tt}^{(v)}
	+H_{dg,1}^{(v)} + H_{dg,2}^{(v)},
\end{align}
where $v=K,K'$.
The $H_{dg,1}^{(v)}$ and $H_{dg,2}^{(v)}$ pieces read
\begin{align}
    H_{dg,\ell}^{(v)}
    =
    \sum_\vr f_{v,\vr}^\dag h_{dg,\ell} f_{v,\vr}.
\end{align}
The matrices are the same for both valleys and given by
\begin{align}
	h_{dg,1}
	&=
	-t_0\mathrm{diag}\Big( 6a^2, \,3(b^2+c^2), \,3(b^2+c^2),\,3d^2,\,3d^2 \Big),
	&
	h_{dg,2}
	&=
	t_0\mathrm{diag}\Big( \bar\mu_z, \,\bar\mu_+, \,\bar\mu_-, \,\bar\mu_A, \,\bar\mu_B \Big) .
\end{align}
Table~\ref{tab:TightBindingParam} lists the parameter values that we use---which are taken from Ref.~\citenum{poFaithfulTightbindingModels2019}---and Fig.~\ref{fig:5bnd_linecut} shows the resulting band structure for  valley $K$.
Once again, the Hamiltonian $H_0^{(K')}$ may be obtained directly from $H_0^{(K)}$ via Eq.~\eqref{eqn:5bndTRS}.
By construction, the flat bands are \emph{completely} flat when $\bar\mu_z=\bar\mu_+=\bar\mu_-=\bar\mu_A=\bar\mu_B=0$,

\subsection{Spin-orbit coupling}\label{sec:5bndSOC}

We add spin-orbit coupling by finding terms that satisfy the symmetries of Rashba and Ising SOC given in Table~\ref{tab:OrdParamSyms}.
Since the weight of the two flat bands resides predominantly on the $p_\pm$ orbitals at the hexagon centres \cite{poFaithfulTightbindingModels2019}, we focus on terms involving these orbitals.

\subsubsection{Rashba SOC}

The simplest Rashba term involving the $p_\pm$ orbitals that satisfies the same symmetries as in Table~\ref{tab:OrdParamSyms} is
\begin{align}
	H_{R,0}^{(K)}&=
	\frac{\lambda_{R,0}}{2}
	\sum_\vr e^{i\varphi_0} f_{K,+,\vr}^\dag \big( s^x - i s^y \Big) f_{K,-,\vr} + h.c.
\end{align}
The Hamiltonian corresponding to the other valley $H^{(K')}_{R,0}$ as usual follows from time reversal.  
In order to preserve mirror symmetry (specifically, $is^x\M_y^*$), the phase must vanish: $\varphi_0=0$.

We further consider two symmetry-allowed Rashba terms involving nearest-neighbour hopping between the $p_+$ and $p_-$ orbitals:
\begin{align}\label{eqn:5bnd_RashbaInterSite}
 	H_{R,1}^{(K)}
 	&=
 	-
 	\frac{\lambda_{R,1}}{3}
 	\sum_\vr
 	e^{i\varphi_1}
 	f_{K,+,\vr}^\dag 
 	s^x
 	\Big[
 		f_{K,-,\vr+\vet}
 		+
 		\omega^* e^{2\pi i s^z/3}
 		f_{K,-,\vr-\veo-\vet}
 		+
 		\omega e^{-2\pi i s^z/3}
 		f_{K,-,\vr+\veo}
 	\Big]
 	+h.c.
 	\nt
 	H_{R,2}^{(K)}
 	&=
 	-
 	\frac{\lambda_{R,2}}{3}
 	\sum_\vr
 	e^{i\varphi_2}
 	f_{K,-,\vr}^\dag 
 	s^x
 	\Big[
 		f_{K,+,\vr+\vet}
 		+
 		\omega^* e^{2\pi i s^z/3}
 		f_{K,+,\vr-\veo-\vet}
 		+
 		\omega e^{-2\pi i s^z/3}
 		f_{K,+,\vr+\veo}
 	\Big]
 	+h.c.
\end{align}
These terms also satisfy the symmetries of Table~\ref{tab:OrdParamSyms}, and the term corresponding to the other valley may similarly be obtained by applying time reversal; $is^x\M_y^*$ is only satisfied when $\varphi_1=\varphi_2=0$.

\subsubsection{Ising SOC}

\begin{figure}
	\centering
	\includegraphics[width=0.99\textwidth]{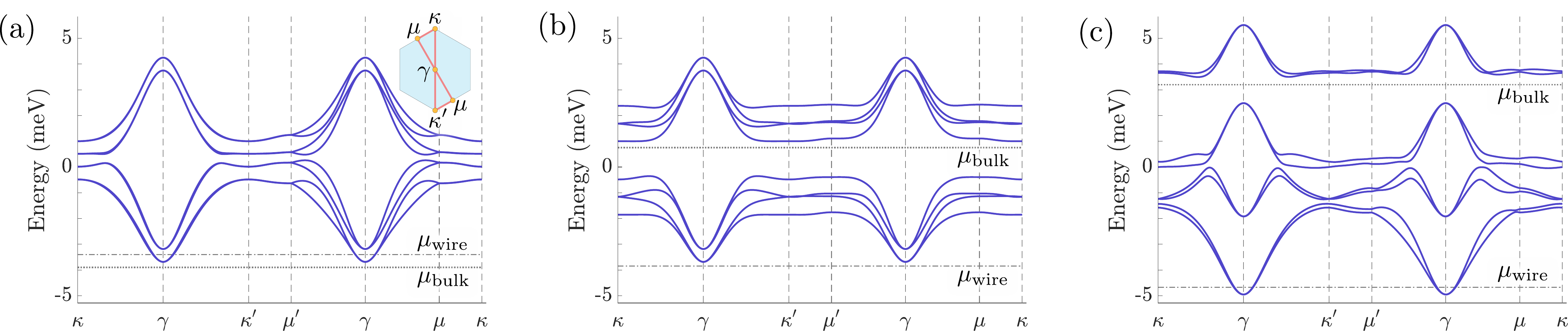}
	\caption{\AppCaptions
	Bulk band structures of trivial state (a), sIVC ordered insulator (b), and AFM IVC-ordered insulator (c).
	The parameters used are provided in Table~\ref{tab:TightBindingParam} and the appropriate columns of Table~\ref{tab:TightBindingParam-2}.
	The dotted and dash-dotted lines respectively denote the bulk and wire chemical potentials ($\mu_\mathrm{bulk}$ and $\mu_\mathrm{wire}$) used in the wire simulation.
	The momentum linecut used to plot the energies is shown in the inset of (a).
	}
	\label{fig:5bnd_bulkenergies}
\end{figure}

At the level of the microscopic graphene lattice, Ising SOC acts simply as a chemical potential whose sign depends on the spin and valley.
It can therefore be translated in a straightforward manner to the five band model as
\begin{align}\label{eqn:5bnd-Ising}
	H_I
	&=
	H_I^{(K)}+H_I^{(K')}
	=
	\frac{\lambda_I}{2}
	\sum_\vr 
	\big(
		f_{K,\vr}^\dag s^z f_{K,\vr}
		-
		f_{K',\vr}^\dag s^z f_{K',\vr}
	\big),
\end{align}
where the orbital and spin indices have both been suppressed.
Unlike the Rashba terms, this expression involves all five orbitals, a choice based largely on convenience.
The Ising contribution as written preserves the mirror symmetry $\M_y^*$.
A mirror-breaking Ising term involving the $p_+$ and $p_-$ orbitals does exist, but requires inter-site hopping (similar to Eq.~\eqref{eqn:5bnd_RashbaInterSite}); we do not consider it here.

\subsubsection{Total SOC contribution}

The total contribution of the SOC terms to the five band Hamiltonian that we consider is
\begin{align}
	H_\mathrm{SOC}
	&=
	\sum_{v=K,K'}\Big(
		H_{R,0}^{(v)}+H_{R,1}^{(v)}+H_{R,2}^{(v)}
		+
		H_I^{(v)}
	\Big).
\end{align}
Figure~\ref{fig:5bnd_bulkenergies}(a) shows the flat band energies with the choice of SOC parameters given on the left side of Table~\ref{tab:TightBindingParam-2}. 
As expected, the twofold spin degeneracy of Fig.~\ref{fig:5bnd_linecut} is lifted by the SOC.

\subsection{IVC order}\label{sec:5bndIVC}

The IVC terms may be obtained in exactly the same fashion as the SOC terms considered above: we find the simplest expressions involving $p_\pm$ that satisfy the appropriate symmetries of Table~\ref{tab:OrdParamSyms}.
For simplicity, here we focus on the singlet IVC (sIVC) case at $\nu=0$ and antiferromagnetic IVC (AFM IVC) case at $\nu=\pm2$.

\subsubsection{Singlet IVC order}

As was the case for Rashba and Ising SOC, there exists a simple onsite term involving the $p_\pm$ orbitals and satisfying the correct symmetries:  
 \begin{align}
 	H_\mathrm{sIVC}
 	&=
 	\frac{\Delta_\mathrm{sIVC}}{2}
 	\sum_{\vr}e^{i\phi_0}
 	\Big[
 		f_{K,+,\vr}^\dagger f_{K^\prime,+,\vr}
 		-
 		f_{K,-,\vr}^\dagger f_{K^\prime,-,\vr}
 	\Big]
 	+h.c.
 \end{align}
 Notably, this term breaks the $\mathrm{U_v}(1)$ symmetry and time reversal, but preserves $\tilde{\mathcal{T}}=i\tau^y\mathcal{K}$.
 While the Hamiltonian is equivalent by definition for every value of $\phi_0$, its value alters which $\mathrm{U_v}(1)$ valley operations must be composed with the physical discrete symmetries.
 The choice consistent with Table~\ref{tab:OrdParamSyms} is $\phi_0=0$.

 Figure~\ref{fig:5bnd_bulkenergies}(b) shows the electronic structure with  $\Delta_\mathrm{sIVC}=\unit[3.0]{\text{meV}}$ and the SOC parameters defined on the left hand side of Table~\ref{tab:TightBindingParam-2}.
 A clear gap at $\nu=0$ is observed.
{\renewcommand{\arraystretch}{1.4}
\begin{table}[t]
\centering
\begin{tabular*}{\textwidth}{@{\extracolsep{\fill} }cccccc|c||c||ccc}
	\multicolumn{7}{c||}{Spin-orbit coupling}
	&
	\multicolumn{1}{c||}{sIVC}
	&
	\multicolumn{3}{c}{AFM IVC}
	\\\hline\hline
	$\lambda_{R,0}$	&	$\varphi_0$
	&
	$\lambda_{R,1}$	&	$\varphi_1$
	&
	$\lambda_{R,2}$	&	$\varphi_2$
	&
	$\lambda_I$
	&
	$\Delta_\mathrm{sIVC}$
	&
	$\Delta_{\mathrm{tIVC},xy}$	&	$\lambda_I^{\mathrm{(MF)}}$
	&   $\frac{1}{2}\mu_BgB$
	\\\hline
	$1.5$~meV	&	$\pi/8$
	&
	$0.1$~meV	&	$\pi/6$
	&
	$-0.25$~meV	&	$-\pi/7$	
	&
	$-0.5$~meV
	&
	$3.0$~meV
	&
	$4.0$~meV	&	$-2.5$~meV  &   $0.23$~meV
\end{tabular*}
\caption{\AppCaptions
	Spin orbit parameters (Sec.~\ref{sec:5bndSOC}) and IVC mean field strengths (Sec.~\ref{sec:5bndIVC}) chosen for five band model simulations of the wire. The resulting band structures and wire profiles are shown in Fig.~\ref{fig:WireSim}.}
\label{tab:TightBindingParam-2}
\end{table}}

 \subsubsection{Antiferromagnetic IVC order}
 
 As discussed in Sec.~\ref{sec:AFM-IVCorder}, the AFM IVC state is obtained by considering \emph{two} order parameters. 
 The first is simply the Ising Hamiltonian of Eq.~\eqref{eqn:5bnd-Ising}.
 To distinguish this interaction-induced term from the TMD-induced term, we denote the strength of the former mean field Ising SOC strength by $\lambda_I^\mathrm{(MF)}$.
 The second order parameter needed to describe the AFM IVC insulator corresponds to two in-plane IVC triplet orders.
 We consider
  \begin{align}
 	H_{\mathrm{tIVC},xy}
 	&=
 	\frac{\Delta_{\mathrm{tIVC},xy}}{4}
 	\sum_{\vr}e^{i\phi_{xy}}
 	\Big[
 		f_{K,+,\vr}^\dagger (s^x-is^y)f_{K^\prime,+,\vr}
 		-
 		f_{K,-,\vr}^\dagger (s^x-is^y)f_{K^\prime,-,\vr}
 	\Big]
 	+h.c.
 \end{align}
The phase choice consistent with the analysis of Sec.~\ref{sec:AFM-IVCorder} is obtained when $\phi_{xy}=0$.
This state preserves the physical time reversal symmetry $\mathcal{T}=is^y\Tns$ and thus necessarily possesses degeneracy at $k=0$, as in Fig.~\ref{fig:ExampleBands}(i) of the main text.
We lift this degeneracy through the application of an in-plane magnetic field $\v{B}=B(\cos\theta_B,\sin\theta_B)$ whose primary effect is as a Zeeman field:
\begin{align}
	H_Z
	&=
	\frac{1}{2}g\mu_BB\sum_{v=K,K'}\sum_\vr
	f_{v,\vr}^\dag
	\big( \cos\theta_B s^x + \sin\theta_B s^y\big)
	f_{v,\vr}.
\end{align}
We restrict our study to $\theta_B=0$.
In Fig.~\ref{fig:5bnd_bulkenergies}(c), the electronic structure with $\lambda_I^\mathrm{(MF)}=\unit[-2.5]{\text{meV}}$,  $\Delta_{\mathrm{tIVC},xy}=\unit[4.0]{\text{meV}}$, and $\frac{1}{2}g\mu_BB=\unit[0.23]{\text{meV}}$ is shown.
A clear gap is seen at $\nu=+2$.
These parameters are recorded in Table~\ref{tab:TightBindingParam-2} and are used in Sec.~\ref{sec:5bnd_wiresim} in the wire simulations.

 \subsection{Wire setup and simulation}\label{sec:5bnd_wiresim}

 \begin{figure}
 \centering
 \includegraphics[width=0.99\textwidth]{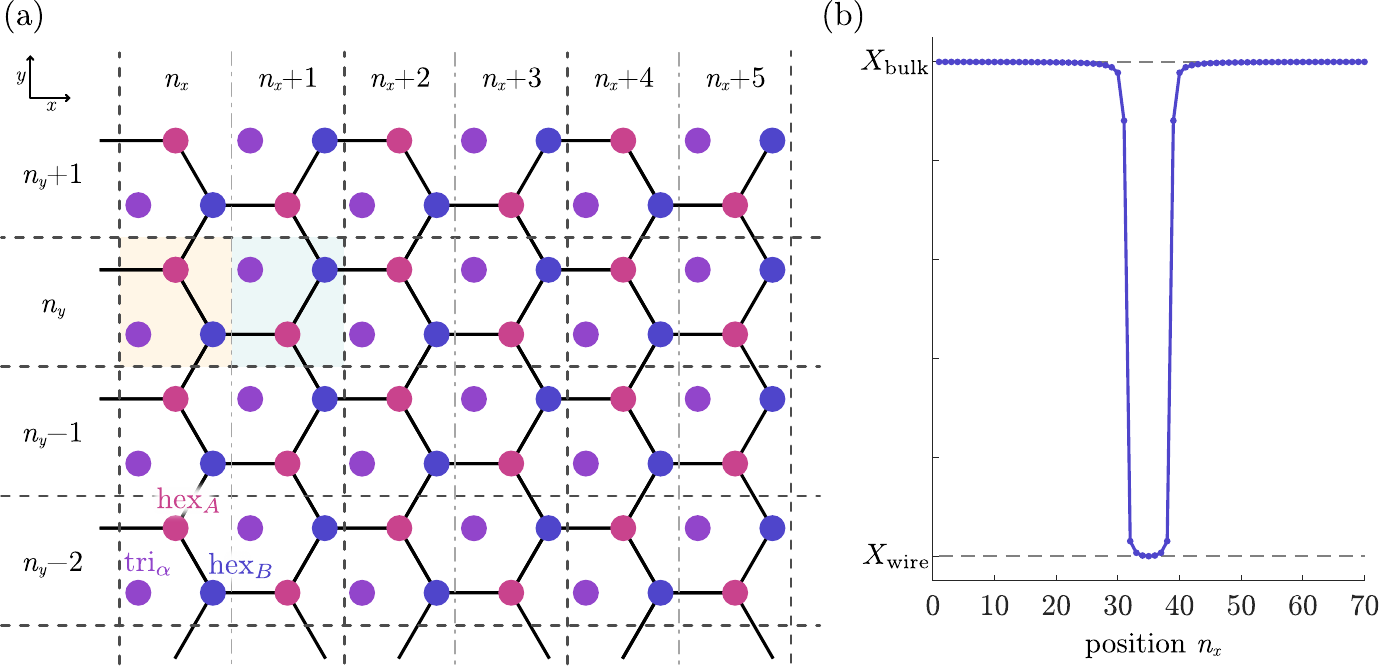}
 \caption{\AppCaptions
	 (a)~Real space representation of wire set up. 
 The orbitals within the moiré lattice are shown using the same colour scheme as in Fig.~\ref{fig:unitcell}(a).
 To simulate the wire, a unit cell that was doubled in the $x$ direction was used; they are outlined with dashed lines.
The two halves of the unit cell, distinguished by even or odd $n_x$, are divided by a dash-dotted line and two such subcells are highlighted in orange and blue for $(n_x,n_y)$ and $(n_x+1,n_y)$, respectively.
(b)~The wire profile chosen for the different varying parameters as a function of position. 
Periodic boundary conditions in which $n_x=71$ is identified with $n_x=1$.
Here, $X_\mathrm{wire/bulk}$ may refer to the chemical potential $\mu$ or the IVC mean field parameters.
For the trivial wire, $\mu_\mathrm{bulk}<\mu_\mathrm{wire}$ and so the true profile would require flipping the vertical axis direction.
Although the different sites within a unit cell $n_x$ are situated a different $x$ coordinates, these fine details were not taken into account: the value of $X$, $X=\mu,\Delta_\mathrm{sIVC},\Delta_{\mathrm{tIVC},xy}$, and/or $\lambda_I^\mathrm{(MF)}$ used was determined solely by $n_x$.
}
 \label{fig:5bndwiresetup}
 \end{figure}
A real-space illustration of the wire setup is shown in Fig.~\ref{fig:5bndwiresetup}(a). 
We arbitrarily chose to consider a wire orienting along the $y$-direction, 
which implies that the Hamiltonian can be written in terms of conserved momenta $k_y$.
This wire configuration requires that we double the unit cell in the $x$ direction. 
As a result, a single unit cell is defined by integers $n_x$ and $n_x+1$. 
Full unit cells are outlined by dashed lines 
in Fig.~\ref{fig:5bndwiresetup}(a), while dash-dotted lines divide the two subcells.
We emphasize again that the hexagonal unit cells depicted in Fig.~\ref{fig:5bndwiresetup}(a) represent \emph{moiré} unit cells that are themselves composed of  $\sim10\,000$ graphene unit cells.

The wire is simulated in a fashion directly analogous to the experimentally proposed setup: we impose a spatially varying chemical potential $\mu(n_x)$ whose values interpolate between the fillings under study.
Note that the spatial separation of the different orbitals within a unit subcell labelled by $n_x$ are not taken into account---any features dependent on such small scale distinctions cannot be trusted, as the five band model fails precisely in such situations (see Sec.~\ref{sec:5bnd_disc}).
Our simulations below assume a system with total length $70$ in the $x$ direction, with periodic boundary conditions. 
Figure~\ref{fig:5bndwiresetup}(b) displays how the chemical potential varies between its bulk and wire values, $\mu_\mathrm{bulk}$ and $\mu_\mathrm{wire}$, in our simulations. 
The precise values of the chemical potential are tuned to return the desired results, as would be done experimentally.

While interactions in the bulk are assumed to generate spontaneous breaking of symmetries defining the sIVC and AFM IVC phases, we input these orders by hand at a mean field level.
The generation of these orders is then treated analogously to the chemical potential.
For instance, for the sIVC phase, $\Delta_\mathrm{sIVC}(n_x)$ interpolates between its mean field value (see Table~\ref{tab:TightBindingParam-2}) in the bulk and zero within the wire using the same profile shown in Fig.~\ref{fig:5bndwiresetup}(b).
This setup differs from a scenario in which the two IVC phases adjoining the wire are not connected in any way.
Arguably, it is possible that the arbitrary $\mathrm{U_v}(1)$ phases could differ on either side of the wire. 
We assume that a valley Josephson effect locks those two phases together, resulting in the system considered here.

\subsubsection{Trivial wire simulations}\label{sec:5bnd_trivialwire}

\begin{figure}
	\centering
	\includegraphics[width=0.99\textwidth]{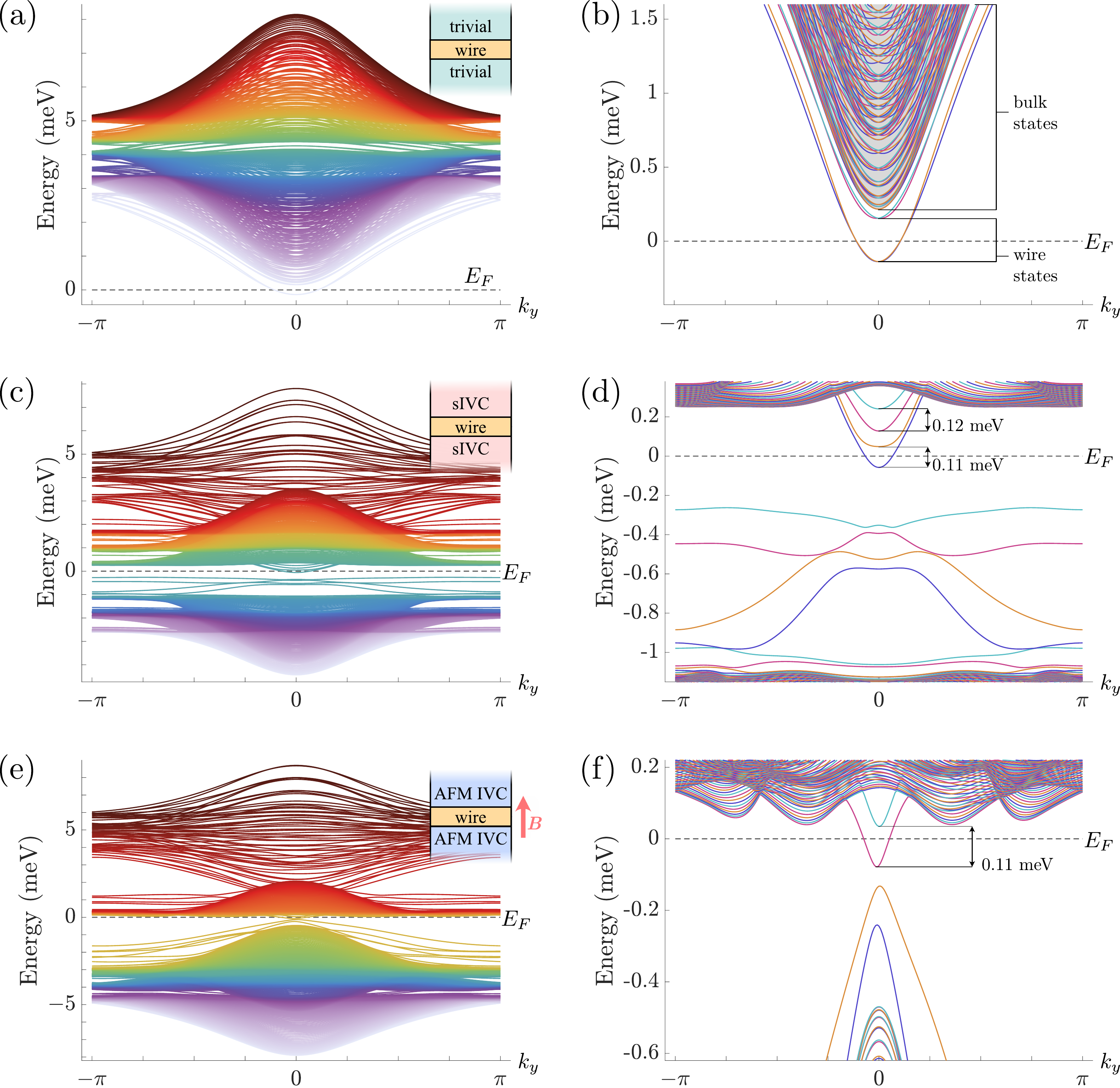}
	\caption{\AppCaptions
	Five band model simulations of the trivial wire (a),(b), the sIVC-proximitized wire (c),(d), and the AFM IVC-proximitized wire (e),(f).
    The five band model is defined through the parameters of Table~\ref{tab:TightBindingParam}, while the SOC and IVC order strengths are as given in Table~\ref{tab:TightBindingParam-2}.
    (a),(c),(e) The electronic structure of the electrostatically defined wire as a function of the conserved momentum $k_y$.
    The Fermi energy $E_F$ is indicated by the dashed line.
	An illustration of the wire setup in each case is shown in the inset.
	Note that the wire simulated in the five band model extends in the $y$ direction, which is the \emph{vertical} direction of Fig.~\ref{fig:5bndwiresetup}(a).
    (b),(d),(f) A zoomed in view around the Fermi energy of (b),(e),(h).
	The bulk and wire states are labelled in (a), and a grey background is added to further clarify the distinction between the two.
    The odd-channel gaps are labelled in (d),(f).
	}
	\label{fig:WireSim}
\end{figure}

The trivial wire is obtained by tuning the chemical potential so that the flat bands within the bulk are completely empty, $\nu_\mathrm{bulk}=-4$, while the chemical potential in the wire region sits just above the band bottom, yielding a filling $\nu_\mathrm{wire}=-4+\delta$ for $\delta$ small.
The values of $\mu_\mathrm{bulk}$ and $\mu_\mathrm{wire}$ considered here are shown relative to the corresponding bulk band structure in Fig.~\ref{fig:5bnd_bulkenergies}(a).
Figures~\ref{fig:WireSim}(a) and~(b) plot the resulting the energy levels of the trivial wire as a function of the conserved momentum $k_y$.
The dashed line at zero energy in both images displays the Fermi energy $E_F$.
In Fig.~\ref{fig:WireSim}(a), all flat band energy states are shown.
To see the confined wire states, Fig.~\ref{fig:WireSim}(b) zooms in on the region close to the Fermi energy.
The wire and bulk states are each labelled in the plot, where the wire states have been identified using the spatial information contained in the wavefunctions.
We have further shaded the bulk states in grey, since there is substantial energetic overlap with some of the wire states.

We can compare the band structure of Fig.~\ref{fig:WireSim}(b) with the phenomenologically obtained effective Hamiltonian recorded in Table~\ref{tab:WireTable} as well as with Fig.~\ref{fig_architecture}(a) of the main text.
We observe that the wire states appear in two nearly degenerate sets.
The first, lowest energy pair of wire states---with negative energies close to $k_y=0$---clearly stand out.
The other set of states appears at slightly higher energies and is almost energetically indistinguishable from the bulk states.
The energetic separation between these two pairs of states follows directly from SOC, which can be ascertained by comparison with Fig.~\ref{fig:5bnd_bulkenergies}(a): at the band bottom, close to $\gamma$, SOC has lifted the spin degeneracy and, further, the chemical potential $\mu_\mathrm{wire}$ is tuned so that it only intersects the lower pair of states.
The SOC, however, was not solely responsible for the lifting of the degeneracy of the wire states in the effective Hamiltonian: the valley-orbit term of Eq.~\eqref{eqn:Ham-1bnd-NoSOC-2d-kexpansion-Odd} is also capable of lifting the degeneracy away from $k_y=0$.
While a detailed study of the states in Fig.~\ref{fig:5bndwiresetup}(b) does reveal splitting, it is clear that the valley-orbit term is very small in our five band simulation.
We can understand the depression of this term through our choice of wire direction.
In particular, for a wire along the $y$ direction, the valley-orbit term vanishes completely in the absence of SOC as a result of the mirror symmetry, as was addressed in the discussion below Eq.~\eqref{eqn:TrivialWireHam} in Sec.~\ref{sec:WireHamNoSOC}.

\subsubsection{sIVC-coupled wire simulations}\label{sec:5bnd_sIVCwire}

We now focus on Figs.~\ref{fig:WireSim}(c) and~(d), which display the flat band energies of an sIVC-proximitized wire; Fig.~\ref{fig:WireSim}(c) shows all flat band energies, while Fig.~\ref{fig:WireSim}(d) provides a zoomed in view on the energy window close to the Fermi level where the confined wire states are expected to appear.
In this case, the wire is defined by interpolating the chemical potential between the values of $\mu_\mathrm{bulk}$ and $\mu_\mathrm{wire}$ shown in Fig.~\ref{fig:5bnd_bulkenergies}(b) \emph{and} by interpolating the sIVC strength $\Delta_\mathrm{sIVC}$ between $\unit[3.0]{\text{meV}}$ in the bulk and zero in the wire.
Both parameters follow the profile shown in Fig.~\ref{fig:5bndwiresetup}(b).
We observe that two odd-channel gaps of magnitude $\sim\unit[0.1]{\text{meV}}$ are obtained---consistent with our previous symmetry-based analysis.
This image should be compared against Fig.~\ref{fig:ExampleBands}(c) of the main text.
Unlike the scenario shown there, the simulated wire here does not realize a non-monotonic conductance.
We can in part attribute this aspect of the simulation to our choice of wire direction, which suppresses the valley orbit term.

Two related features may be noted in this system.
Firstly, in addition to the confined modes of interest, a number of subgap states are also present in Figs.~\ref{fig:WireSim}(c) and~(d).
Following the arguments at the end of Sec.~\ref{sec:sIVCWireHam} and in Sec.~\ref{sec:IVCTopoAspects}, these states originate from the topological nature of the sIVC state, which may be viewed as a 2$d$ topological insulator protected by the emergent time reversal-like symmetry $\tilde{\mathcal{T}}=i\tau^y\mathcal{K}$ when SOC is not present.
Here, inter-mode scattering mediated by the SOC promotes a gap, as does proximity of the two sets of edge modes arising from the two sIVC phases neighbouring the wire.
(Note that some degree of intervalley scattering may be present in physical system with a sufficiently steep chemical potential gradient; such effects are not included in our model.)
It's nevertheless clear that the gaps that arise are smaller than the bulk gap, resulting in the subgap states observed in Fig.~\ref{fig:WireSim}(d).
We have accordingly verified that the corresponding wavefunctions are spatially localized to the wire.
As outlined in Sec.~\ref{sec:IVCTopoAspects} (see also Sec.~\ref{sec:TrivialWireC3MyBreaking}), these states may also be used to obtain the odd channel regime necessary to realize Majorana zero modes. 
Although not relevant for the parameters used to generate Figs.~\ref{fig:WireSim}(c) and~(d), it is not difficult to tune $\mu_\mathrm{bulk}$ and $\mu_\mathrm{wire}$ to regions where the subgap states yield substantial odd channel regimes.

\subsubsection{AFM IVC-couple wire simulations}

The AFM IVC-proximity coupled wire is simulated in a fashion largely analogous to the trivial and sIVC cases above.
Unlike the sIVC state, the AFM IVC insulator preserves the physical time reversal symmetry $\mathcal{T}=is^y\Tns$, meaning that a magnetic field must be applied to realize the odd channel regime needed to obtain Majorana zero modes.
We consider a magnetic field perpendicular to the wire ($x$ direction) with strength $\frac{1}{2}g\mu_BB=\unit[0.23]{\text{meV}}$ (which corresponds to $B=\unit[4.0]{\text{T}}$ when $g=2$).
The bulk and wire are now distinguished by three parameters: the chemical potential $\mu_{\mathrm{bulk/wire}}$, the in-plane tIVC order strength $\Delta_{\mathrm{tIVC},xy}$, and the `mean field' Ising term $\lambda_I^{(MF)}$.
Following the profile of Fig.~\ref{fig:5bndwiresetup}(b), the latter two parameters, $\Delta_{\mathrm{tIVC},xy}$ and $\lambda_I^{(MF)}$, interpolate between the values provided in Table~\ref{tab:TightBindingParam-2} in the bulk and zero in the wire.
Note that the TMD-induced Ising SOC is $\lambda_I=\unit[-0.5]{\text{meV}}$, which is present within both the bulk and wire.
The total Ising SOC parameter of the wire and bulk are thus $\lambda_{I,\mathrm{wire}}=\lambda_I=\unit[-0.5]{\text{meV}}$ and $\lambda_{I,\mathrm{bulk}}=\lambda_I+\lambda_I^\mathrm{(MF)}=\unit[-3.0]{\text{meV}}$.
In Fig.~\ref{fig:5bnd_bulkenergies}(c), the chemical potentials $\mu_\mathrm{bulk}$ and $\mu_\mathrm{wire}$ are shown in relation to the bulk AFM IVC band structure in the presence of a magnetic field.

The resulting flat band energies are shown in Fig.~\ref{fig:WireSim}(e), where the dashed line at zero energy indicates the Fermi energy.
Figure~\ref{fig:WireSim}(f) zooms in on the states close to the Fermi energy, showing two confined wire states descending from the bulk manifold. 
They are separated by a gap of $\unit[0.11]{\text{meV}}$.  
In this case, the remaining two modes are obscured by the bulk states.
As in the previous section, the AFM IVC state possesses pseudo-topological edge modes gapped by both backscattering and SOC.
Such states are present in Fig.~\ref{fig:WireSim}(f) as hole-like curves and provide an additional odd channel regime.

Comparing against the phenomenological band structure shown in Fig.~\ref{fig:ExampleBands}(j) of the main text, aside from the absorption of two of the wire modes into the bulk, the most salient distinction is the five band simulation's absence of non-monotonicities.
We again ascribe this difference to our choice of wire direction.

Although our model assumes that the system extends infinitely in the $y$-direction, in a physical sample, the AFM IVC insulator would possess gapless edge modes at all of its edges in the presence of time reversal symmetry.
A gap is only opened along the boundary by the applied magnetic---and the bulk chemical potential must be tuned to lie within this gap to have true localized Majorana zero modes residing at the sample boundary.

\subsection{Discussion} \label{sec:5bnd_disc}

The wire simulation of the five band model is limited in a number of respects.
Arguably, the most important issue is that we have not attempted to include the effects of interactions in even a mean field sense, instead opting to include the interaction-induced order parameters by hand.
It would be both interesting and useful to reproduce the wire simulations presented here, but with the IVC orders included in a self-consistent Hartree-Fock calculation.
Although one may be concerned about adapting the interacting Hamiltonian, $H_\mathrm{cont}+H_\mathrm{int}$, to the five band model, previous mean field calculations performed using phenomenological tight-binding models are largely in agreement with more sophisticated calculations on the continuum model, at least in the single flavour limit \cite{bultinckGroundStateHidden2020,LiuS21,choiElectronicCorrelationsTwisted2019}.

We also mention although our wire simulations do yield gaps, they are likely depressed compared to what one would find in a more general calculation.
One issue is that we chose the wire to lie along a highly symmetric direction.
For instance, as mentioned in Sec.~\ref{sec:5bnd_trivialwire}, the valley-orbit term vanishes at the order we consider for the $y$-directed wire studied here. 
Inspecting Table~\ref{tab:kexpansion}, we see that the mirror-symmetric, Ising SOC term, $k_x(k_x^2-3k_y^2)s^z$, also vanishes when projected onto the even parity wire states $\bar\phi_{k_y}(y)$ (obtained in a fashion identical to the function $\phi_{k_y}(x)$ of Sec.~\ref{sec:WireHamNoSOC}).
The remaining Ising SOC term, $k_y(3k_x^2-k_y^2)s^z$, only arises when the mirror symmetry is broken;
however, our choice of effective Ising SOC (Eq.~\eqref{eqn:5bnd-Ising}) preserves the mirror symmetry.
It would therefore be worthwhile to investigate the dependence of the band splittings as a function of wire direction.

A related issue is that the five band model cannot account for changes occurring within the moiré unit cell.
Throughout both the main text and this appendix, we have treated  $\mathrm{U_v}(1)$ as a proper symmetry of the theory.
In reality, this symmetry is only emergent and relies on the relative absence of scattering occurring at the scale of the graphene lattice.
A strong enough chemical potential gradient or inhomogeneities introduced by the gate's physical boundaries may explicitly break this symmetry, which would further promote the formation of the gaps necessary to realize the odd channel regime.
We emphasize, however, that this mechanism is not expected to dominate the physics---in fact, if the wire profile were able to induce substantial scattering between valleys, the odd channel regime needed to realize Majoranas could be realized without IVC order.


	{\renewcommand{\arraystretch}{1.3}
	\begin{sidewaystable}
	\centering
	\begin{tabular*}{\textwidth}{@{\extracolsep{\fill}}CCCCC c l c p{29.5ex}cp{29.5ex} c p{36ex}cp{36ex}}
		\hline\hline
		\multirow{2}{*}{{\rotatebox[origin=c]{90}{\parbox{9.1ex}{\scriptsize\raggedright trivial}}}}
		&\multirow{2}{*}{{\rotatebox[origin=c]{90}{\parbox{9.1ex}{\scriptsize\raggedright sIVC}}}}
		&\multirow{2}{*}{{\rotatebox[origin=c]{90}{\parbox{9.1ex}{\scriptsize\raggedright tIVC}}}}
		&\multirow{2}{*}{{\rotatebox[origin=c]{90}{\parbox{9.1ex}{\scriptsize\raggedright FM IVC}}}}
		&\multirow{2}{*}{{\rotatebox[origin=c]{90}{\parbox{9.1ex}{\scriptsize\raggedright AFM IVC}}}}&&
		\multirow{2}{*}{Order}	
		&\rule{0pt}{4ex}&
		\multicolumn{3}{c}{Even}
		&$\phantom{xz}$&
		\multicolumn{3}{c}{Odd}
		\\
		&&&&&&
		&\rule{0pt}{4ex}&
		$\M_y$ (subgroup) preserving &$\phantom{xz}$& $\M_y$ (subgroup) breaking	
		&&	$\M_y$ (subgroup) preserving    
		&$\phantom{xz}$& 
		$\M_y$ (subgroup) breaking \\
		\hline
		\colA 	&	\colB	&	\colC	&	\colD	&	\colE
		&&
		No SOC	
		&&  $\id$, $k_x^2+k_y^2$ &&	
		&&	$k_x(k_x^2-3k_y^2)\tau^z$
		&&	$\{k_y(3k_x^2-k_y^2)\tau^z\}$
		\\\hline
		\colA	&	\colB	&	\colC	&	\colD	&	\colE
		&&
		Rashba (R)	
		&&	$\tau^z\left(\frac{1}{2}\big(k_x^2-k_y^2\big)s^x-k_xk_ys^y\right)$	
		&&	$\tau^z\left(k_xk_ys^x+\frac{1}{2}\big(k_x^2-k_y^2\big)s^y\right)$
		&&	$k_xs^x+k_ys^y$
		&&	$k_ys^x-k_xs^y$
		\\
		\colA	&	\colB	&	\colC	&	\colD	&	\colE
		&&
		Ising (I)	&&
		$\tau^zs^z$	&&	
		&&	$k_x(k_x^2-3k_y^2)s^z$
		&&	$k_y(3k_x^2-k_y^2)s^z$
		\\\hline
			&	\colB	&		&	\colD	&	
		&&
		sIVC	&& 	&&
		&&	$k_y(3k_x^2-k_y^2)\tau^x$
		&&	$\{k_x(k_x^2-3k_y^2)\tau^x\}$
		\\
		&	\colB	&		&	\colD	&	
		&&
		sIVC + R	
		&& $\tau^x\left(k_xk_ys^x+\frac{1}{2}\big(k_x^2-k_y^2\big)s^y\right)$, $[\tau^ys^z]$	
		&&	$\tau^x\left( \frac{1}{2}(k_x^2-k_y^2)s^x - k_xk_y s^y \right)$	
		&&
		\\
		&	\colB	&		&	\colD	&	
		&&
		sIVC + I	
		&&
		&&	$\tau^xs^z$	
		&&	
		\\\hline
		&	&	\colC	&	\colD	&	
		&&
		tIVC	
		&& 	&&
		&& $k_y(3k_x^2-k_y^2)\tau^x s^z$
		&&	$\{k_x(k_x^2-3k_y^2)\tau^xs^z\}$
		\\
		&	&	\colC	&	\colD	&	
		&&
		tIVC + R	
		&& $[\tau^y]$	&&
		&& 	$\tau^y\big(k_xs^x+k_ys^y\big)$
		&&	$\tau^y\big(k_ys^x-k_xs^y\big)$
		\\
		&	&	\colC	&	\colD	&	
		&&
		tIVC + I	
		&&		&&	$\tau^x$ 
		&& 
		\\\hline
		&	&	&	\colD	&	
		&&
		FM IVC
		&&	$s^z$	&&
		&&	$k_x(k_x^2-3k_y^2)\tau^zs^z$,
			\newline
			$k_y(3k_x^2-k_y^2)\tau^x s^z$
		&&	$\{k_y(3k_x^2-k_y^2)\tau^zs^z\}$,
			\newline
			$\{k_x(k_x^2-3k_y^2)\tau^xs^z\}$
		\\
		&	&	&	\colD	&
		&&
		FM IVC + R
		&& 
		\\
		&	&	&	\colD	&
		&&
		FM IVC + I
		&&	$\tau^z$		&&
		&&	$k_x(k_x^2-3k_y^2)\id$
		&&	$k_y(3k_x^2-k_y^2)\id$
		\\\hline
		&	&	&	&	\colE
		&&
		AFM IVC
		&&	$\tau^zs^z$	&&
		&&
		{$k_x(3k_x^2-k_y^2)s^z$}, \newline$k_y(k_x^2-3k_y^2)\big(\tau^xs^x+\tau^ys^y\big)$
		&& $\{k_y(3k_x^2-k_y^2)s^z\}$,\newline $\{k_x(3k_x^2-k_y^2)\big(\tau^xs^x+\tau^ys^y\big)\}$
		\\
		&	&	&	&	\colE
		&&
		AFM IVC + R
		&&	{$\frac{1}{2}(k_x^2-k_y^2)\tau^x-k_xk_y\tau^y$}
		&&	 {$k_xk_y\tau^x+\frac{1}{2}(k_x^2-k_y^2)\tau^y$}
		&&
		$\tau^x(k_ys^x+k_xs^y)+\tau^y(k_xs^x-k_ys^y)$, 
		\newline$(k_y\tau^x-k_x\tau^y)s^z$
		&&
		$\tau^x(k_x s^x-k_ys^y)-\tau^y(k_y s^x+k_ys^y)$,
		\newline
		$(k_x\tau^x+k_y\tau^y)s^z$
\end{tabular*}
	\caption{\AppCaptions
	List of allowed terms to lowest order in $\vk$ away from the $\gamma$ point for each of the symmetry breaking scenarios shown in Table~\ref{tab:OrdParamSyms} (with the exception of the first line, in which all symmetries are preserved).
	The terms are sorted according to their parity and whether or not they respect $\M_y$/$\M_y^*$ or one of its subgroups.
    As discussed in Sec.~\ref{sec:MirrorSymScenario}, the $\M_y$ symmetry (or a subgroup) may be re-established for SOC, which is used to obtain the mirror action in the rows that include SOC.
    Without SOC, the IVC orders and trivial wire preserve the mirror symmetry or one of its subgroups and therefore we should strictly speaking exclude all $\M_y$-breaking terms from these rows.
    Their presence is no longer prohibited with the inclusion of SOC, which  explicitly breaks $\M_y$; however, this effect only occurs beyond first order in $\lambda_{R/I}$. 
    To emphasize these subtleties, we include such terms but indicate their subleading status through curly braces `$\{\cdot\}$.'
    Similarly, in the rows denoted ``sIVC + R'' and ``tIVC + R,'' the terms $\tau^zs^z$ and $\tau^y$ appear in brackets to indicate that, while allowed by symmetry, they only arise at a higher order in $\lambda_R$ and are therefore not included in Table~\ref{tab:WireTable}.
	The row labelled ``FM IVC + R'' is left intentionally blank as no new terms are added by these terms.
	The colours on the left-hand side indicate which rows should be included for the various phases.
	For instance, the dark blue bar only extends through the first three rows, implying that the terms listed in those rows are the only ones allowed within our symmetry analysis for the trivial wire scenario.
	Including the terms for each of these five scenarios and reducing the theory to 1$d$ as first described in Sec.~\ref{sec:TrivialWireHam} returns the results of Table~\ref{tab:WireTable}.
	The $\vk$-dependence of the terms listed no longer helpful when the system breaks the $\C_3$ or $\M_y$ either as a result of strain or interactions. 
	Nevertheless, both the allowed matrices $\tau^\mu s^a$ and the parity of the terms multiplying them ($t_{\mu,a}(\vk)$) are unchanged.
	}
	\label{tab:kexpansion}
	\end{sidewaystable}
	}

\end{document}